\documentclass[fleqn, useAMS,usenatbib]{mnras}
\usepackage{amssymb, graphicx, verbatim}
\usepackage{amsmath}
\usepackage{ulem}
\usepackage[T1]{fontenc}

\newcommand{\e}[1]{\times 10^{#1}}

\newcommand{\msun}{M$_\odot$}

\def\ni {$^{56}$Ni}

\def\co {$^{56}$Co}

\def\kms {km~s$^{-1}$}
\def\ergs {erg s$^{-1}$}

\newcommand{\arj}[1]{\textcolor{black}{{#1}}}
\newcommand{\arja}[1]{\textcolor{black}{{#1}}}

\title[Emission line models for the lowest-mass core collapse supernovae]{Emission line models for the lowest-mass core collapse supernovae. I: Case study of a 9 $M_\odot$ one-dimensional neutrino-driven explosion.}
\author[Anders Jerkstrand]{A. Jerkstrand$^{1}$\thanks{E-mail:anders@mpa-garching.mpg.de}, 
T. Ertl$^{1}$, H.-T. Janka$^{1}$, E. M\"uller$^1$, T. Sukhbold$^2$, S. E. Woosley$^3$\\
$^{1}$Max-Planck Institut f\"ur Astrophysik, Karl-Schwarzschild Str. 1, D-85748 Garching, Germany\\
$^{2}$Department of Astronomy and Center for Cosmology \& Astro-Particle Physics, The Ohio State University, Columbus, Ohio 43210, USA\\
$^{3}$Department of Astronomy and Astrophysics, University of California, Santa Cruz, CA 95064, USA
}
\begin{document}

\date{}

\pagerange{\pageref{firstpage}--\pageref{lastpage}} \pubyear{2017}

\maketitle

\label{firstpage}

\begin{abstract}
A large fraction of core-collapse supernovae (CCSNe), 30-50\%, are expected to originate from the low-mass end of progenitors with $M_{ZAMS}= 8-12$ \msun. However, degeneracy effects make stellar evolution modelling of such stars challenging, and few predictions for their supernova light curves and spectra have been presented.
Here we calculate synthetic nebular spectra of a 9 \msun~Fe CCSN model exploded with the neutrino mechanism. The model predicts emission lines with FWHM$\sim$1000 \kms, including signatures from each deep layer in the metal core. We compare this model to observations of the three subluminous IIP SNe with published nebular spectra; SN 1997D, SN 2005cs, and SN 2008bk. The prediction of both line profiles and luminosities are in good agreement with SN 1997D and SN 2008bk. The close fit of a model with no tuning parameters provides strong evidence for an association of these objects with low-mass Fe CCSNe. 
For SN 2005cs, the interpretation is less clear, as the observational coverage ended before key diagnostic lines from the core had emerged.
We perform a parameterised study of the amount of explosively made stable nickel, and find that none of these three SNe show the high $^{58}$Ni/$^{56}$Ni ratio predicted by current models of electron capture SNe (ECSNe) and ECSN-like explosions. Combined with clear detection of lines from O and He shell material, these SNe rather originate from Fe core progenitors.
We argue that the outcome of self-consistent explosion simulations of low-mass stars, which gives fits to many key observables, strongly suggests that the class of subluminous Type IIP SNe is the observational counterpart of the lowest mass CCSNe.
% DC: All ok
\end{abstract}


\section{Introduction} % ===========================================================

%\paragraph{CCSNe from the 8-11 range, overview}
Stars more massive than about 8 \msun~end their evolution as core-collapse supernovae. Due to the steepness of the initial mass function (IMF), over 40\% of all CCSNe occur in the 8-12 \msun~range and 25\% in the 8-10 \msun~range. If a significant fraction of stars over 20 \msun~would in addition fail to produce supernovae and collapse directly to black holes, the fraction in the 8-12 \msun\ range could be as high as 50-60\% \citep{Sukhbold2016}.

Stars in the 8-12 \msun~range have a qualitatively different core structure than at $>12$ \msun, with significantly lower compactness \citep{Sukhbold2016}. They are typically easier to explode, and some explode even in 1D neutrino simulations \citep{Kitaura2006, Melson2015,Radice2017}.
However, the evolution of such stars is significantly more complex to model than at $>$12 \msun~due to degeneracy effects \citep[see e.g.][for an early review]{Miyaji1980}. Thermal pulses develop that are numerically challenging to follow, and mass loss is highly uncertain in the later phases. There have therefore been few models presented in the literature. This has in turn led to few models for their corresponding supernova light curves and spectra, and currently a fuzzy picture which observed events belong in this category.

% Mass division points
The values of critical mass division points between final outcomes, between CO and ONe white dwarfs (WDs) ($M_{up}$), between ONe WDs and ECSNe ($M_n$), and between ECSNe and Fe CCSNe ($M_{mas}$) all shift by about 2 \msun~with uncertain overshooting efficiencies \citep{Siess2007}. There is also some dependency on metallicity \citep{Umeda1999,Eldridge2004}. ``Super-AGB'' stars represent the final phase for stars between $M_{up}$ and $M_{mas}$, that can end as either ONe WDs or ECSNe.

% M_up (OC WD --> ONe WD)
In the literature one finds values for $M_{up}$ of $<9$ \msun \citep{GarciaBerro1997}, $7.5-9$ \msun \citep{Poelarends2007}, $\le$8.9 \msun\ \citep{Siess2007}. There is close agreement in the necessary CO core mass needed to ignite carbon of 1.06 \msun.
% M_n (ONe WD --> ECSN)
Although the value of $M_n$ varies between studies (e.g. 8.7 \msun\ in \citet{Jones2013}), there is close agreement in the predicted ONe core mass needed to produce an ECSN; 1.37 \msun \citep{Nomoto1984,Takahashi2013}. 
% Lionel : Siess 2007 and Poaelarends 2008 do not compute 1.37 value
% Lionel : EC occurs because degenerate electrons have large Fermi energies --> requires degenerate conditions..but similarity to Chandrasekhar mass is coincidence. EC can occur on Ne and Mg but not C or O. Mg too rare to be important compared to Ne.
% M_mas (ECSN --> CCSN)
For $M_{mas}$, values have been reported of $\ge$ 8.8 \msun~\citep{Nomoto1984}, $\ge$11~\msun \citep{Ritossa1999, Takahashi2013}, $8.8-9.5$ \msun \citep{Jones2013}, 9.25 \msun \citep{Poelarends2008}, $\le$ 9 \msun~\citep{Woosley2015}, and 8.7 \msun \citep[][with overshooting]{Siess2007}.
% Ritossa1999 does not follow long enough?
% 
\citet{Woosley2015} also warn that the behaviour in this mass range is sometimes non-monotonic with mass, implying that such as grouping scheme may be oversimplified.

% ECSN progenitors
The mass range that theoretically leads to ECSNe, $M_n-M_{up}$, is relatively small, of order 1 $M_\odot$ \citep{Siess2007}. It is almost certainly not larger than 1.5 \msun, and may be much smaller than 1 \msun~ \citep{Poelarends2008,Doherty2015} or non-existent, depending on uncertainties in mass loss and 3d dredgeup. An interval width of 1 \msun\ still corresponds to 15\% of all CCSNe with a standard Salpeter IMF, dropping to 8\% for 0.5 \msun. An important current research topic is whether ECSNe exist at all, and if so what fraction of CCSNe they constitute.
% Lionel priv comm. : Mainly mass loss and 3d dredgeup that gives uncertainty in Mn-Mup.

%ECSN progenitors
Progenitor structures at the point of core collapse for ECSNe have been calculated by \citet{Nomoto1984, Nomoto1987, Takahashi2013, Jones2013}. Models halted earlier, but predicted to evolve to ECSNe, have been calculated by \citet{Ritossa1996,GarciaBerro1997,Iben1997,Ritossa1999,Siess2006,Woosley2015}.  In these models the He layer has typically been reduced to almost nothing by a 2nd dredge-up/dredge-out \citep{Nomoto1987,Ritossa1996}.
The ONe core is surrounded by a thin O/C shell of mass a few$\times$0.01 \msun, followed by a dilute H/He envelope. As the ONe core and part of the O/C shell collapse to a neutron star, the outer O/C shell gets burnt to iron-group elements \citep{Mayle1988,Kitaura2006,Janka2008}. 
Longer-term simulation of the neutrino wind \citep{Pllumbi2015} indicated no major modifications to the yields obtained after the explosive phase \citep{Wanajo2009,Wanajo2011}. An important property of ECSNe from single star progenitors is therefore that they have no O-rich shell in the ejecta, and also no He shell. The spectrum of such an event will be that of a H/He envelope, plus possibly signatures of explosively made iron-group elements. It is unclear whether the innermost Fe-rich layer is massive enough to be seen against the $\sim$8 \msun~H/He envelope emission. If it is, the unusually high neutron excess may produce unique signatures, e.g. of stable nickel and zinc \citep{Wanajo2009}.
Follow-up simulations in 2D demonstrated that for ECSNe dimensionality plays a relatively minor role for the dynamics but a more significant role for the nucleosynthesis \citep{Janka2008, Wanajo2011,Muller2016}.

%Fe CCSN progenitors.
For higher masses, neon may ignite degenerately in shell flashes \citep{Nomoto1984}. These flashes may or may not propagate to the center \citep{Jones2014}. If they do, the star eventually forms an iron core.
But O and Si also ignite quasi-degenerately, and it is unclear if these flashes may eject the outer layers of the star and maybe even prevent neutron star formation. Modern 1D models use subgrid flame propagation calculations, but a full understanding requires 3D simulations. These degenerate  flashes occur up to about 10.3 \msun, above which more normal non-degenerate evolution occurs \citep{Woosley2015} Recently a grid of iron core progenitors below 12 \msun~was presented by \citet{Woosley2015}, who obtained iron core formation down to 9 \msun.

%Explosion simulations of 8-11 range : Fe CCSNe
Similar to ECSNe, the lowest mass iron CCSNe have steep density gradients outside their degenerate cores,
and explode with similar energy and explosive nucleosynthesis \citep{Janka2012,Muller2016, Wanajo2017}, although multi-D effects are here more important for the explosion energy \citep{Melson2015,Radice2017}. This raises the question whether ECSNe and low-mass Fe CCSNe can be spectroscopically distinguished. The major distinction of the ejecta is that Fe core progenitors have layers of Si, O, and (perhaps) He that, although thin, will not be fully burnt to iron-group elements,
and may provide additional lines of Si, S, Ca, O, C, He to spectra. So ECSNe may be identified by \textit{lack} of lines rather than presence of lines.

% -----------------
The main observational candidates for low-mass CCSNe are Type II SNe with low plateau luminosities and low expansion velocities, so called ``subluminous Type IIP SNe''. This class has about a dozen reasonably well-observed members, including SN 1997D \citep{Turatto1998, Benetti2001}, SN 2003gd \citep{Hendry2005, Meikle2007}, SN 2003Z \citep{Spiro2014}, SN 2004eg \citep{Spiro2014},  SN 2005cs \citep{Pastorello2006, Pastorello2009}, 2006ov \citep{Spiro2014}, SN 2008bk \citep{vanDyk2012, Maguire2012}, SN 2008S \citep{Botticella2009}, and SN 2009md \citep{Fraser2011}. \citet{Pastorello2004} presented data on further candidates SN 1994N, SN 1999br, SN 1999eu and SN 2001dc.  Of these, SN 1997D, SN 2005cs, and SN 2008bk were spectroscopically observed at nebular phases (t$\gtrsim$200d). Recently, Tomasella et al. (in prep) presented data on two additional events, SN 2013am and SN 2013K, that have properties between subluminous and normal IIP SNe.
% 2008in is a transitional object in Roy 2011

Taking a more global view of the progenitors, strong mass loss is expected in the super-AGB phase, e.g. for ECSN progenitors. Thus, this kind of SNe may conceivably become Type IIn SN. It is also possible that ECSNe and low-mass Fe CCSNe occur from He stars that have lost their H envelopes to a companion. In that case a Type Ib/c SN follows that is fast and very faint due to the low ejecta mass of $\sim$ 0.1 \msun\ \citep{Tauris2015,Moriya2016}. The existence of such transients is difficult to establish due to the difficulty of detecting them.

%Progenitor imaging
Three of the subluminous IIP SNe (SN 2003gd, SN 2005cs, and SN 2008bk) have progenitor detections, and these have all pointed to values in the $M_{ZAMS}=8-15$ \msun~range. \citet{vanDyk2003} estimated $M_{ZAMS}=8-9$ \msun~ for SN 2003gd. The optical disapperance of the progenitor could later be confirmed \citep{Maund2009}. \citet{Maund2005} obtained $M_{ZAMS}=9_{-2}^{+3}$ \msun~for SN 2005cs, and \citet{Takats2006} found a similar estimate, with confirmed disappearance by \citet{Maund2014}. \citet{Eldridge2007} argued that upper limits in the NIR for SN 2005cs rule out a super-AGB progenitor, and thereby the possibility of an ECSN. \citet{Mattila2008} estimated $M_{ZAMS}=8.5 \pm 1$ \msun~for SN 2008bk. The progenitor was later confirmed to have  optically disappeared \citep{Maund2014}, with a revised progenitor mass estimate $M_{ZAMS}=13\pm 2$ \msun.
Note that \citet{Fraser2011} identified a source with  $M_{ZAMS} = 8.5_{-1.5}^{+6.5}$ \msun~for SN 2009md, but this was later shown not to have disappeared and was thus not the SN progenitor \citep{Maund2015}.

Thus in three cases have subluminous IIP SNe been linked to the low-mass end of progenitors, all with confirmed optical disappearances. While the formation of optically thick dust  clumps in SN ejecta may completely block out any surviving progenitor system in the optical, it is unlikely that this would have happened in all three cases as typically inferred covering factors are $\sim$0.5 \citep[e.g.][]{Lucy1989}. Any high-mass progenitor would also presumably have easily been detected, given the detection of low-mass stars (although extinction is again a problem). Progenitor imaging therefore provides strong arguments against a scenario of high-mass progenitors for this class.

%Models of light curves and spectra for 8-11 range
Light curve modelling has also been undertaken. The first subluminous IIP SN, SN 1997D, was only observed from the very end of the plateau, and the plateau light curve itself can therefore not significantly constrain explosion models. However, both the low and high-mass models initially proposed \citep{Turatto1998, Chugai2000} had short plateaus of 50-60d. Later on, SN 1997D-clones have been discovered that all have much longer plateaus ($\gtrsim 100$d), and it is therefore likely that these original models are unsuitable.

\citet{Chugai2000} demonstrated that at the conditions of high density and low \ni~mass, Rayleigh scattering has some effect on the photospheric spectral formation. Particularly for models with very high mass, this effect produced worsening of the spectral fits for SN 1997D, although the effects were quite minor. \citet{Zampieri2003} used a semi-analytic approach and favoured high ejecta mass (14-17 \msun) models for SN 1997D (now with assumed 110d plateau) and SN 1999br. 
\citet{Utrobin2008} found a high mass for SN 2005cs ($M_{pre-SN} = 17 \pm 2$ \msun) using advanced light curve models,
while \citet{Pastorello2009} found $M_{pre-SN} \sim 12$ \msun~with a semi-analytic approach.
\citet{Spiro2014} found $M_{pre-SN}=10$ \msun~for SN 2005cs and $M_{pre-SN}~\sim$ 15 \msun~for SN 2008in (a transitional object studied by \citet{Roy2011}). \citet{Pumo2017} performed modelling for SN 2003Z, SN 2008bk, and SN 2009md, estimating main sequence masses in the $11-15$ \msun~range. Thus, light curve modelling indicates $M_{ZAMS}\sim 10-20$ \msun, in dissonance with a hypothesis based on the progenitor imaging that these stars would come from 8-11 \msun~progenitors.

The light curve models described above do not simulate explosions of evolved progenitors, but use polytropes. Adding some focus on specific evolutionary models, 
\citet{Tominaga2013} computed light curves of thermal bomb explosions of H/He envelopes ($2.0-4.7$ \msun) attached to a ECSN progenitor core, using STELLA, and \citet{Moriya2014} extended this work to include CSM interaction. These two works pointed out that, with the very low \ni~mass, other energy sources such as a central pulsar or CSM interaction may dominate the light curve after a few hundred days.

% Spectral models
A third modelling approach is spectral modelling, which is the topic of this paper.
The last few years have seen the production of the first grids of spectral models for Type IIP SNe over $M_{ZAMS}$ \citep{J12,Dessart2013b,Jerkstrand2014}. As demonstrated e.g for SN 1987A \citep{Fransson1989,Kozma1998I,Kozma1998II}, models for the nebular phase of the SN allow diagnostics of the nucleosynthesis products, which produce emission lines that become distinct after about 200d. Constraints on these products allows inferrence of the progenitor, as well as on its explosive nucleosynthesis and mixing dynamics. However, these grids have so far only been computed down to $M_{ZAMS}=12$ \msun~\citep{J12, Jerkstrand2014, Lisakov2017}.
For application to subluminous IIP SNe, a 12 \msun~model has been shown to produce relatively good agreement with SN 2008bk \citep{Maguire2012,Lisakov2017}. 
No spectral calculations exist for the $8-12$ \msun~range,
a gap in the current theoretical literature we aim to address with this paper.

\subsection{Objectives and outline of paper}
Here we seek to further advance our knowledge of the lowest-mass SNe by modelling late-time spectra. Specifically, the goals of this paper are

\begin{itemize}
\item Provide the first nebular-phase (200d-600d) spectroscopic models for Type II SNe from a low-mass ($<10$ \msun) progenitor.
\item Understand the basic spectral formation processes for such SNe.
\item Assess the viability of sub-luminous Type II SNe to be linked to 8-11 \msun~progenitors by their nebular spectra.
\item Analyze whether ECSNe can be spectroscopically distinguished from low-mass Fe CCSNe.

\item Assess whether any stellar evolution/explosion physics can be constrained from current nebular observations.
\item Provide an outlook for things to be learned from future 3D models.
\end{itemize}

\section{Modelling} % --------------------------------------------------------------------------------------------------------
\label{sec:modelling}
We use the SUMO spectral synthesis code \citep{J11,J12} to model the physical state of the ejecta and to compute emergent spectra. SUMO computes the temperature and NLTE excitation/ionization solutions in each zone of the SN ejecta, taking a large number of physical processes into account. Specifically, the modelling consists of the following computational steps. (i) Emission, transport, and deposition of radioactive decay products (gamma-rays, X-rays, leptons). (ii) Determination of the distribution of non-thermal electrons created by the radioactivity \citep[by an expanded][method]{Kozma1992}. (iii) Thermal equilibrium in each compositional zone. (iv) NLTE ionization balance for the 20 most common elements. (v) NLTE excitation balance for about 50 atoms and ions. (vi) Radiative transfer through the ejecta. The solutions are generally coupled to each other and global convergence is achieved by iteration.

The radiative transfer was calculated at a spectral resolution of $\Delta \lambda/\lambda = 3.3\e{-4}$, corresponding to 100 \kms, in 4137 frequency bins covering 400-25,000 \AA. This guarantees resolution of the most narrow lines from core regions expanding with just a few 100 \kms.

\subsection{Ejecta model}

% Overview
We use the 9.0 \msun\ progenitor model of \citet{Sukhbold2016}. This progenitor is a solar metallicity, non-rotating star, evolved with KEPLER \citep{Weaver1978} as described in \citet{Woosley2015} although with a larger network. KEPLER uses efficient semi-convection, which leads to iron core formation down to $M_{\rm ZAMS}=9$ \msun, the lowest mass followed through all burning stages in the grid. The KEPLER calculations employed an adaptive subgrid to follow flame propagations, as well as a large nuclear network of 365 isotopes.

The progenitor is highly degenerate, and ignites Ne, O, and Si off-centre in flashes. The Si flash leads to a strong deflagration, otherwise mostly encountered in the $9.8-10.3$ \msun\ range.
%The main difference in the pre-SN evolution between the 9 and 11 models is that the 9 model is more degenerate, and ignites Ne, O, and Si off-centre in flashes. The Si flash leads to a strong deflagration, otherwise mostly encountered in the 9.8-10.3 \msun\ range. The 11 model, on the other hand, ignites all fuels at the centre. \textit{Consequences for pre-SN structure?}
The star, with a pre-collapse mass of 8.75 \msun,  formed an Fe core of 1.32 \msun, an O-rich shell of 0.08 \msun, and a He shell of 0.17 \msun. Outside this was a 7.18 \msun\  H/He envelope. The composition, density, temperature and $Y_e$ profiles are plotted in Fig. \ref{fig:progenitor} in the appendix.

\subsubsection{Explosion and nucleosynthesis}
The progenitor was exploded with two different methods in \citet{Sukhbold2016}; with a neutrino simulation using P-HOTB \citep{Janka1996}, and with a piston using KEPLER, with properties adjusted to mimic the neutrino simulation. We use as our standard case the P-HOTB simulation, which was rerun to include the effects of radioactive decay, as well as to extend to the homologous phase (we run to 100d). We calculate spectra of this model at 200, 300, 400, 500 and 600d. Comparison with the KEPLER model is given in the appendix. The P-HOTB explosion uses a 4-parameter inner boundary condition to describe the ``neutrino engine'', where the parameters for the 9 \msun\ model were chosen to reproduce the kinetic energy and \ni\ mass of the Crab nebula when applied to the Z9.6 progenitor model (see \citet{Sukhbold2016} for details). These parameters also give good agreement with the main properties of an ab-initio`` 3D simulation \citep{Melson2015}. %\textcolor{red}{Is there a unique link between 56Ni mass and explosion energy with 4 free parameters?}

\begin{figure} 
\includegraphics[width=1\linewidth]{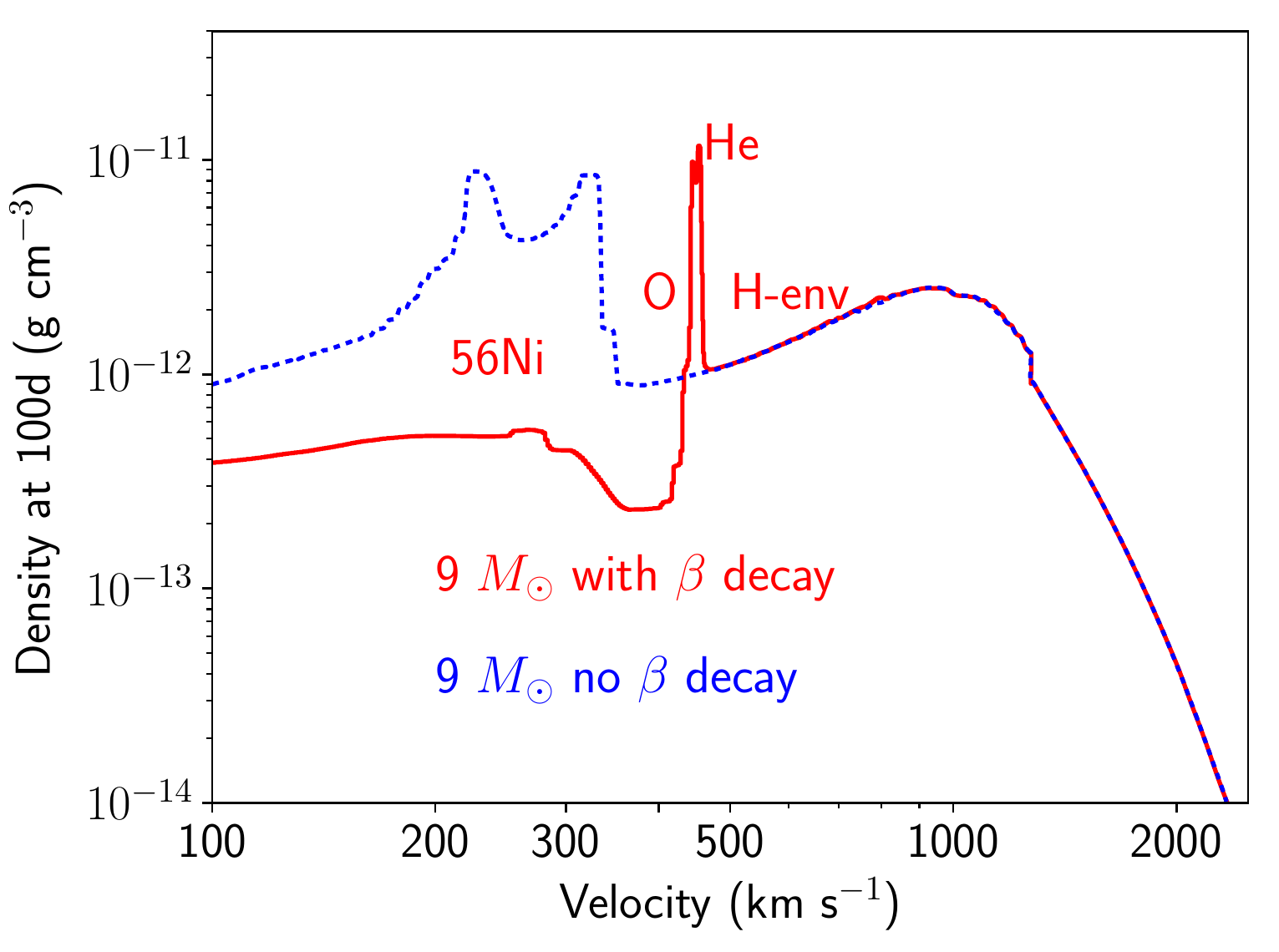} % plotdensity.py
\caption{Density profile of ejecta at 100d post-explosion. The standard model with $\beta$ decay is the red solid line, and the model without decay is blue dotted}.
% DC comparison with KEPLER that H envelopes are very similar
\label{fig:density}
\end{figure}

% Density
With this prescription for the explosion, the ejecta obtained a kinetic energy of 0.11 Bethe and a \ni~mass of $6\e{-3}$ \msun. %\in good agreement with \citet{Wanajo2011}}.
The homologous density profile at 100d is shown in Fig. \ref{fig:density}. The helium core (consisting of the \ni, O, and He zones)  expands with a maximum velocity of 460 \kms. % in the 9 model and 500 \kms\ in the 11 model. 
The radioactive decay has carved out a low-density inner region \citep[the ``$^{56}$Ni bubble'',][]{Herant1992}, and pushed the O and He layers from an initial $200-300$ \kms~into a shell at $430-460$ \kms.
Outside this shell, the envelope density shows an increase to a peak density at about 1000 \kms\, followed by a rapid decline.%, whereas the 11 model shows a monotonic decline.

\begin{figure}
\includegraphics[width=1\linewidth]{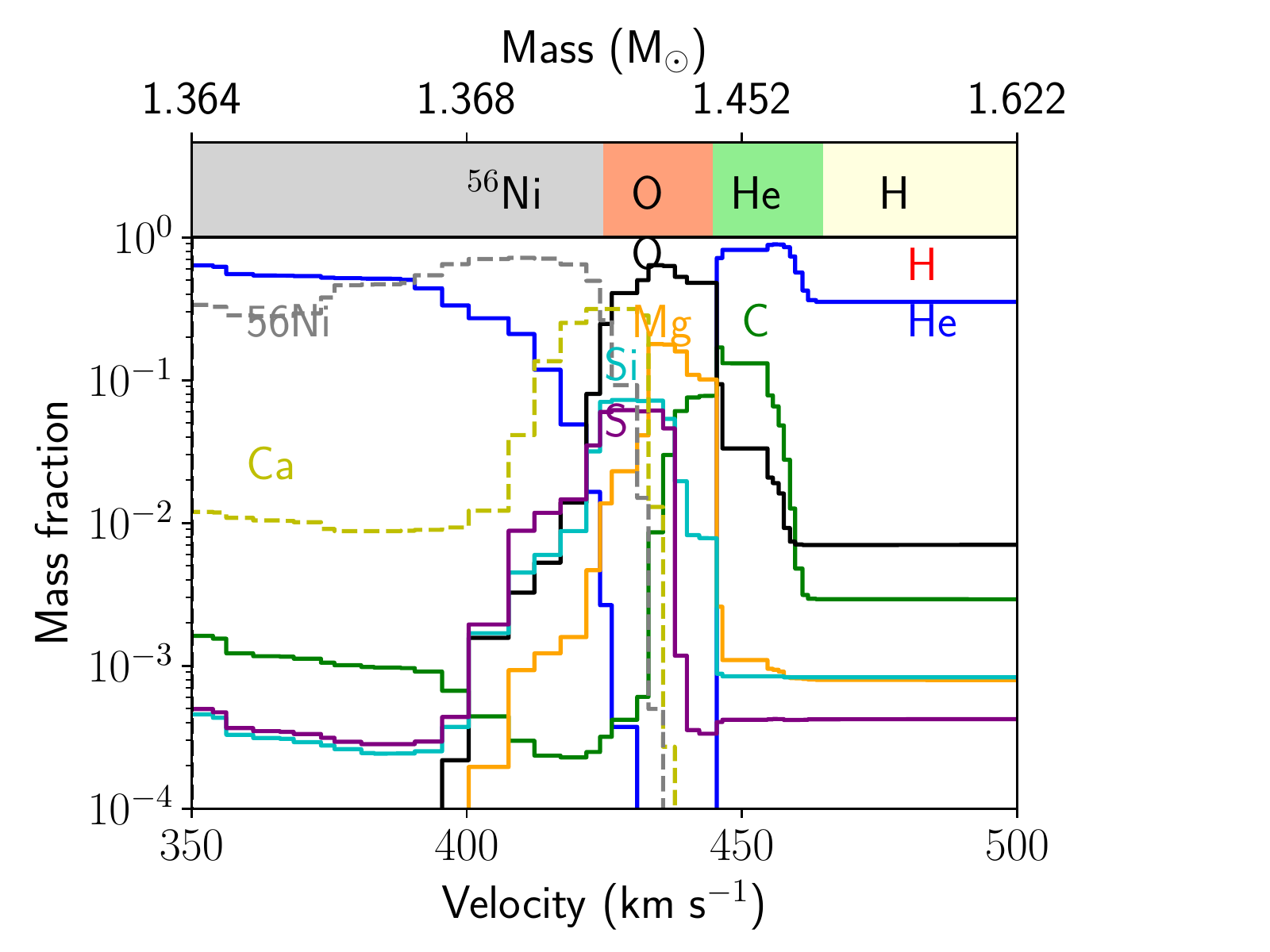} 
\caption{Composition (mass fractions) of the explosion model. The rough division into four compositionally distinct zones is illustrated by the top bar}.
\label{fig:comp}
\end{figure}

% Composition and zoning
The P-HOTB explosion models track the composition of H, $^4$He, $^{12}$C, $^{16}$O, $^{20}$Ne, $^{24}$Mg, $^{28}$Si, $^{32}$S, $^{36}$Ar, $^{40}$Ca, $^{44}$Ti, $^{48}$Cr, $^{52}$Fe, $^{56}$Ni, and tracer nuclei\footnote{These replace \ni\ as product in reactions in neutron-rich environment (electron fraction $Y_e < 0.49$)}, which we took as stable Fe for mapping into the spectral code (this mass was much smaller than the \ni\ mass). In sections \ref{sec:sensnucleo} and Appendix \ref{sec:KEPLER} sensitivity to composition is studied. Inspection of the composition distribution (Fig. \ref{fig:comp}) shows that the ejecta can be divided into 4 compositionally rather distinct regions; a \ni-dominated inner layer ($0-425$ \kms), an O-dominated layer ($425-445$ \kms), a He-dominated layer ($445-465$ \kms), and a H-dominated envelope ($>$465 \kms). %The division between these occur roughly at 200, 240, and 350 \kms.
%and at 210, 270 and 500 \kms~in model 11. %These layers contain masses masses of 0.02 (0.3\%), 0.035 (0.5\%), 0.17 (2.3\%) and 7.2 Msun (97\%) on model 9, and 
For the spectral calculations, the ejecta were divided into shells of 10 \kms\ thickness in the He core, and 100 \kms\  in the envelope . A higher velocity resolution than 10 \kms\ becomes problematic as the code uses the Sobolev approximation, which requires coherence over a Sobolev line width, which is of order a few \kms.
%What is explosive burning and what is hydrostatic? check woosley}

Solar composition values for N, Na, Al, K, Sc, Ti, V, Cr, Mn, Co, Ni (species that SUMO can treat but were lacking in the explosion model), were added in the He and H envelopes. Note that signatures for these elements from the deeper layers will be missing in the model.

\label{sec:ejectamodels}
\begin{table*}
\centering
\caption{Properties of ejecta used in the modelling. The masses of selected elements are listed. $M(O)$ (column 5) is the total oxygen mass, and column 6 is the newly synthesized oxygen mass (resides in CO core). 
Of the original 9.0 \msun, 0.25 \msun\ is lost to stellar winds, and 1.35 \msun~becomes the neutron star.}
\begin{tabular}{|c|c|c|c|c|c|c|c|}
\hline
Model  & $M_{ZAMS}$ & $M_{ejecta}$ & $E_{kin}$ & M(O) & M(O) in CO core &  M($^{56}$Ni) \\
       & $(M_\odot)$               & $(M_\odot)$  &  $\left(10^{51}~\mbox{erg}\right)$ & $(M_\odot)$ & $(M_\odot)$  & $(M_\odot)$ \\
       \hline
9  & 9.0  & 7.4  & 0.11 & 0.072  & 0.016 (22\%) & 0.0062 \\
%Ertl11.0 & 11.0 & 9.2 & 0.31 & 0.21    & 0.16 (76\%)   & 0.018  & 200,400,600\\
% O masses printed in plot.py in models/ertl. Change vel limit to get from CO core only.
\hline
\end{tabular}
\label{table:ejectaprop}
\end{table*}

\subsection{Models with dust}
\label{sec:dust}
\arja{There is indication of dust formation in one of the observed candidates we compare the model to (see Sec. \ref{sec:2008bk}). Suitable
models for this object therefore need to consider dust effects in some manner. Here we follow earlier methodology \citep[e.g.][]{J12} and apply a uniform absorption coefficient from the center of the nebula to some maximum velocity $V_{dust}$.
We choose $V_{dust}$ to coincide with the edge of the He core, so $V_{dust}=465$ \kms. The amount of dust is chosen by its total radial optical depth $\tau_d$}.

\section{Physical conditions} % ============================================
\label{sec:physical}
\arja{The physical conditions and processes in the ejecta are now discussed}.

\subsection{Gamma-ray deposition}
\label{sec:gammadep}

\begin{figure}
\includegraphics[width=1\linewidth]{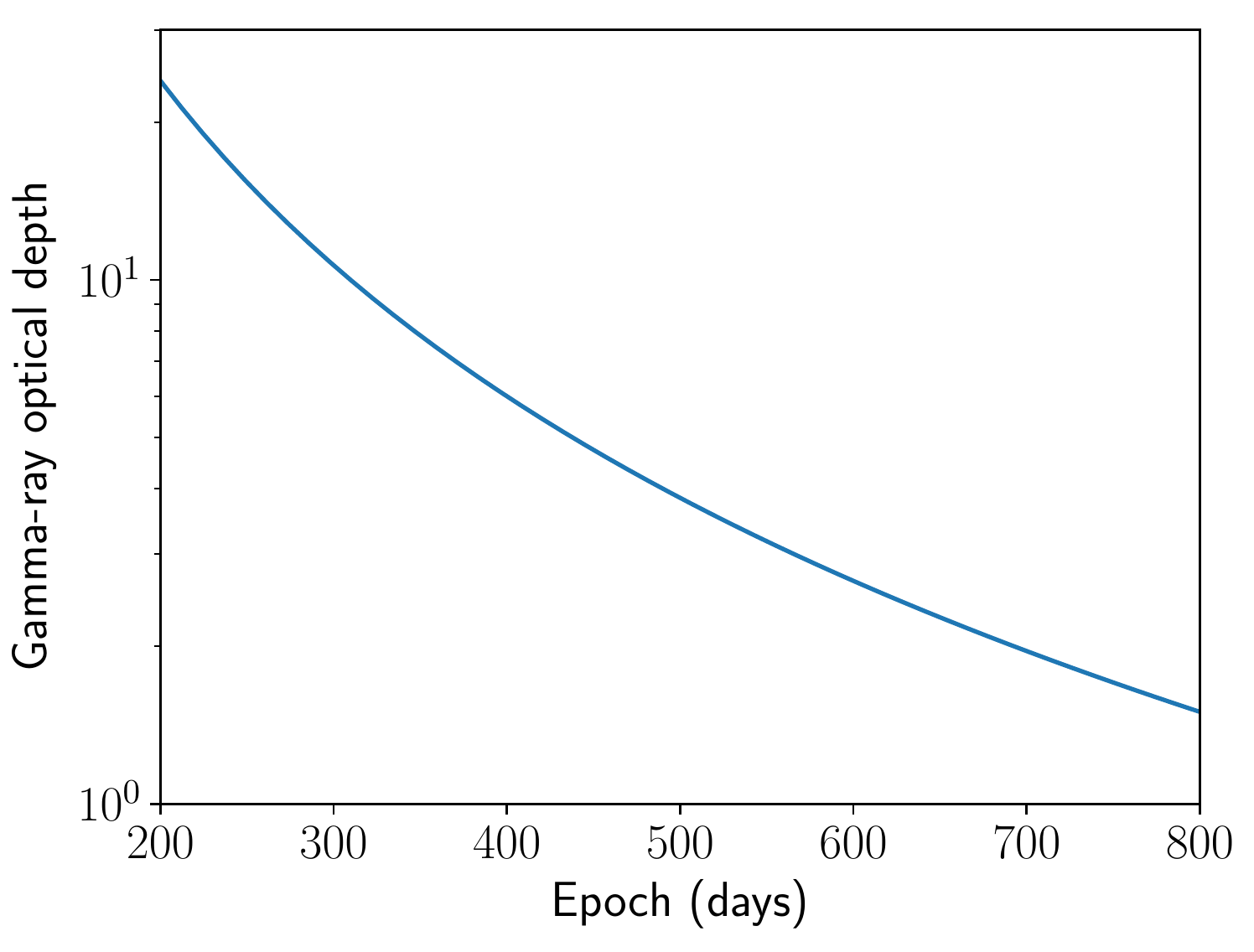} % timeseries.py M1 M2 M3
\caption{\arja{Gamma-ray optical depth as function of time.}}
% DC M^2/E follow pattern of bigger jump between He80 and He100 than between He100 and He130.
\label{fig:gammatau}
\end{figure}

\arja{Figure \ref{fig:gammatau} shows the radial gamma-ray optical depth versus time. The low explosion energy makes these SNe dense and gamma rays are fully trapped for several years ($\tau \gg 1$)}.

\begin{figure}
\includegraphics[width=0.9\linewidth]{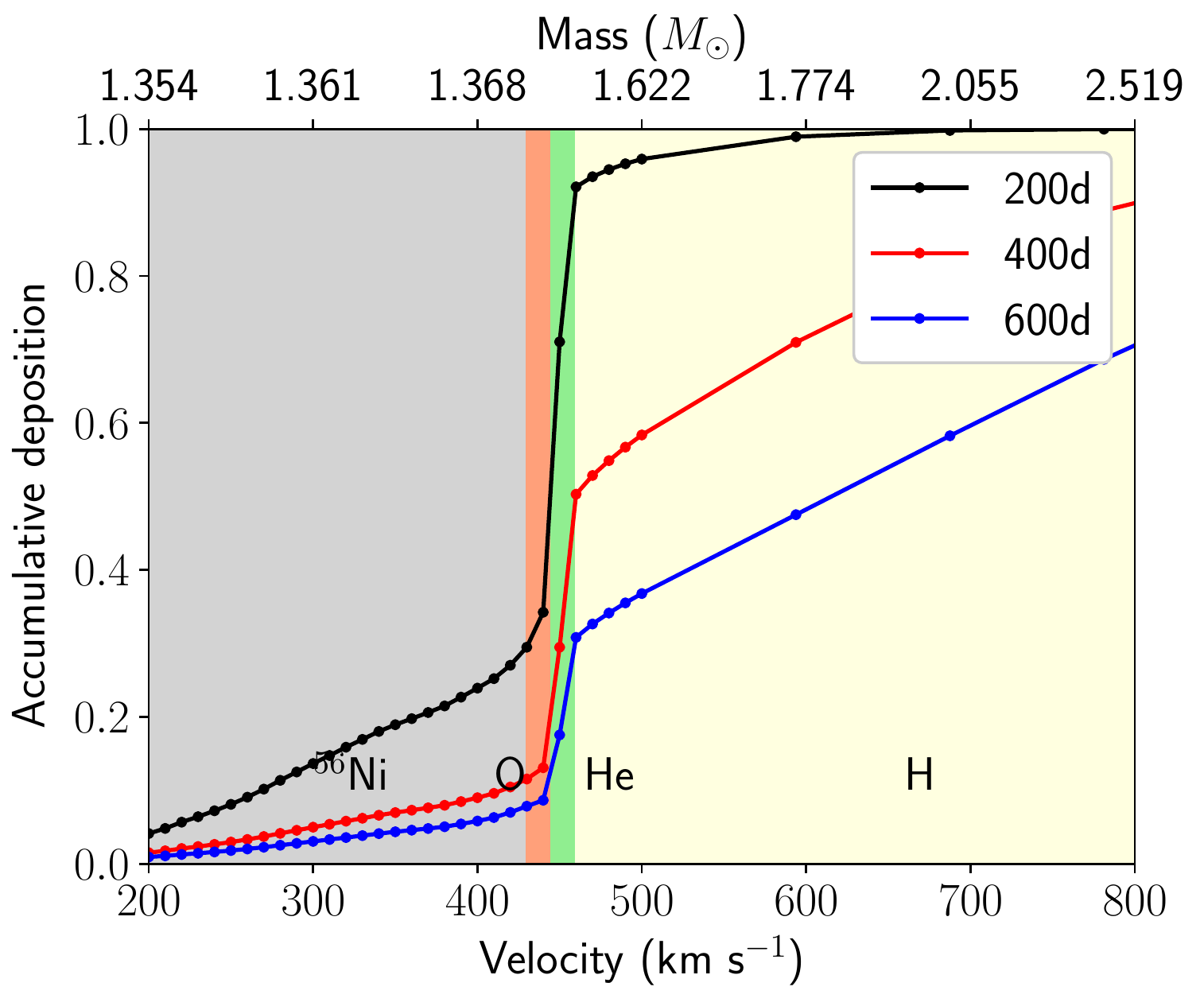} % plotovermass.py figure(12)
\caption{\arja{Accumulative energy deposition from radioactive decay products. The zones are labelled with their most abundant element, colored as gray (\ni), pink (O), green (He), and yellow (H).}}
\label{fig:gammaaccum}
\end{figure}

\arja{Figure \ref{fig:gammaaccum} shows the accumulative gamma ray deposition as it evolves with time from 200d to 600d. At 200d the gamma rays are mostly trapped in the He core, with 30\% in the \ni \ layer, 20\% in the O layer,
and 40\% in the He layer. Only about 10\% penetrate into the H envelope. With time, the fraction reaching the H zone increases though to 50\% at 400d and 70\% at 600d. Virtually all deposition occurs in the innermost 1000 \kms\ at all times, and thus all line profiles should have HWZI$<$1000 \kms}.

% ---- GAMMA DEP ACCUMULATION

\subsection{Temperature}

\begin{figure}
\includegraphics[width=0.9\linewidth]{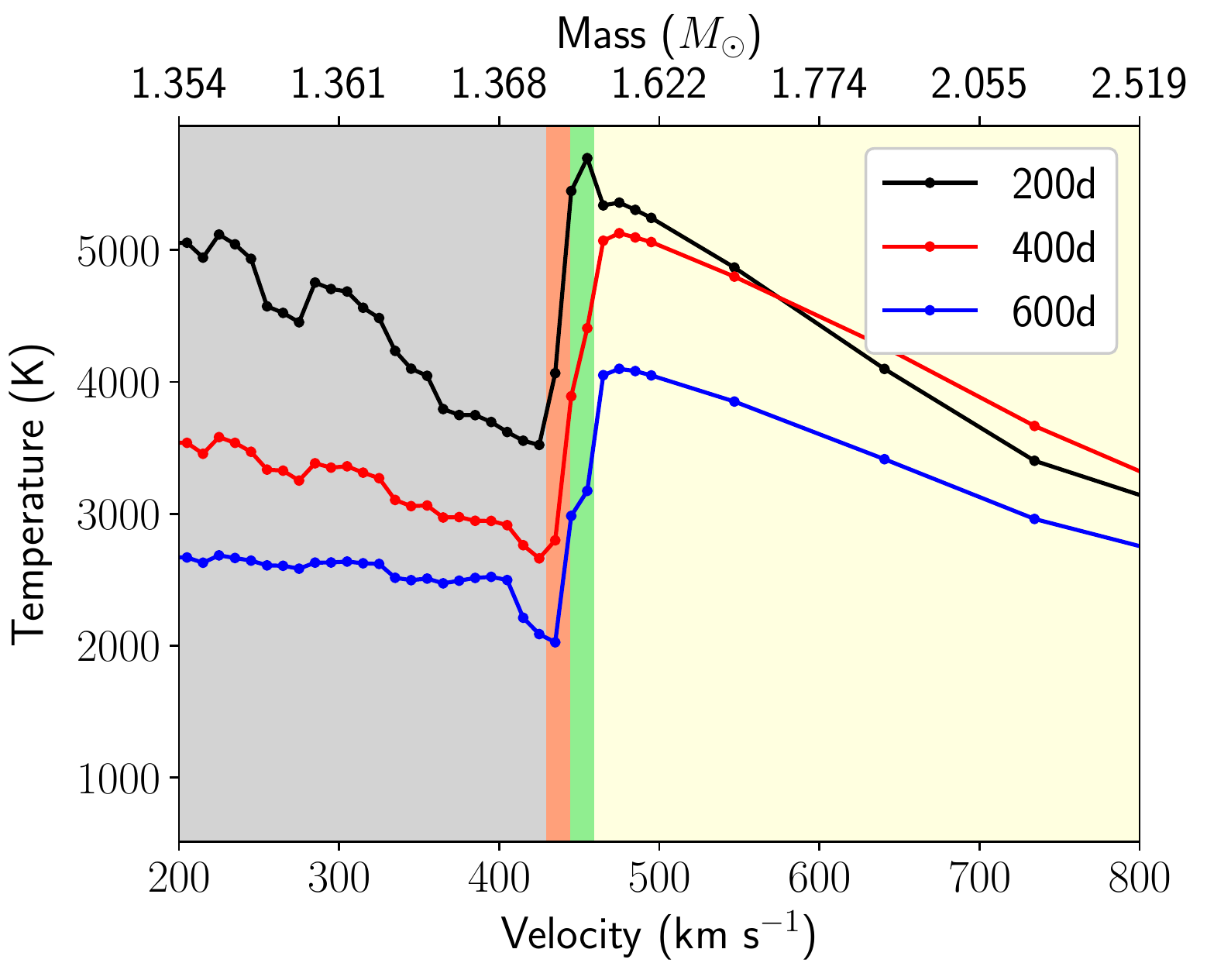} % plotovermass.py figure(1)
\caption{\arja{Temperature in the models. The coordinates at zone midpoints are used.}}
\label{fig:temps}
\end{figure}

\arja{Figure \ref{fig:temps} shows the temperature evolution. The He core is at $T\sim 5000$ K at 200d, decreasing to $\sim$4000 K at 400d and $\sim$3000 K at 600d. At each epoch there is however variation of about 1000 K depending on location. The core is thus relatively hot despite the low \ni \ mass, and one can expect most of the radiative cooling in the optical regime. The inner H envelope is at $T\sim$5000 K at 200 and 400d, and $\sim$4000 K at 600d. 
Note that the temperature peaks around 400d in some parts of the envelope, as the increasing fraction of gamma-ray energy deposited there grows faster with time than the \co\ decline up to around this time}.

\subsection{Ionization}

\begin{figure}
\includegraphics[width=0.9\linewidth]{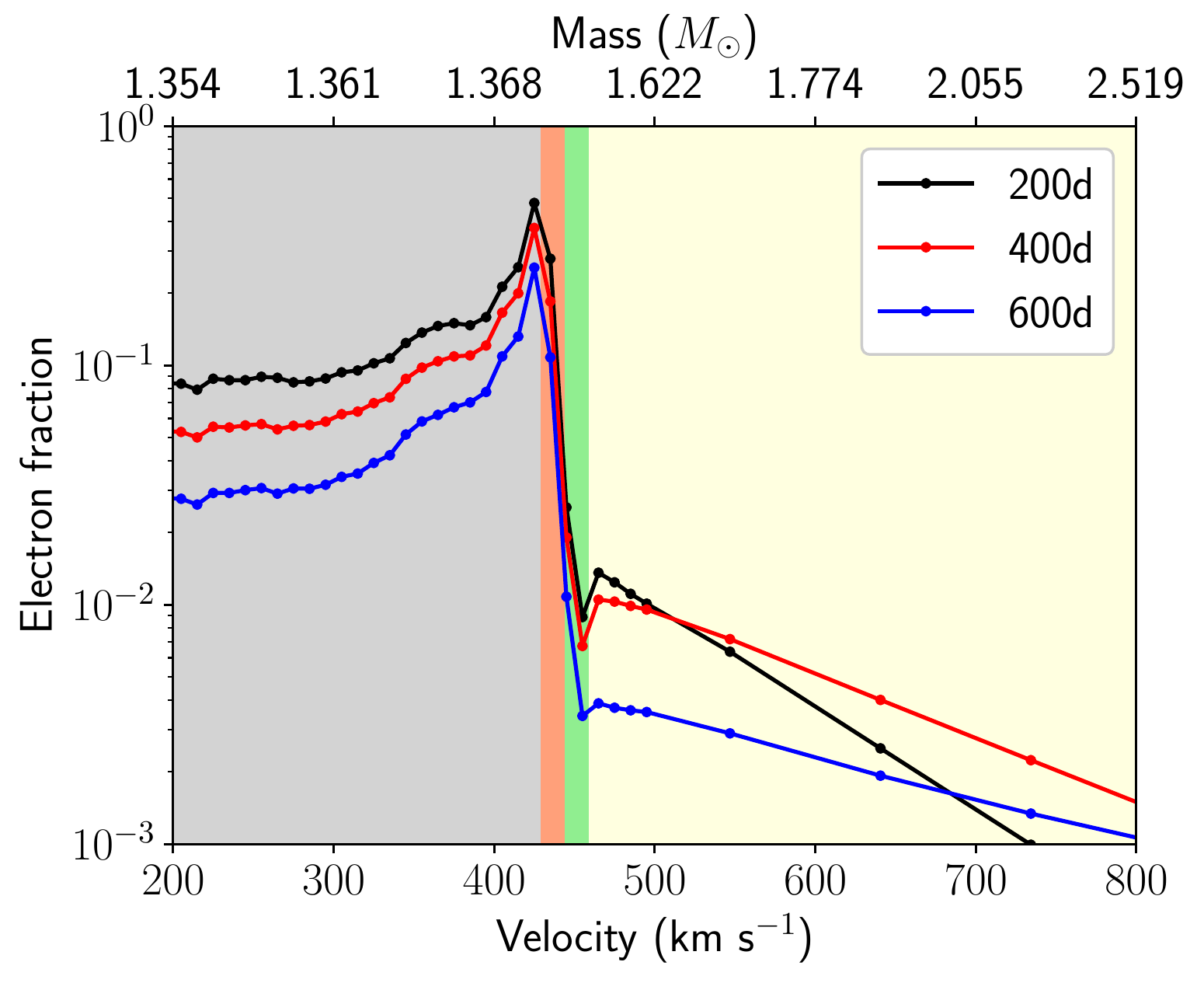} % plotovermass.py figure(2)
\caption{\arja{Electron fraction in the models. The coordinates at zone midpoints are used.}}
\label{fig:xe}
\end{figure}

\arja{The ionization is shown in Fig. \ref{fig:xe}. The electron fraction $x_e = n_e/n_{nuclei}$ is around 10\% in the inner \ni~layer, and about 1\% in the O/He shells, similar to standard Type IIP models. %The O shell shows a factor $\sim$ 2 higher ionization that the other surrounding zones. 
The lower value in the O/He shells is due to the high density contrast created by the $\beta$~decay, and the higher ionization potentials compared to iron-group composition of the \ni~region. 
The low gamma deposition in the outer H envelope gives a very low degree of ionization, with $x_e$ falling below $10^{-3}$  outside 800 \kms}.

\arja{Because the electron fraction is low, the spectra will be dominated by neutral and low ionization-potential singly ionized species; thus H I, He I, C I, O I, Mg I, Ca I/Ca II,  Si I, S I, Fe I/Fe II can be expected}.

% IONIZATION 

\subsection{Electron scattering}
\arja{The electron (Thomson) scattering optical depth at 200d is 0.35, with similar contribution (1/3 each) by \ni~core, O/He shells, and H envelope. Thus, some fraction of emitted photons will experience at least one scattering on their way out. %This should be kept in mind when analyzing which regions create different lines, as plotting the last zone may convey scattering rather than creation (this holds true for line scattering for all epochs as well). 
The Thomson scattering creates a mild distortion of line profiles, although this is hardly discernible when convolved with telescope point spread functions since the emitting gas moves so slowly ($\sim$ 500  \kms). At 400d the Thomson optical depth has dropped to 0.08 and has neglegible influence}.
%\textcolor{red}{Compare models with and without electron scattering - effects on line profiles, especially at 200d.}
% python thomson apa 1 006_200

\subsection{Steady-state verification}

\arja{SUMO is limited to the assumption of steady state in the ejecta, e.g. that physical processes occur on a shorter time-scale than the dynamic time or the $^{56}$Co decay time. In the nebular phase the $^{56}$Co decay time (111d) is the shorter of these. These time-scales can be checked as follows}.

\subsubsection{Recombination}
\begin{figure}
\includegraphics[width=0.9\linewidth]{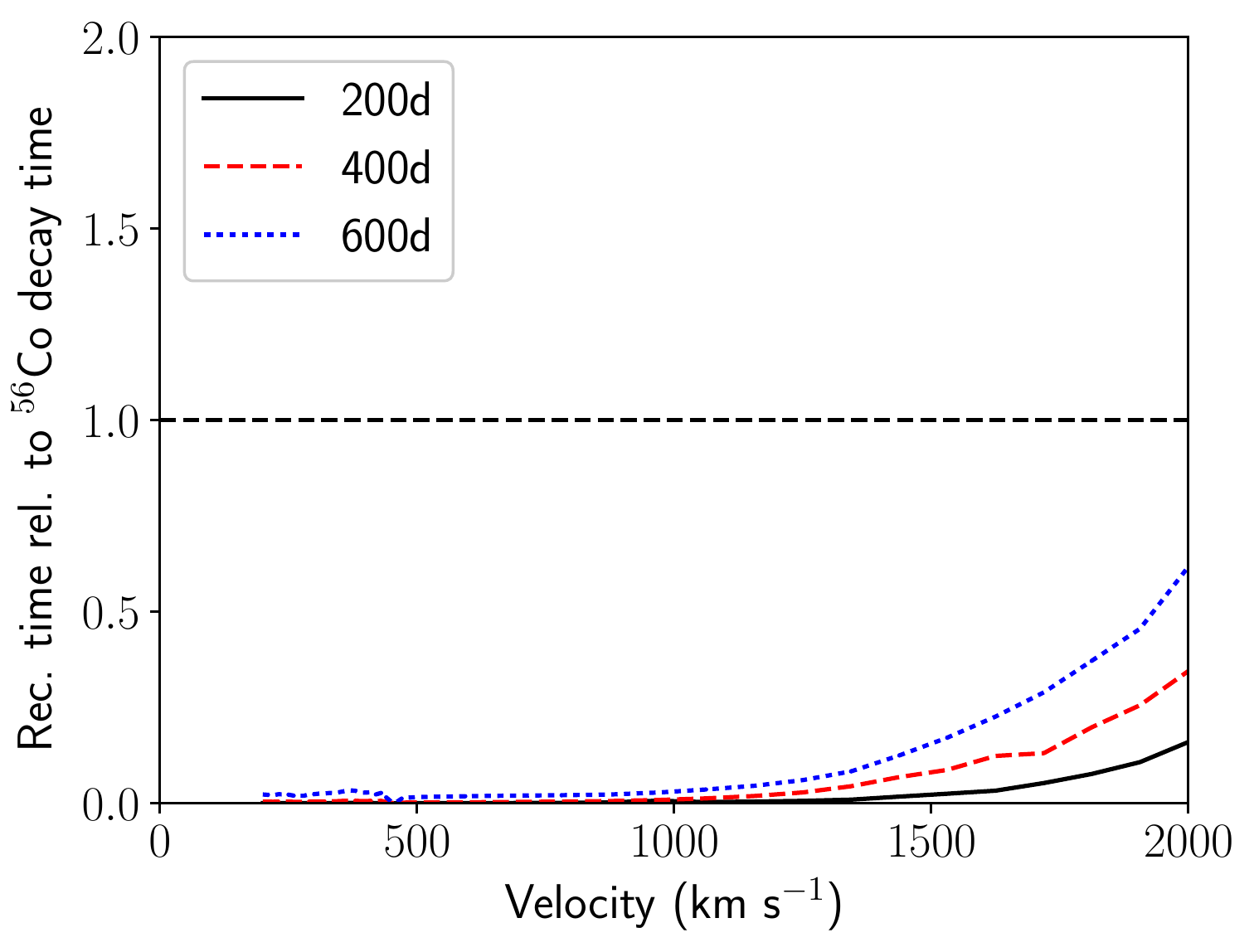} % plotovermass.py figure(3)
\includegraphics[width=0.9\linewidth]{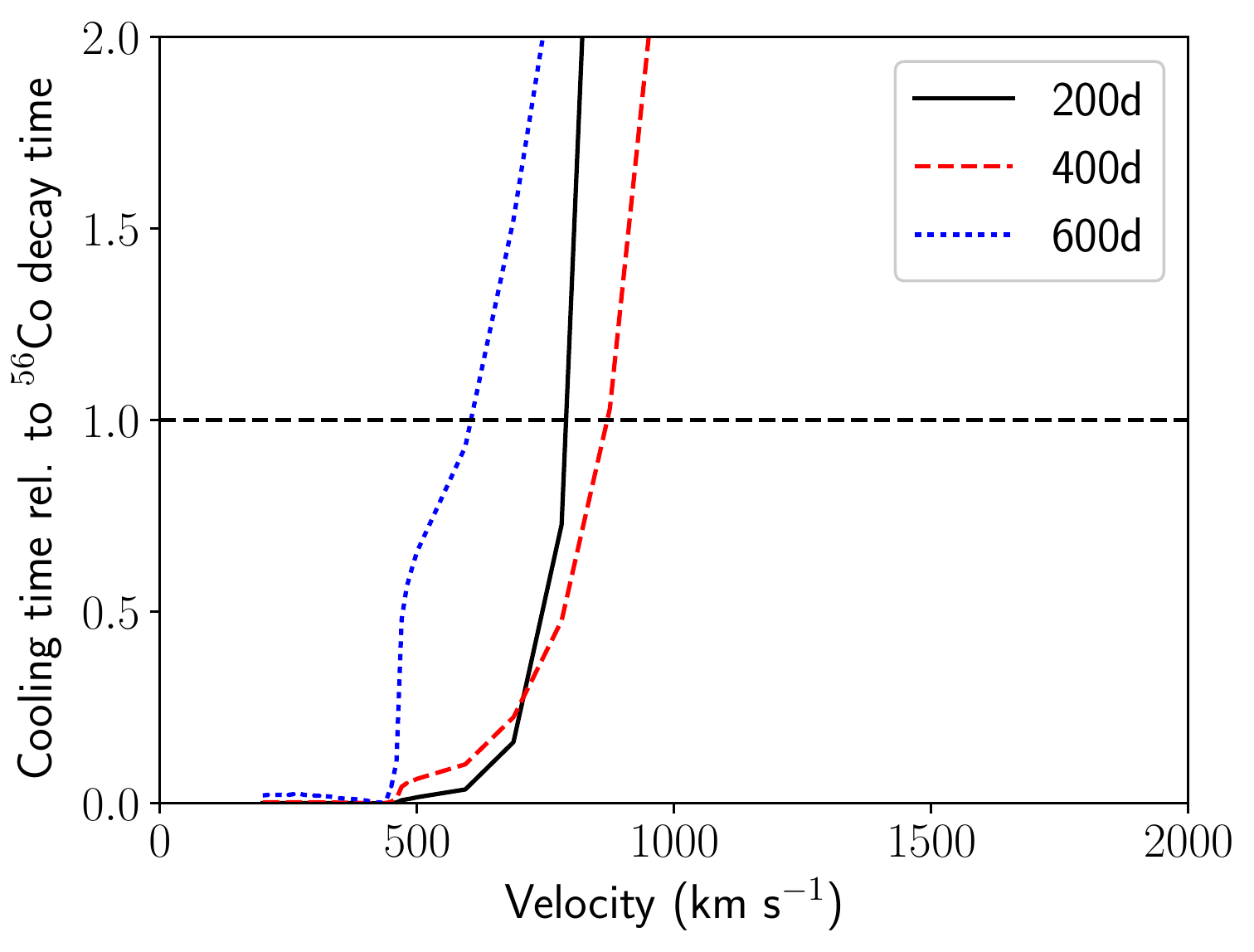} % plotovermass.py figure(4)
\caption{\arja{Recombination timescale (top) and cooling time-scale (bottom) relative to the radioactive decay time scale.}}
\label{fig:rectime}
\end{figure}

\arja{The recombination time-scale is 
\begin{equation}
\tau_{\rm rec} = \frac{1}{n_e \alpha(T)}
\end{equation}
\arja{where $\alpha(T)$ is the recombination coefficient. Figure \ref{fig:rectime} (top) shows the relative recombination time scale $\tau_{\rm rec}/\tau_{\rm 56Co}$, where $\alpha(T)$ for the most common element in the zone was used. Steady-state ($\tau_{\rm rec}/\tau_{\rm 56Co} \ll 1$) is fulfilled for the models at all times}}.%, except for a small outer region in model 9 at 200d.

% DC 2017-07-06 the correct cooling quantity is plotted: 3/2 nkT/c(T).

\subsubsection{Cooling} % ------------------------------------------------------------------------------
\arja{The radiative cooling time-scale is}
\begin{equation}
\tau_{cool} = \frac{\frac{3}{2}nkT}{c(T)}
\end{equation}
\arja{where $n$ is the total particle density and $c(T)$ is the radiative cooling rate. Figure \ref{fig:rectime} (bottom), shows the relative cooling time scale $\tau_{\rm cool}/\tau_{\rm 56Co}$.  Steady-state is fulfilled out to about 500-750 \kms, covering the He core and inner H envelope. This region is responsible for the vast majority  of gamma ray energy reprocessing (Fig. \ref{fig:gammaaccum}), and as such one can expect little difference in a time-dependent model}.

%\subsubsection{Tests against time-dependent model}
%\arja{A new version of the code which includes time-dependence is under preparation, and will be presented elsewhere. With a preliminary code version some models were run for comparison tests. This verified that the steady state model gave in all aspects close agreement with the time-dependent model}.

\section{Spectra : the 9 \msun\ model at 400d} % =============================================================

\arja{As foundation for the discussion about the spectral formation, we take an in-depth look at the model at 400d. Variations with respect to other epochs will be discussed in Sect. \ref{sec:timeevol}}.%, and variations between the 9 and 11 models will be discussed in Sect. \ref{sec:Mzamsdep}.

\begin{figure*}
\includegraphics[width=1\linewidth]{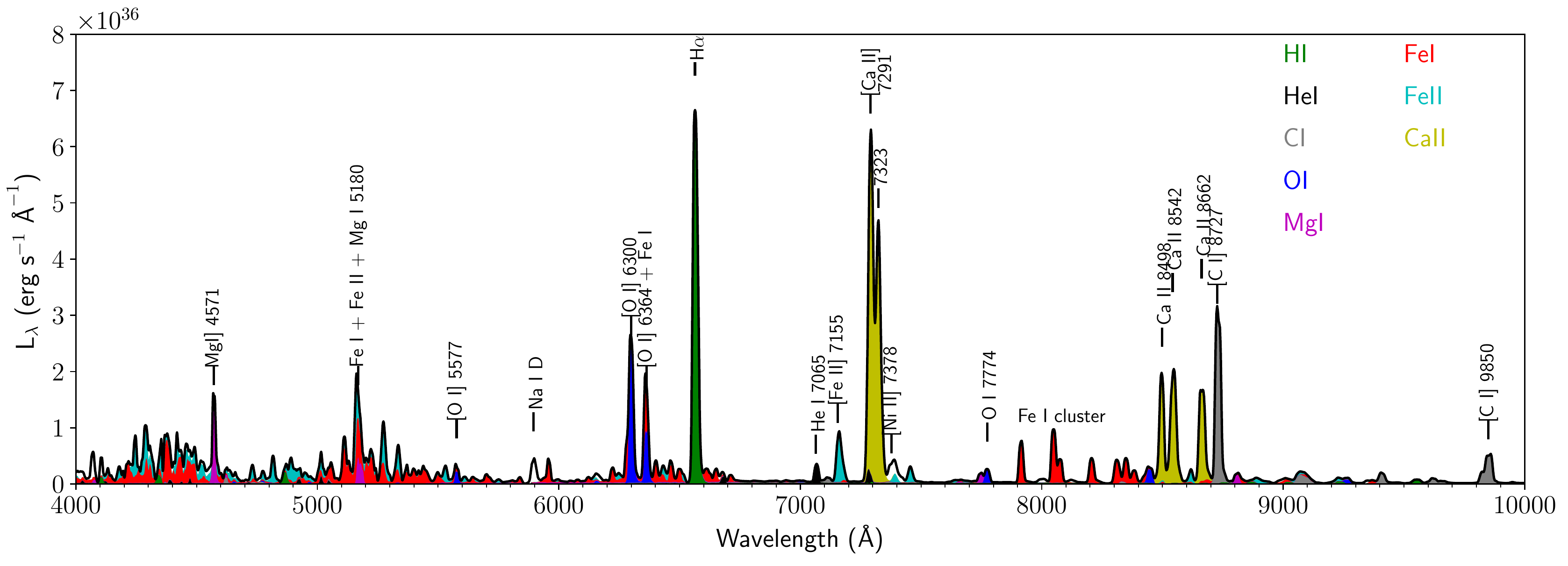} % plotspectra.py 
\caption{\arja{Model spectra of 9 \msun\ model at 400d. Contributing elements are color coded.}}
\label{fig:spectra400}
\end{figure*}

\begin{figure*}
\includegraphics[width=0.8\linewidth]{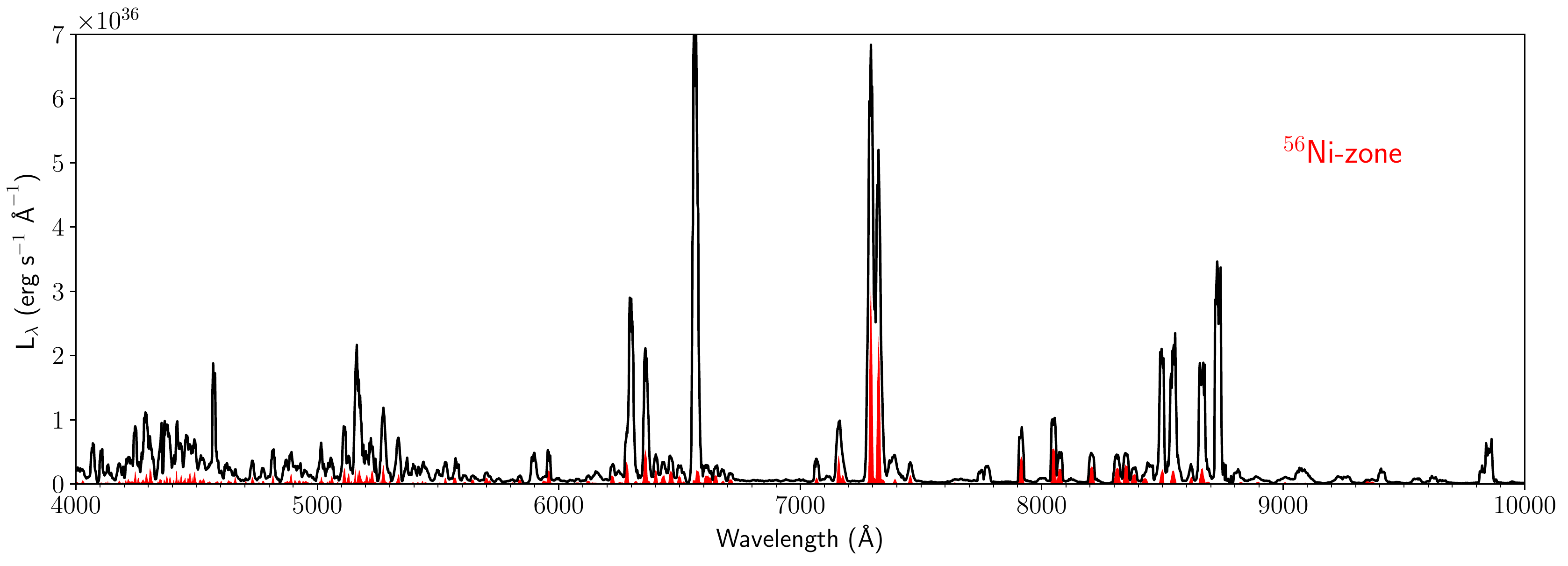} % plotspectra.py 
\includegraphics[width=0.8\linewidth]{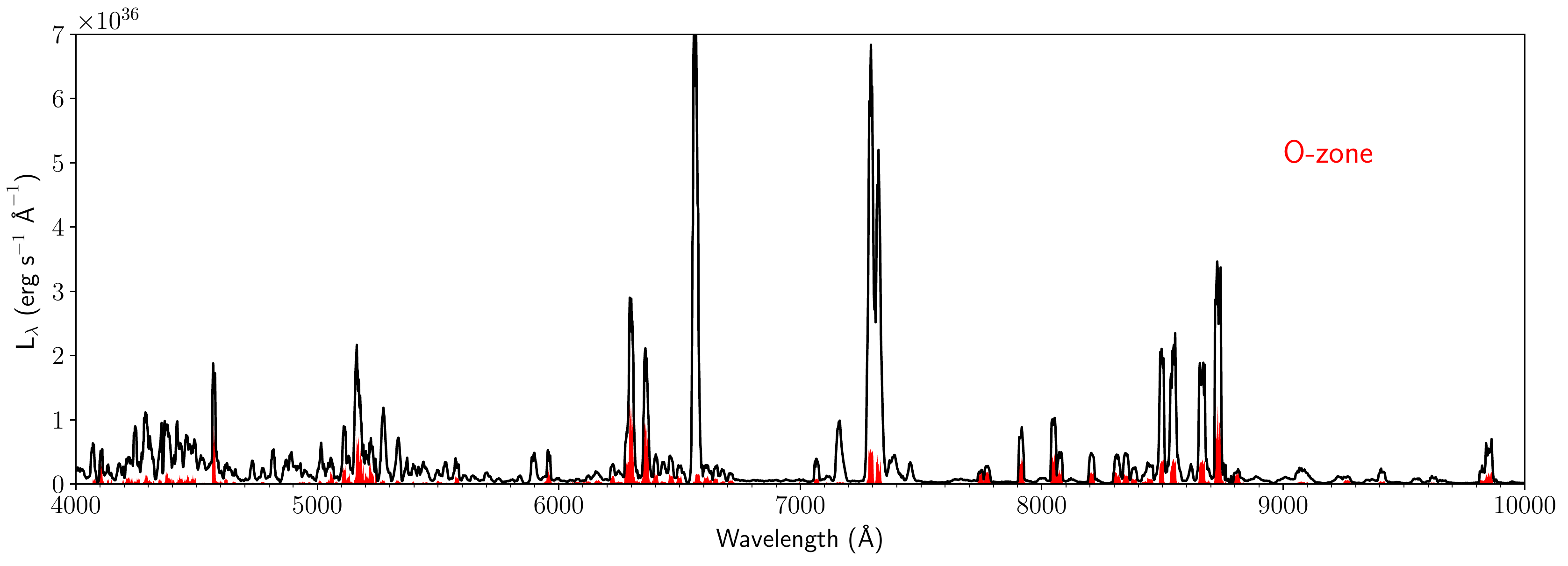} 
\includegraphics[width=0.8\linewidth]{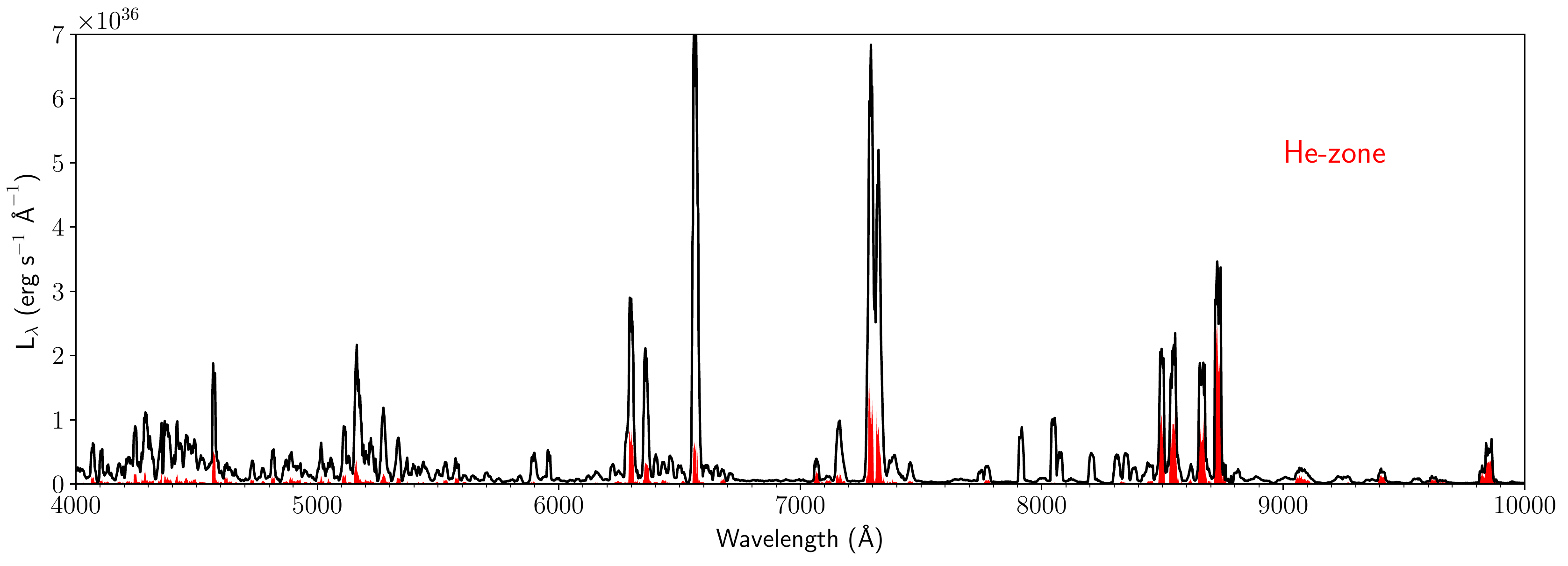} 
\includegraphics[width=0.8\linewidth]{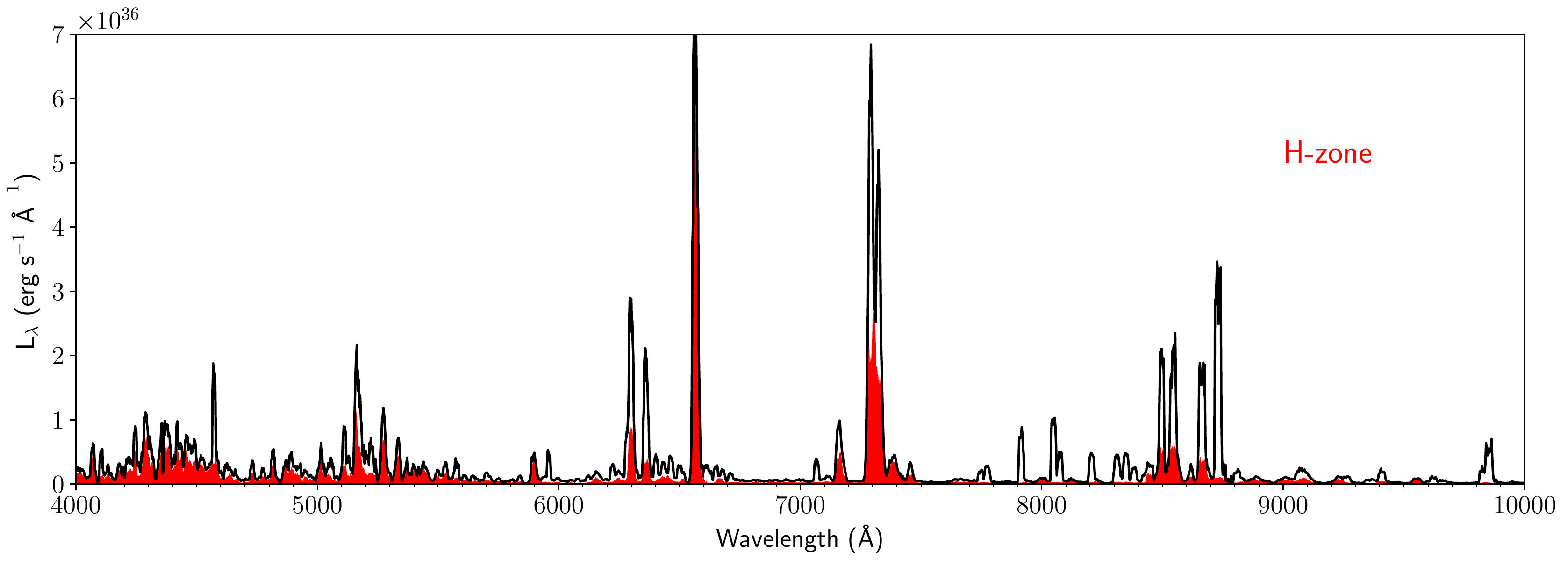} 
\caption{\arja{Contribution by different zones to the last interaction of photons, 9 \msun\ model at 400d. The total spectrum is the black line, and red filled are zone contributions}.}
\label{fig:zones9} 
\end{figure*}

\arja{The spectrum of this model, together with contributions by various elements, is shown in Fig. \ref{fig:spectra400}. Note that in this section the spectra are not convolved to any telescope resolution.
To aid in the interpretation, we plot also in Fig. \ref{fig:zones9} the last interaction zone of the photons as divided into \ni\ core (0-430 \kms), O-shell (430-450 \kms), He shell (450-470 \kms) and H envelope ($>$470 \kms). As composition varies over small velocity intervals that is comparable to the binning, this division is somewhat ``smeared'', and emission from these specific compositions is not as ``clean'' as in higher $M_{ZAMS}$ models. For example, some \ni~emission becomes part of the O-zone emission. A relatively clean Si/S region normally exists, but here the Si/S is always mixed with a larger amount of O}.
 
\arja{As emission comes mainly from the metal core and inner H envelope, line profiles are narrow with FWHM$\lesssim$1000 \kms. Table \ref{table:FWHMmodel} shows a summary of FWHM values in the model spectrum. The layered structure of the model is reflected by the increase from 800 \kms~for the iron core lines to 1100 \kms~for H$\alpha$. Below comments on emission from each species is given}.

\begin{table}
\center
\begin{tabular}{c|c|c}
\hline
Line & FWHM (\kms) & Comment\\
\hline
H$\alpha$      & 1100 & \\ % DC 2017-07-06 1100 (manual plotting)
He I 7065       & 900  & box-like \\ % DC 2017-07-06 1100
{[}C I{]} 8727  & 900 & box-like\\ % DC 2017-07-06 1100
{[}O I{]} 6300, 6364 & 900\\ % DC 2017-07-06 1100
Mg I{]} 4571 & 900 & box-like\\ % DC 201%7-07-06 1100
{[}Ca II{]} 7291, 7323 & 900 &\\ % not box-like  DC 201%7-07-06 1100
{[}Fe I{]} 7912 & 800 & box-like \\
\hline
\end{tabular}
\caption{\arja{FWHM widths in model at 400d.}}
\label{table:FWHMmodel}
\end{table}

\textit{Hydrogen.}
\arja{Hydrogen displays only one line in the optical; H$\alpha$. H$\alpha$ emits mainly by recombination. Ionization is dominated by Balmer photoionization, rather than non-thermal.  H$\alpha$ is optically thick out to about 1700 \kms, and so a clearly seen absorption trough is also produced. The inner $\pm$500 \kms\ is flat-topped as there is no H in the core.}
% H inner env. zones ~28-34 (model 028_400)

\textit{Helium.}
\arja{Helium displays a weak but distinct He I 7065.
%The low expansion velocities here provide a unique opportunity to see He I 5876 which otherwise usually scatters in Na I D.
This line is produced mainly by the He shell, and thus provides an interesting diagnostic of this layer. However, there is also a contribution ($\sim1/3$) from freeze-out He in the \ni\ core. Other He lines are too weak to be discernible.}
% DC 2017-07-06 : He 2/3 from He shell, 1/3 from 56Ni core

\textit{Carbon.} % He shell 26 and 27. CI cooling is 10 and 20%.
\arja{Carbon displays [C I] 8727, which is one of the strongest lines in the spectrum, and [C I] 9850, significantly weaker but also unblended. As He I 7065, these lines come from the (C-rich) He zone. The high C abundance here (13\%) makes carbon an important cooling agent of this zone ($\sim20$\% of total cooling), which reprocesses $\sim$20\% of the gamma energy. With a heating fraction of $\sim$40\%, a total of $0.2\times 0.4\times 0.2 = \sim$2\% of the radioactive energy emerges in [C I] 8727. Carbon is mainly neutral ($>$90\%) which gives robustness for the C I emission. The optical depth is around unity}.

\arja{This line gives promise to be used both as a signature of these low-mass models, and/or to test the amount of C that exists in the He layer, in turn a diagnostic of He shell burning.
There is a distinction here to higher-mass CCSN models, where this line is formed mainly in the C/O zone \citep{J15a}, which is prone to CO molecule formation and ensuing quenching of atomic emission lines. CO formation is quenched in a He-dominated zone \citep{Lepp1990}, and the [C I] lines are therefore likely robust predictions. Note that the He/C zone mass is 0.16 \msun~in the model, whereas the C/O zone is $<$0.01 \msun~(see KEPLER model in Appendix)}.
% Hr shell mainly 26 and 27

\textit{Oxygen.} % O shell 25 and 26
\arja{Oxygen displays [O I] 6300, 6364 (although the 6364 \AA\ component is blended with a strong Fe I line at a similar wavelength), and weak but distinct [O I] 5577 and O I 7774 lines. While O I 7774 comes exclusively from the O shell,  [O I] 6300, 6364 have also some contribution from the He zone (where the O abundance is 3\%) and the H zone. %The contaminating Fe I line comes mainly from the \ni\ zone.
In the O shell, oxygen cooling is at a moderate level (5-10\%) as there are relatively high abundances of \ni, Ca, Si, S, C that are also efficient coolants. The [O I] lines are here marginally optically thin, $\tau_s=0.1-1$ at 400d.}
%\arj{[O I] 6300 has FWHM 950 \kms, and [O I] 6364 910 \kms}.

\textit{Magnesium.} \arja{Magnesium shows Mg I] 4571, coming to a large part from the O-shell. Although not the dominant contributor the blend at 5180 \AA, Mg I 5180 also emits at significant levels. Mg I provides a few percent cooling of the O shell. Mg I] 4571 is optically thick ($\tau_s \sim 10^3$), and its emission is therefore not strongly sensitive to the specific abundance of magnesium in the zone.}

\textit{Calcium.}
\arja{Calcium shows [Ca II] 7291, 7323 and the IR triplet (Ca II 8498, 8542, 8662).
%\subsubsection{Ca II 7291, 7323}
Both multiplets have contributions from all four regions, linked to the strong cooling capability of calcium even at low abundance. The H envelope provides a broad base with about half the total luminosity. The \ni~region provides about 20\% with the two components of [Ca II] 7291 and 7323 clearly separated. The O and He shells provide minor ($\sim$10\% of luminosity each), but distinct, contributions as well.}

\textit{Iron.}
\arja{Iron is prominent throughout the optical spectrum, with Fe I lines providing most of the flux in the ``quasi-continuum''. Virtually all flux below 6000 \AA~is due to Fe I lines, as well a a cluster of strong lines between 7900-8500 \AA.
Inspection of the zone origins show that below 6000 \AA~most photons had their last emission point in the H envelope. At longer wavelengths this contribution is smaller and emission from the iron core dominates, including the 7900-8500 \AA~cluster of narrow lines. Coming from the innermost region, the Fe I lines have the smallest FWHM at $\sim$800 \kms}.

% Specific line analysis
\arja{Fe II produces [Fe II] 7155  as the only distinct line, although a few of lines in the blue region are also dominated by Fe II. This line has contributions from several of the zones, especially the H zone, and is therefore not a direct diagnostic of the \ni~core}.
%(FWHM 1050 \kms, but has a doublet nature)

\subsection{Signature of the He core}
\label{sec:Hecore}
% at 300d, the comparison with H-zine model looks similar, although O I 7774 is now also produced by H zone model.
% Still 7900-8500 range, [C I] 8727 and [9850] and He I 7065 are strong signals.

\arja{Plotting the last interaction zone of the photon packets (Fig. \ref{fig:zones9}) gives a good first idea about the origin of various lines. However, scattering both in lines and with free electrons partially breaks the link to true emissivity creation by thermal processes. Therefore, complementary models runs where zones are 'switched off' or interchanged can help in understanding the spectral formation and identifying signatures of each layer}.

\begin{figure*}
\includegraphics[width=1\linewidth]{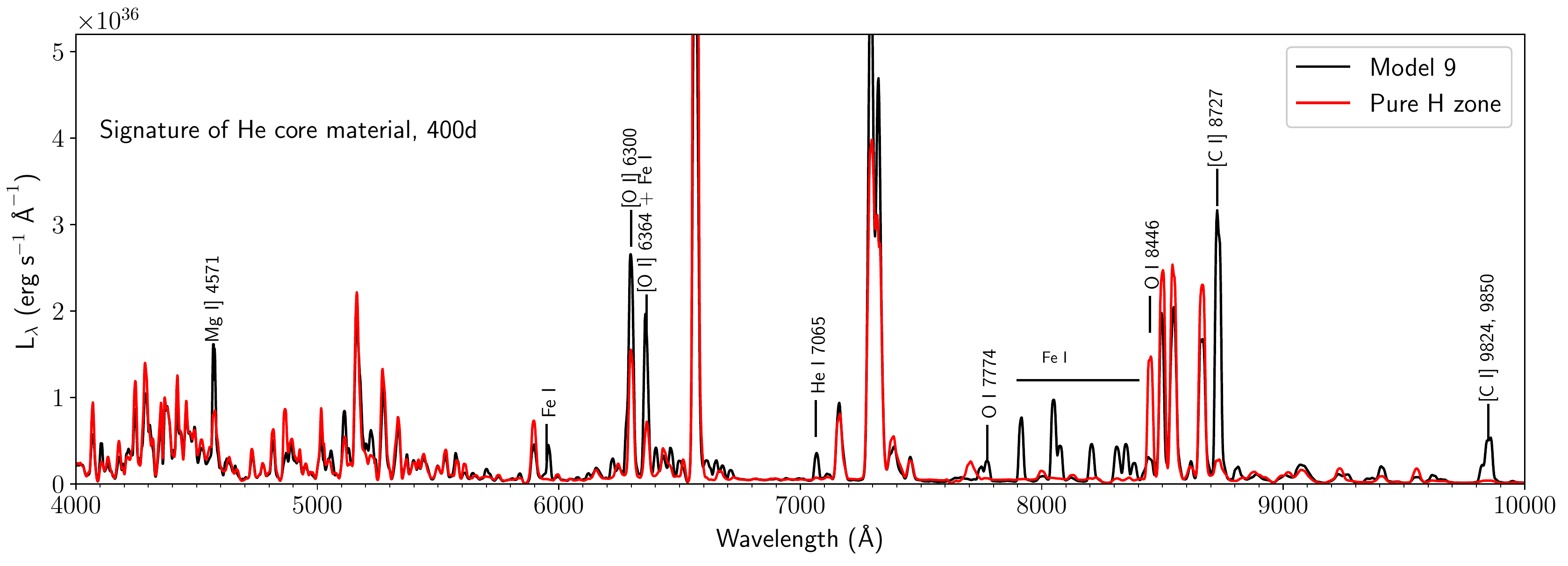}  
\caption{\arja{Signature of He core, as illustrated by comparing the standard 9 \msun\ model (black) with a model where all composition is that of H-zone material (red), at 400d. The main lines differing are labeled.}}
\label{fig:spectraHecore}
\end{figure*}

\arja{Figure \ref{fig:spectraHecore} compares the spectra of the 9 \msun\ model with a model where the composition has been replaced by H-zone material throughout (i.e. a pure H-zone nebula).
Thus, this comparison illustrates the sensitivity to the presence of He core materials (\ni, O, and He zones), for the same density profile and energy deposition (held fixed)}.

%Below 5800 \AA, there is no distinct signal from the He core, and the only real difference to the pure H model is that the H model has a stronger H-beta line. At longer wavelengths, some signals start to appear. 

\arja{The comparison confirms that Mg I] 4571, [O I] 6300, 6364 + Fe I 6364, He I 7065, O I 7774, Fe I 5950, the Fe I cluster at 7900-8500 \AA, and [C I] 8727 and [C I] 9850 give signatures of the He core, at least for this epoch of 400d. In particular do the He I 7065, Fe I 7900-8500 \AA\ and [C I] 8727 + 9850 lines provide distinct diagnostics, with almost no emission at all in the pure H nebula. O and Mg lines become stronger in the He core model, but the lines are still there in the H-zone model. Note that synthesized sodium is missing in the model, so the Na I D line in both cases is from primordial sodium.}

%\paragraph*{The O I 8446 line.}
\arja{The O I 8446 (and O I 1.13 $\mu$m, not shown) is stronger in the H-zone model. This flux comes from a resonance in primordial O I. Emission in Ly$\beta$ at 1025.72 \AA\ is absorbed directly in (primordial) O I 1025.76. Fluorescence into 1.13 $\mu$m, 8448 \AA, and 1602 \AA~follow with close to 100\% efficiency. The main implication is that observation of strong O I 8448 and O I 1.13 $\mu$m are not indication of synthesized oxygen. On the contrary, high fluorescence requires emissivity in Ly$\beta$, so rather H-rich gas. The model shows that primordial (solar) abundances of O is sufficient to make a strong line}.

\section{Time evolution} % ===============================================
\label{sec:timeevol}

\begin{figure*}
\includegraphics[width=1\linewidth]{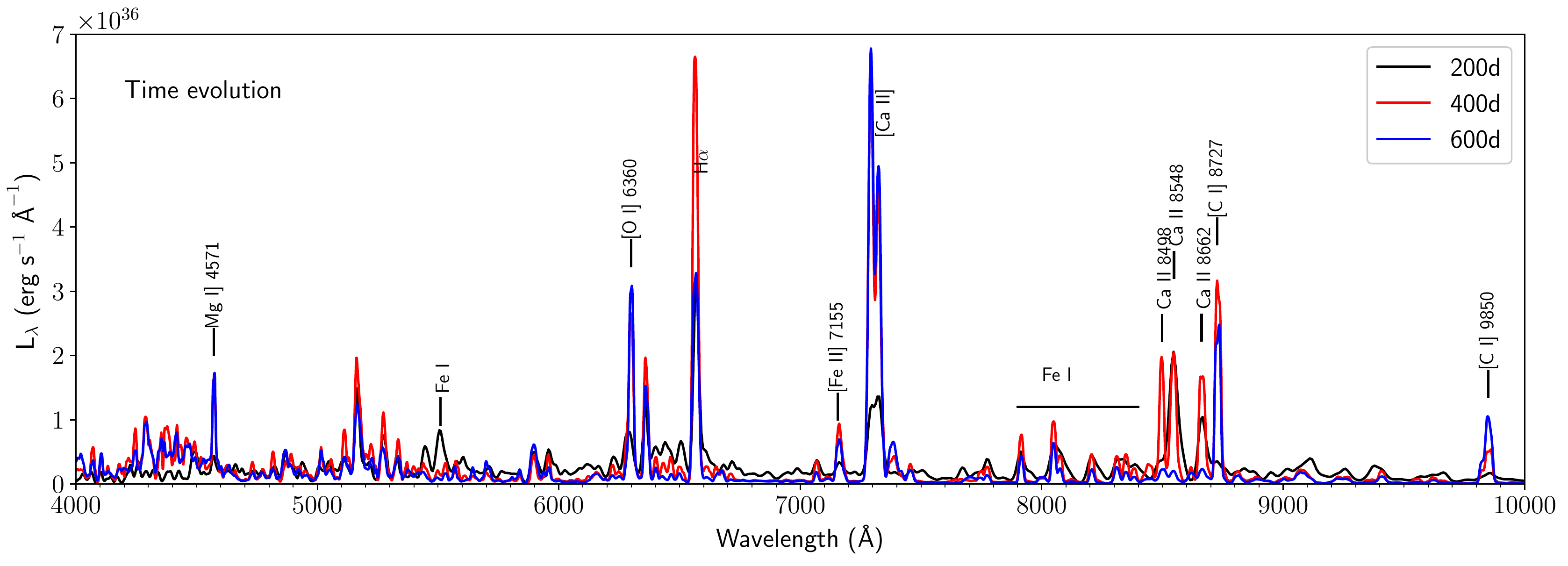} % plotspectra.py 
\caption{\arja{Comparison of the 9 \msun\ model at 200d (black), 400d (red) and 600d (blue). The 200d model has been scaled down by $\exp{\left(-200/111.\right)}$ and the 600d model has been scaled up by $\exp{\left(200/111.\right)}$, to normalize all flux levels to 400d.}}
\label{fig:spectra_11_400}
\end{figure*}

\arja{Figure \ref{fig:spectra_11_400} compares the spectra of model 9 at 200d, 400d, and 600d. Here renormalization in flux levels has been done to emphasize changes in spectral shape, so the 200d model is multiplied with $\exp{\left(-200/111\right)}$ and the 600d model with $\exp{\left(+200/111\right)}$, to synchronize powering levels to 400d}.

%200d.
\arja{At 200d the spectrum is relatively feature-less, with emission lines not yet particularly distinct. 
Mg I] 4571, [O I] 6300, [Ca II] 7291, 7323, and [C I] 8727, [C I] 9850 have yet to distinctly emerge.
Fe I and Ca II almost exclusively make up for the optical lines. As discussed in Sect. \ref{sec:gammadep}, at 200d a significant fraction of gamma rays (30\%) are absorbed by the \ni\ layer. This makes the Fe I cluster at 7900-8500 \AA~particularly distinct at this epoch, one of few signatures of the core structure of the SN at this epoch.}

\arja{Of the bolometric luminosity of $1.45\e{40}$ \ergs, $<$1\% is in the UV ($<$3000 \AA), 68\% is in the optical  ($3000-10,000$ \AA), 25\% is in the NIR (10,000-25,000 \AA), 7\% in MIR}. % output when making spec14.pdf

%400d.
\arja{At 400d Mg I] 4571, [O I] 6300, 6364, [Ca II] 7291, 7323, Ca II 8498, and [C I] 8727, and [C I] 9850 all ``breaking out'' far above the quasi-continuum. 
Most of these lines show a relatively uniform growth between 200 and 400d. The UV, optical, NIR and MIR fractions are now 1.4\% 78\%, 13\% and 7\%. Note that the UV and optical fractions have increased, as reduced line opacity allows more escape of photons at these wavelengths}.

%600d.
\arja{Relatively little evolution is seen  between 400d and 600d. H$\alpha$ has reduced its prominence, the Fe I cluster at 7900-8500 \AA~is now less distinct, and the Ca II NIR triplet has become very weak. The UV, optical, NIR and MIR fractions are now 9.4\% 71\%, 11\% and 9\%}.

\arja{In summary, the spectrum is predicted to gradually switch on to a nebular characteristic sometime between 200d and 400d, with Mg I] 4571, [O I] 6300, 6364,  [Ca II] 7291, 7323, Fe I 7900-8500, and [C I] 8727 all becoming very distinct. Some of these activations are due to delayed gamma ray penetration into the He and H layers (e.g. for [C I] 8727) whereas some are more due to changing physical conditions and reduced optical depths}.

\section{Comparison with observational candidates}
\label{sec:compdata}

\arja{Three low-velocity Type II SNe have nebular spectra published; SN 1997D \citep{Turatto1998, Benetti2001}, SN 2005cs \citep{Pastorello2009}, and SN 2008bk \citep{Maguire2012}. Here, we compare our models in turn to these three.} 

\subsection{SN 1997D}
\arja{SN 1997D was discovered on January 14 1997 (MJD 50462) in NGC 1536 by \citet{deMello1997}. It was then already at the end of the plateau phase, so the explosion epoch is ill constrained}.

%\textit{Extinction.} 
\arja{No distinct Na I D could be seen \citep{Turatto1998}, and we use only the galactic extinction component $E(B-V)=0.02$ mag \citep{Zampieri2003} here.
% No mention of extinction in Benetti2001, seems to assumed 0.
% Zampieri 2003: E(B-V) = 0.02
%Distance. 
NGC 1536 has only a single distance measurement in NED\footnote{https://ned.ipac.caltech.edu}, at 13.4 MPc \citep{Tully1988}, which we use. %One should note that the stated distance error is $\delta \mu=0.8$mag, which means a factor 1.5}.
%Recession velocity. 
We assume a recession velocity of 1217 \kms (NED). %The Hubble flow distance at 1217 \kms~is 17.5 MPc, factor 1.3 higher than the Tully-Fischer value}.
%\ni~mass}. 
Regarding \ni~mass, \citet{Turatto1998} and \citet{Benetti2001} estimate $2\e{-3}$ \msun~by comparing optical photometry to SN 1987A, with what appears to be an assumed discovery age of 50d. This is a factor 3 lower than in our model, and we therefore expect to have to scale all spectra by a factor $\sim$1/3 to reach flux agreement. This is larger than desired, but we demonstrate in the appendix by comparing P-HOTB and KEPLER models that lowering by the \ni~mass by a factor 2 has only limited impact. We note, however, that \citet{Zampieri2003} inferred $8\e{-3}$ \msun~with an assumed discovery age of 100d.}

\textit{Nebular data set.} \arja{Three nebular-phase spectra were presented in \citet{Benetti2001}, on 1997-05-01 (MJD 50569), 1997-09-21 (MJD 50712) and 1998-02-02 (MJD 50846). Assuming an explosion epoch of 100d before discovery (MJD 50362), giving a typical plateau length of 110d, this implies epochs of +207d, +350d, and +484d. The spectral resolution varied between 7-17 \AA~over the data sets and the wavelengths. Taking 12 \AA~at 6000 \AA~as representable numbers, the resolution is about 600 \kms}.
% Zampieri 2003: Comparison to 1999br suggests 1997D detected at 90-100 days post explosion. Gets 56Ni mass of 0.008.

\arja{The +207d and +350d spectra show reasonable agreement with contemporary photometry in \citet{Benetti2001}. The +484d spectrum is, however, a factor 4.5 brighter than the contemporary V-band photometry, presumably due to some issue with the flux calibration. The spectrum was therefore scaled down by a factor 4.5. This produced consistency in the amount of rescaling needed for the model comparisons.}

\begin{figure*}
\includegraphics[width=1\linewidth]{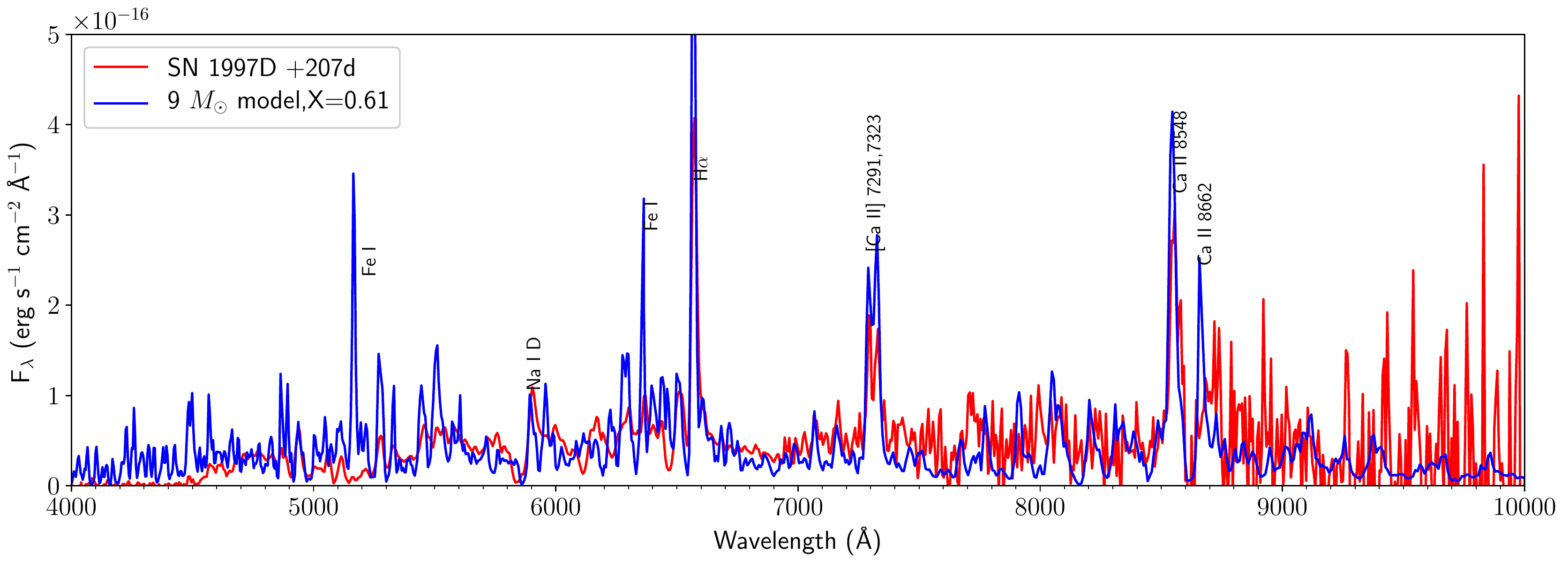}
\includegraphics[width=1\linewidth]{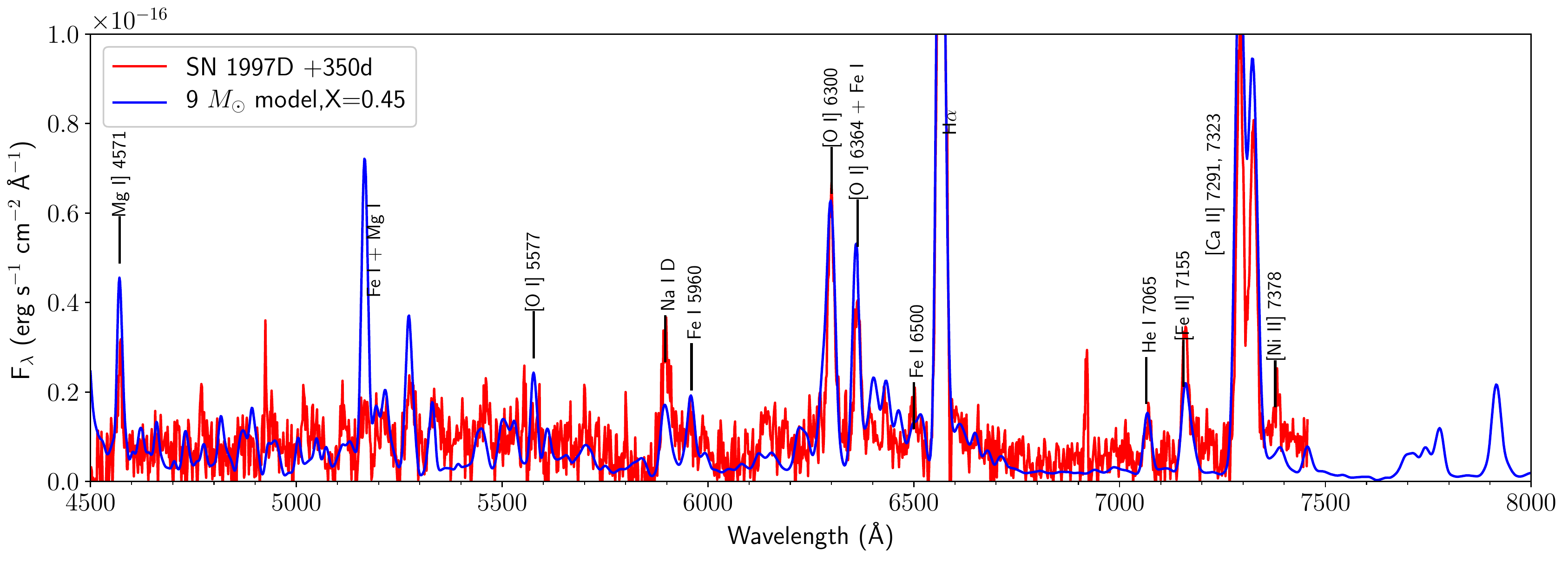}
\includegraphics[width=1\linewidth]{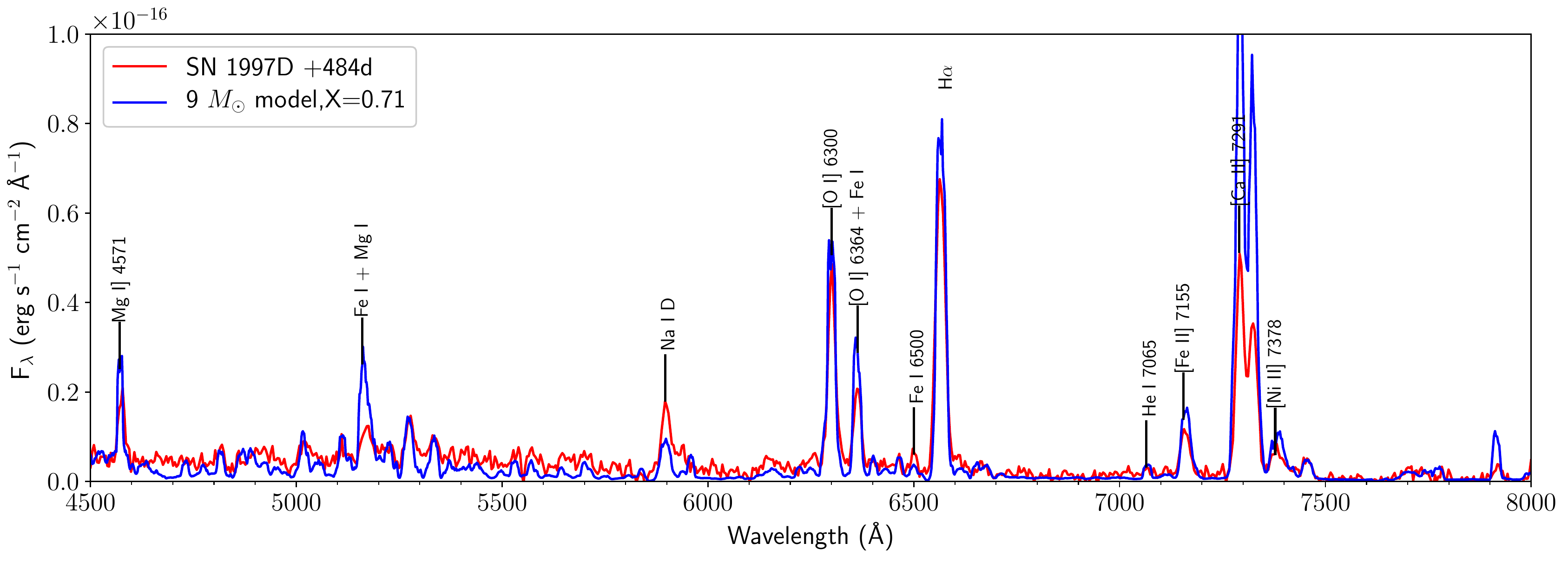}
\caption{\arja{Top: SN 1997D at +207d (red) compared to the 9 \msun\ model (blue), at 200d rescaled with $\exp{\left(-7/111\right)}$ and a factor $X=0.61$ (to compensate for the lower \ni~mass). Middle: SN 1997D at +350d (red) compared to the 9 \msun\ model (blue), at 300d rescaled with $\exp{\left(-50/111\right)}$ and $X=0.45$. Bottom: SN 1997D at +484d (red) compared to the 9 \msun\ model (blue), at 500d rescaled with $\exp{\left(+16/111\right)}$ and X=0.71}. The model spectra have been convolved with a Gaussian with FWHM=600 \kms~to match the telescope resolution.}%...Well reproduced lines are labeled in black text, overproduced in blue, and underproduced in red.}}
\label{fig:1997D}
\end{figure*}

\arja{Figure \ref{fig:1997D} shows comparisons of the three nebular spectra with the 9 \msun\ model. Here the model has been convolved with a Gaussian with FWHM=600 \kms~to match the telescope resolution.}

\begin{figure*}
\includegraphics[width=0.32\linewidth]{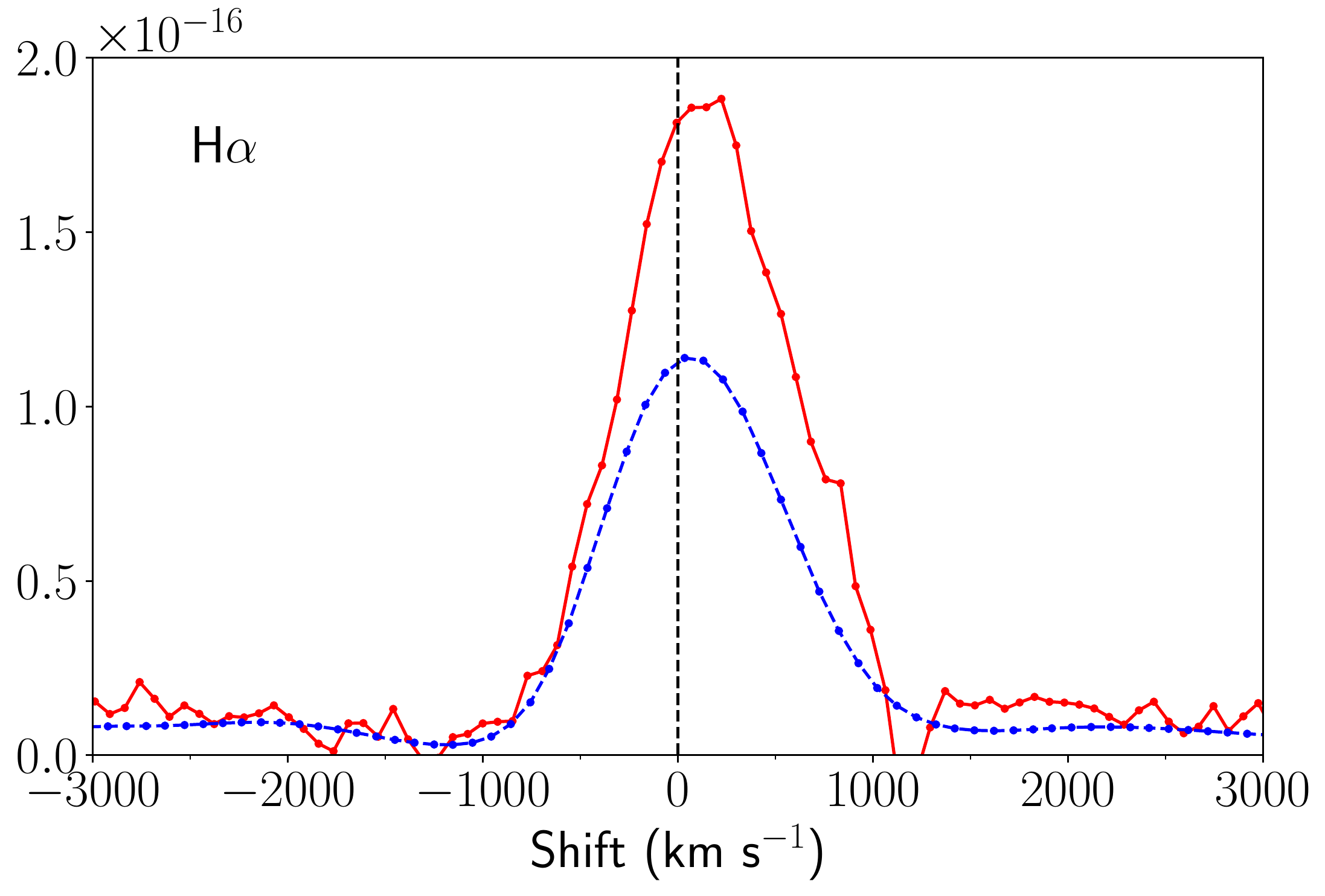} 
\includegraphics[width=0.32\linewidth]{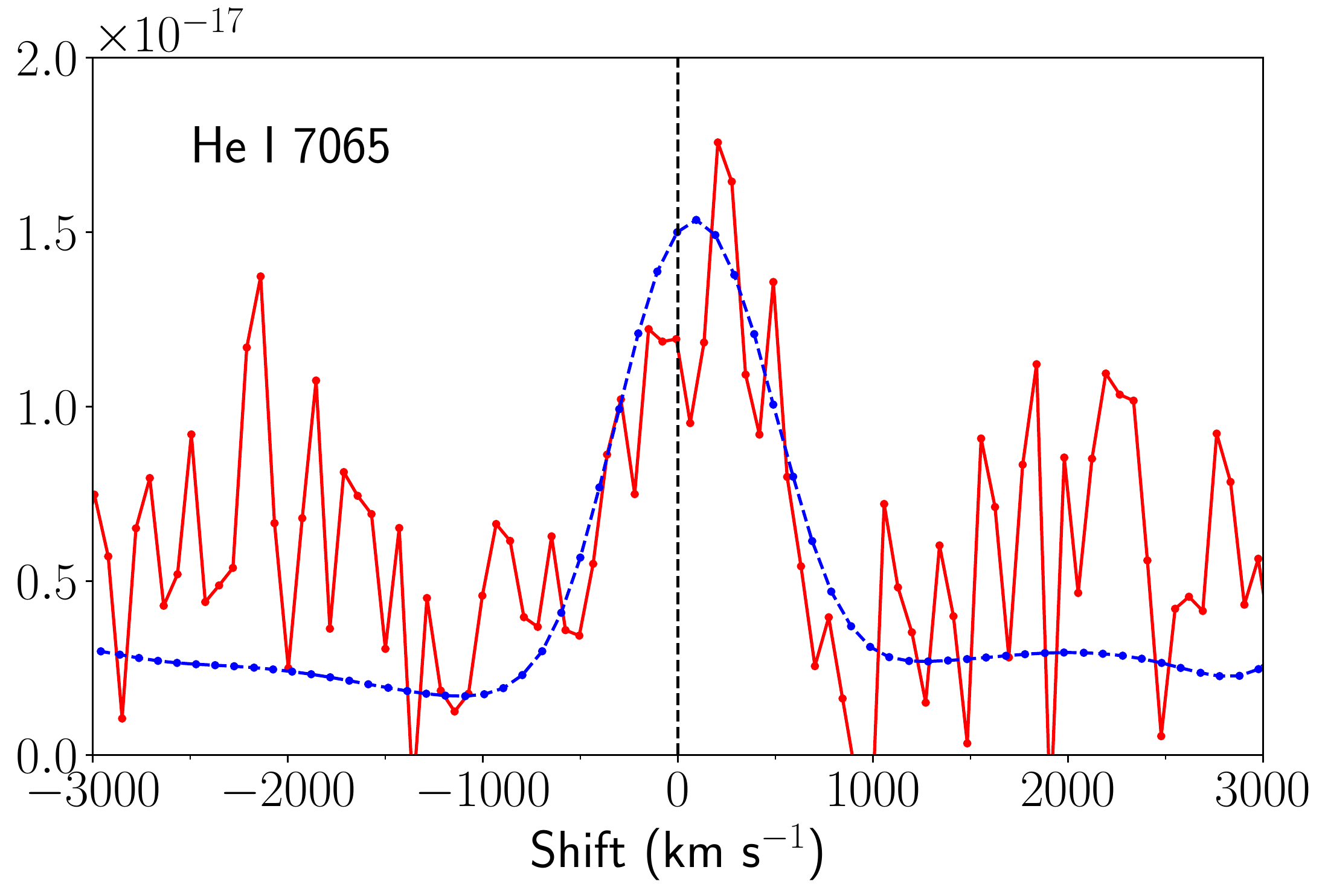} 
\includegraphics[width=0.32\linewidth]{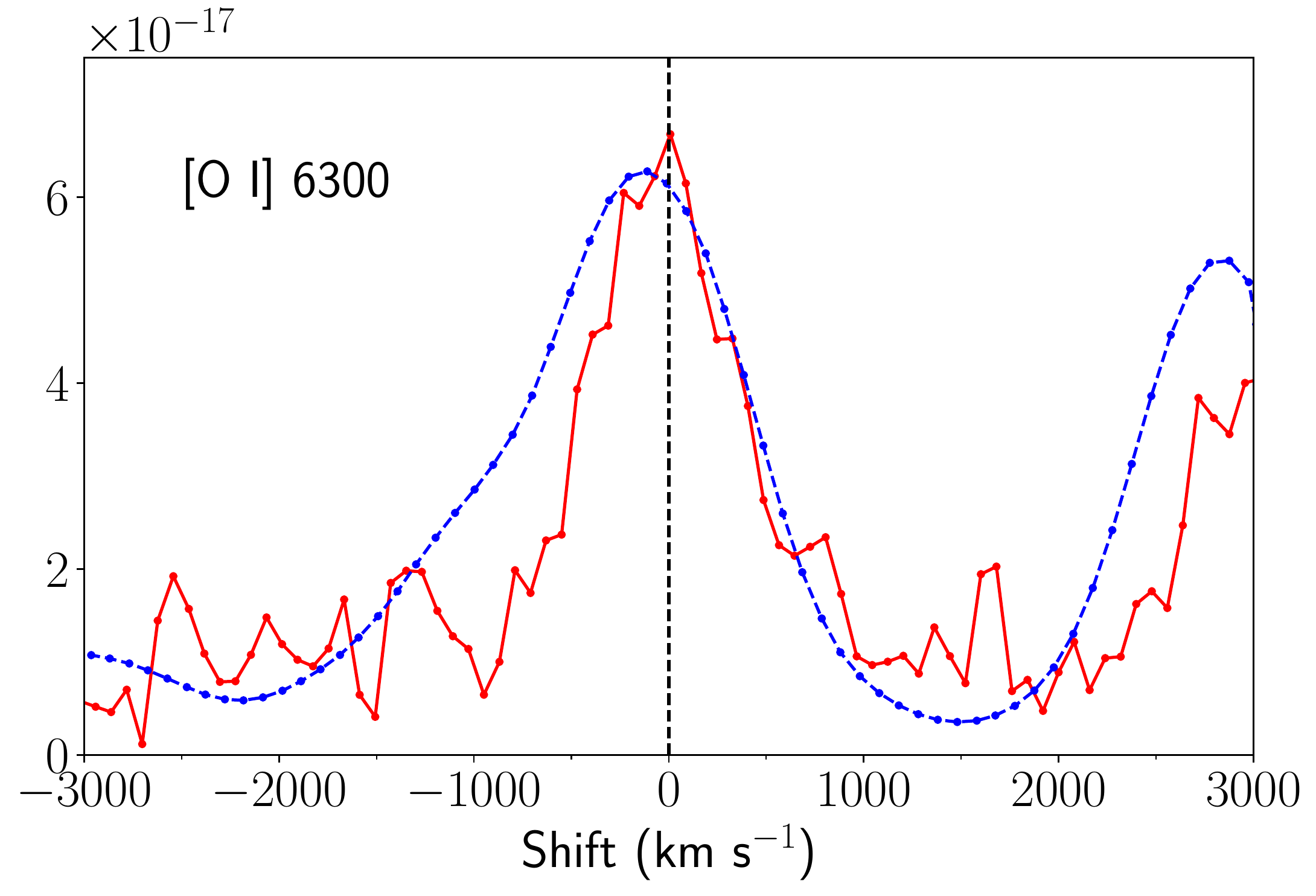} 
\includegraphics[width=0.32\linewidth]{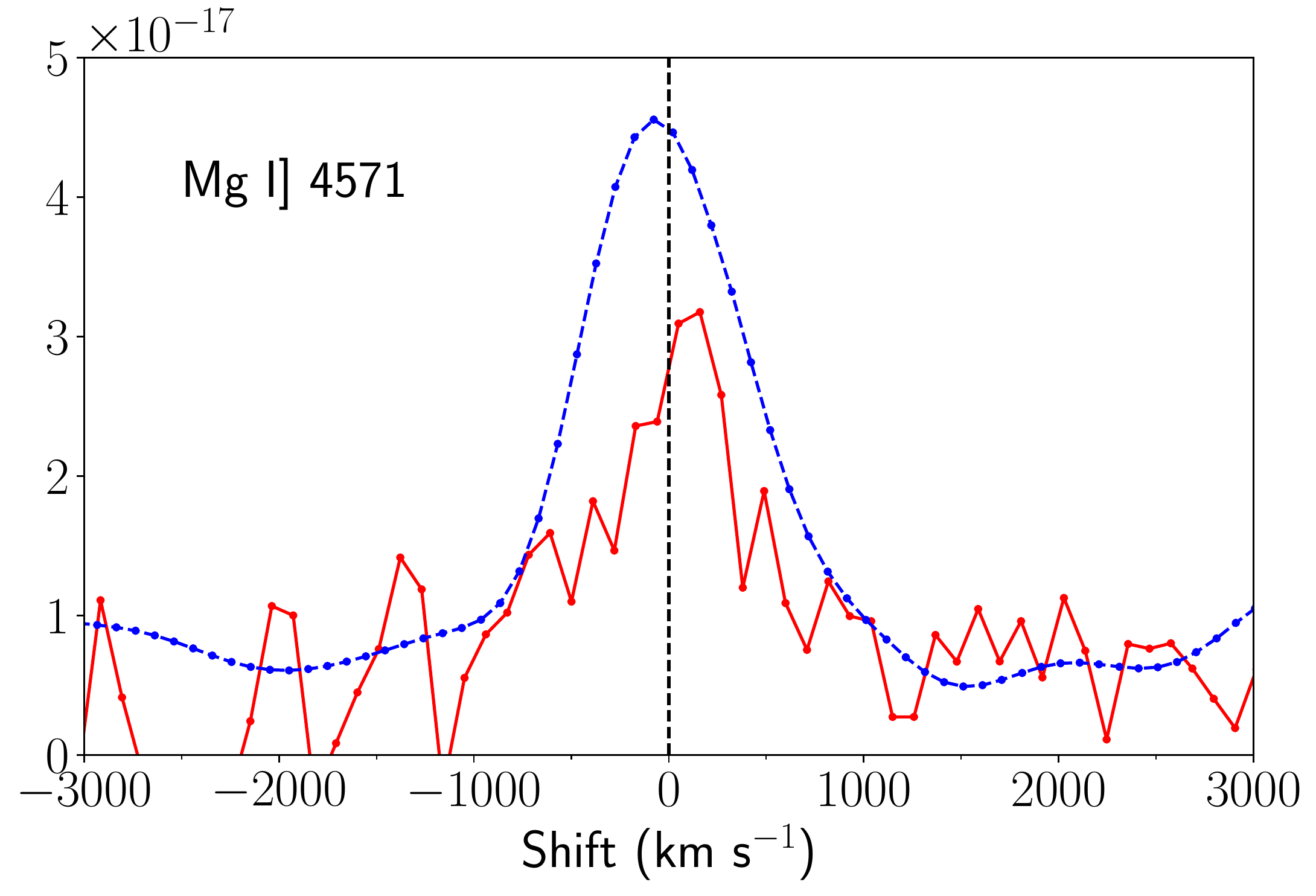} 
\includegraphics[width=0.32\linewidth]{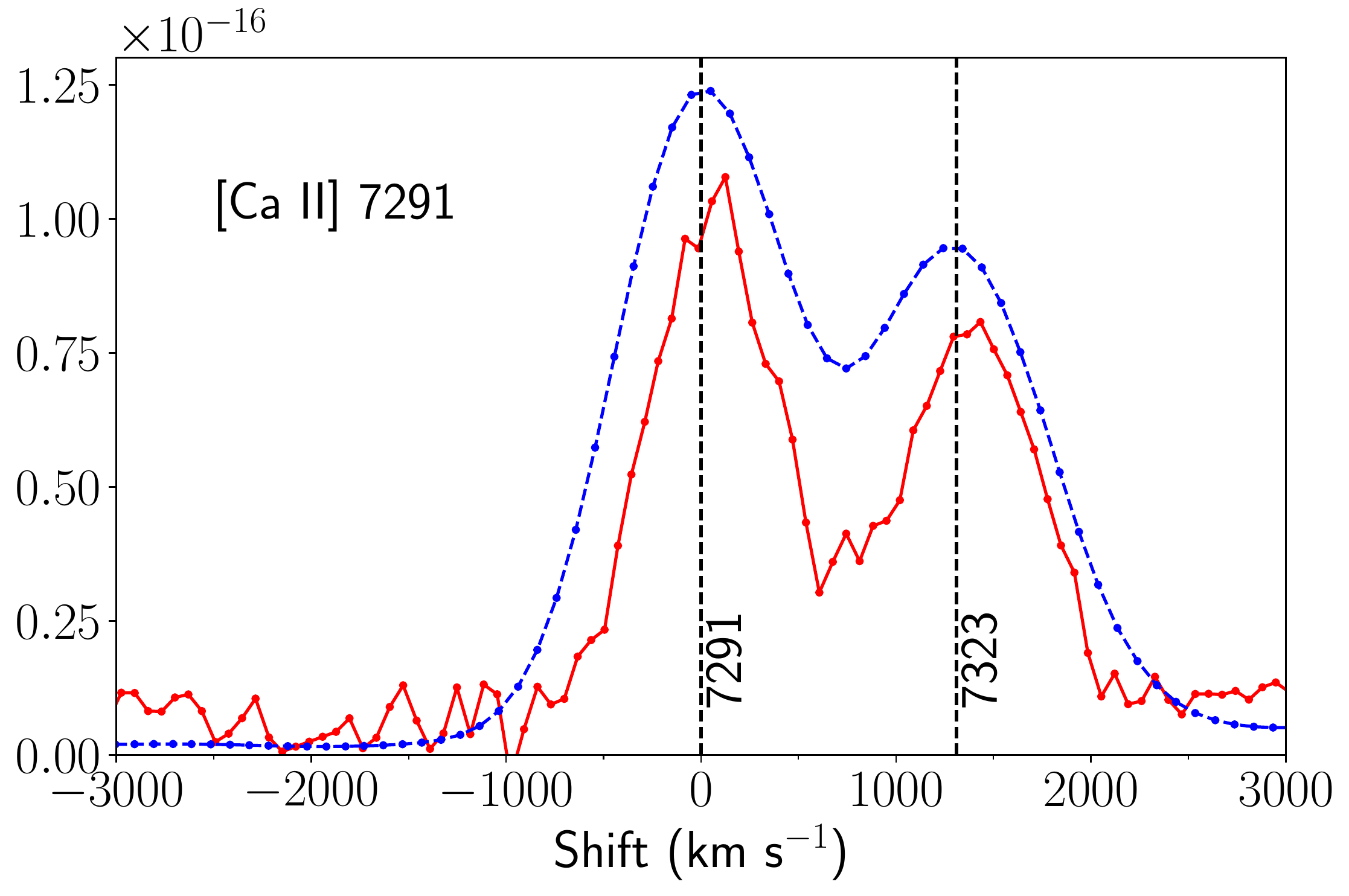} 
\includegraphics[width=0.32\linewidth]{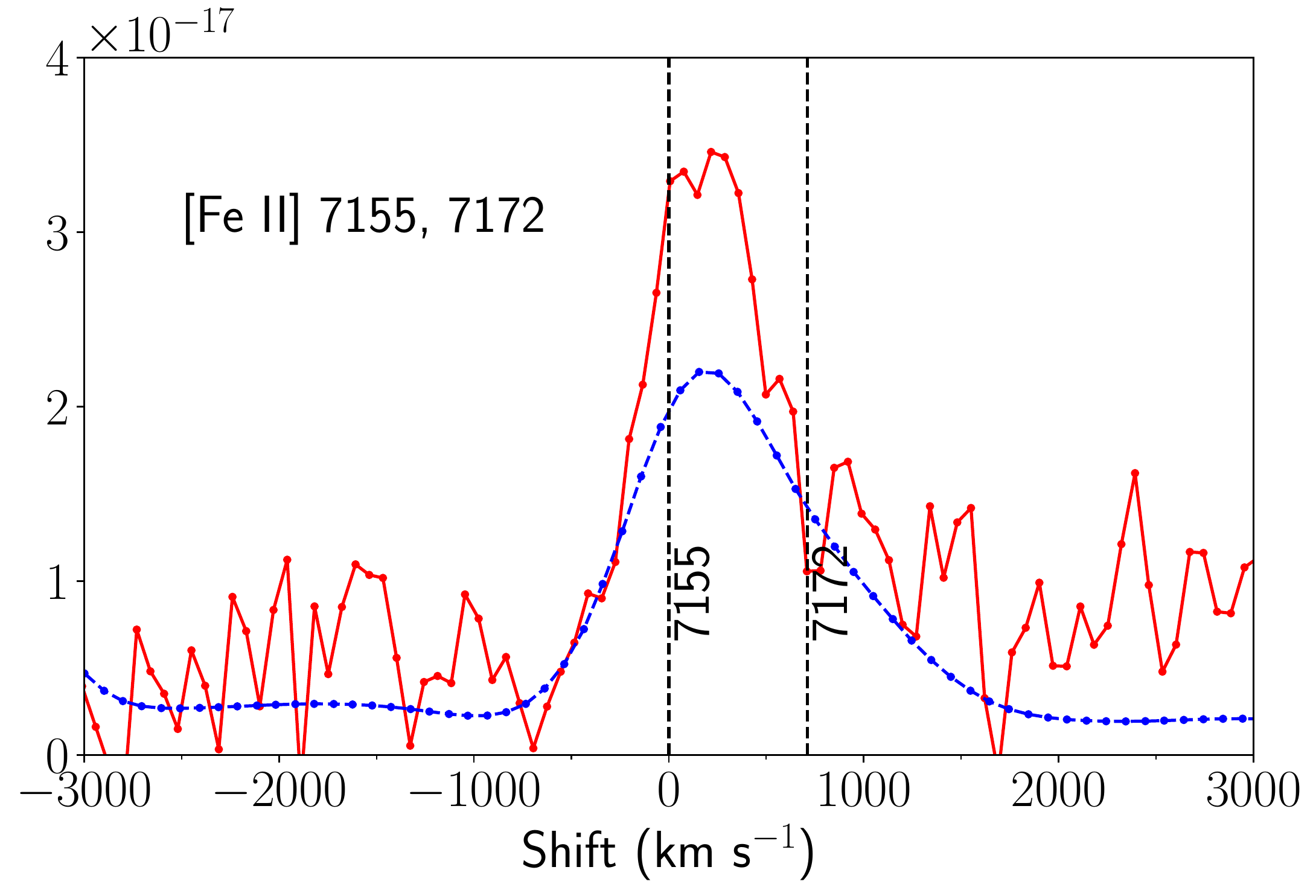} 
\caption{\arja{Zoom-in on line profiles at +350d in SN 1997D (red, solid) and in the 9 model (blue, dashed). The model spectrum has been convolved to the telescope resolution of 600 \kms.}}
\label{fig:zoom1997D}
\end{figure*}

\subsubsection{+207d}
\arja{At 207d a model scaling of $X=0.61$ is required to match the overall optical luminosity, which corresponds to a \ni~mass of $3.8\e{-3}$ \msun.
The color agreement between model and data is reasonably good, although the model is somewhat bluer. There is also reasonably matching line profiles and luminosities for H$\alpha$, [Ca II] 7291, 7323, and Ca II 8548, 8662. Note how Ca II 8498 is suppressed both in the model and in the SN. The 1:1 ratio of [Ca II] 7291 and 7323 means the lines are still optically thick}.

\arja{The observed spectrum is noisy beyond 7500 \AA, and there are not many more clearly identified emission lines. At this epoch [O I] and Mg I] lines have not clearly emerged yet, neither in model or in the observed spectrum. In the blue region, $\lesssim 5500$ \AA, there is somewhat too much ``fine-structure'' in the model, with too many strong narrow lines. The model makes two strong iron lines at 5180 and 6364 \AA\ that are not observed}.

\subsubsection{+350d}
\arja{At 350d a scaling $X=0.45$ is required (\ni\ mass $2.8\e{-3}$ \msun). Mg I] 4571 and [O I] 6300, 6364 have now emerged and the model predictions match the observed spectrum well. Detection of He I 7065 is now clearly confirmed (marginally seen at 207d). All these lines clarify the presence of He core material. Also [Ni II] 7378 is distinct in this observed spectrum}.

\arja{Figure \ref{fig:zoom1997D} shows zoom-ins of selected strong lines at this epoch. H$\alpha$ fits well, just slightly too broad in the model. The flat-top has been washed out by the 600 \kms\ convolution, and a rounded peak is seen also in the data. The emitting region is clearly bounded at about 1000 \kms, which supports the model calculation that gamma rays cannot penetrate further}.
% Direct measurement at Y=0.5 --> FWHM = 670 - (-320) = 1020 km/s. Model : 1100

\arja{He I 7065 shows good agreement. From the observed line profile the He emitting region is smaller, $V\lesssim 600$ \kms, supporting the stratified structure with the He shell interior to the H envelope}.
% Direct measurement at Y=0.5 --> FWHM = 580 - (-375) = 950 km/s. Model : 900

\arja{While the red wing of [O I] 6300 shows agreement, there is contamination in the model by an Fe I line on the blue side that is not observed.}
% Direct measurement at Y=0.5 --> FWHM = 437 - (-500) = 940 km/s. Model : 900

\arja{The [Ca II] lines are about 25\% too broad in the model. The contribution at high velocities in the model is due to the inner H envelope, but such a component is not seen in the data. This is a difficult discrepancy to resolve since [Ca II] is normally an important cooler of H gas.}
% Direct measurement at Y=0.5 --> FWHM = 570 - (-160) = 820 km/s. Model : 900

\arja{Finally, the observed [Fe II] 7155 line has formally a lower FWHM (730 \kms) than the previously discussed lines, which supports a significant contribution by synthesized iron in a region interior to O, He and H layers}.

\arja{Table \ref{table:1997DFWHM} summarizes the observed FWHM, and comparison with the model ones. Here, the ``deconvolved'' FWHM is obtained from $\mbox{FWHM}_{obs}^2 = \mbox{FWHM}_{dec}^2 + \mbox{FWHM}_{teles.}^2$, which holds if both intrinsic line and instrument PSF are Gaussians. Overall the model gives somewhat too broad lines.}
% Direct measurement at Y=0.5 --> FWHM = 570 - (-160) = 730 km/s. Model : 800

\subsubsection{+484d}
\arja{At 484d there is continued matching to Mg I and [O I] lines. There is no strong change in the rescaling required ($X=0.69$), which means that there is no indication of dust formation or gamma ray escape yet. Note that the [Ni II] 7378 line is made by primordial nickel in the model. One discrepancy is that, whereas the observed [O I] 6300 to 6364 ratio is now approaching the optically thin 3:1 ratio, the model ratio is still close to 1:1 due to Fe I line contamination. While a factor $\sim$2 too high \ni~mass in the model is one factor in this overproduction, there likely remains other discrepancies that makes too strong Fe I lines in the model}.

\subsubsection{Discussion}
Overall, this 9 \msun~model, with no tuned parameters at all to fit the data, provides a convincing spectral fit to SN 1997D, which strengthens the hypothesis of an origin as a low-mass star. Both line luminosities and line profiles are in good agreement with the model.
The estimated FWHM in Table \ref{table:1997DFWHM} implies an emitting region $\sim$ 400-800 \kms, close to the size of the gamma-trapping region in the model.

Although there are no observations of the spectral region containing carbon lines, He core material is convincingly established by the presence of He I 7065, Mg I] 4571 and [O I] 6300, 6364, with the right expected luminosities. The presence of these lines suggest that SN 1997D was an Fe CCSN, not an ECSN.

Taking the average scaling factor $X$ we need to match the model flux level to each observed spectrum, our estimate for the \ni~mass in SN 1997D becomes $3.6\e{-3}$ \msun. The model, with $6.2\e{-3}$ \msun, gives slightly too broad lines, which is consistent with a higher \ni~mass than observed, as this is linked to a higher explosion energy.

\begin{table}
\centering
\begin{tabular}{cccc}
\hline
Line & FWHM  & $\mbox{FWHM}_{dec.}$  & Model\\
        & (\kms)              & (\kms)          & (\kms) \\
\hline
H$\alpha$ & 1020 & 820 & 1100\\
He 7065 & 950 & 740 & 900\\
O I 6300,6360 & 940 & 720 & 900\\
Ca II 7291 & 820 & 560 & 900\\
Fe II 7155 & 730 & 420 & 800\\
\hline
\end{tabular}
\caption{Observed line profile widths in SN 1997D, at +350d, compared to the model (unconvolved) values.}
\label{table:1997DFWHM}
\end{table}

\subsection{SN 2005cs} % =====================================================
SN 2005cs was discovered on June 28 2005 (MJD 53549) by \citet{Kloehr2005} in the Whirlpool galaxy (M51), with pre-explosion limits of about 1 day earlier \citep{Pastorello2006}. We take the explosion epoch as June 27 2005 (MJD 53548).

\textit{Nebular data set.} \citet{Pastorello2009} present nebular spectra at intervals of $\sim$2 months up to +333d, all available on Wiserep \citep{Yaron2012} \footnote{http://wiserep.weizmann.ac.il}. To resolve the intrinsic narrow-line structure of these low-velocity SNe, it is beneficial to use the highest-resolution spectra taken. These (FWHM$\sim$300 \kms) correspond to the observations on 2006-01-07 (+194d), 2006-03-25 (+270d), 2006-04-01 (+277d), and 2006-04-05 (+281d). The similar epochs of the last three makes it sufficient to study one, which we choose as day +277. The spectra at +226d and +333d are at lower resolution. A NIR spectrum exists at +281d, which we also study. We convolve the model spectra with matching Gaussians of FWHM=300 \kms~for the comparisons.

The raw nebular spectra are heavily contaminated by background light (Pastorello, priv. comm.). The extraction therefore involved the subtraction of a blackbody that was fit to the estimated photometry of the contaminating light (from difference in synthetic photometry on total spectra and template-subtracted photometry of the SN). The combination of a large contrast between background/SN light, and lack of a template spectrum of the region means there is significant uncertainty in the resulting SN spectrum.
For example, at +277d the contamination is 90-95\% of the flux in the blue, and $\sim$70\% in the red (Pastorello, priv. comm.). While errors due to continuum subtraction is moderate in the red, in the blue the uncertainty is much larger. For example, a 5\% error in the continuum estimate (say from 90 to 95\%) results in a factor 2 error in the SN residual (from 5 to 10\%). The consequence is that we have to treat the spectral data in the blue with caution, but can trust it more in the red. Agreement between synthetic photometry on the spectrum and template-subtracted photometry is generally good, but slit losses or flux calibration errors cannot be checked with this method, and any such loss would be compensated by added background light. Focus for model comparison should therefore be on line luminosities, to factor $\sim$2, but less on spectral shape, particularly in the blue. 

Figure \ref{fig:97D_05cs} in the appendix compares the first SN 2005cs spectrum at +194d to SN 1997D at a similar epoch. While SN 1997D also has a declining blue tail, it is less dramatic, and the overall spectrum is bluer. There is also more fine-structure in the blue in SN 1997D. Figure 7 in \citet{Pastorello2009} shows that subluminous IIP SNe differ color-wise to normal IIP SNe mainly in the early tail phase (120-200d), where they are 0.5-1 mag redder in $B-V$ and $V-R$. SN 2005cs is the most extreme case in $V-R$, while SN 1997D has similar values for $B-V$. The spectrum does have the correct $V-R$ value.
 
\textit{Extinction.} \citet{Maund2005} estimated $E(B-V)$ = 0.16 mag from Na I D lines, and $E(B-V)$ = 0.12 mag from analysis of surrounding red supergiants. Other authors have found similar values $E(B-V) = 0.05-0.2$ mag from the Na I D lines  \citep{Li2006, Pastorello2006}. \citet{Baron2007} found $E(B-V) = 0.035-0.05$ mag from spectral modelling, and \citet{Brown2007} and \citet{Dessart2008} had similar results. A reasonable choice to use is the average value of these estimates, $E(B-V) \sim 0.1$ mag, which we use here. All three methods consistently give $E(B-V) < 0.2$ mag which robustly tells us that the extinction is small/moderate and does not introduce any major uncertainty.% Brown 2007, Dessart 2008 : support Baron 2007 conclusion of low extinction

\textit{Distance and recession velocity.} We use the median NED distance of 8.1 Mpc (standard deviation 1.2 MPc). The distance has been estimated by PNLF, SBF, TRGB, EPM, SEAM, and Tully-Fisher methods with relatively consistent results. Apart from SN 2005cs, also SN 1994I and SN 2011dh exploded in this galaxy, giving multiple occassions to use the EPM and SEAM methods.
Finally, we Doppler correct the spectra with 600 \kms (NED). % Extinction: 

It should be mentioned here that the estimated bolometric luminosity of the SN declined slower than $^{56}$Co decay \citep{Pastorello2009} in the tail phase, thus making it unclear whether residual energy from the explosion or other energy sources were important at late times. We do not attempt to consider any other sources here, but mention this caveat/uncertainty. \citet{Utrobin2007} find from models a possible explanation to be residual energy from the shock passage still leaking out in the early tail phase, although in the models this effects last for a few weeks rather than a few months as observed in SN 2005cs. After 330d, the observed decline rate steepened somewhat, perhaps going onto a true $^{56}$Co tail \citep{Pastorello2009}, although with only two observed epochs between 330 and 400d, it is hard to tell, and the
bright background also introduces significant uncertainties even in template-subtraction approaches.

This deviation from a $^{56}$Co decline rate makes it unclear if there even is any $^{56}$Ni in SN 2005cs. One cannot a-priori rule out that the late-time luminosity is generated by circumstellar interaction, for instance. This issue makes it important to see whether we can identify any lines unambigously coming from a $^{56}$Ni region in the SN.

\begin{figure*}
\includegraphics[width=1\linewidth]{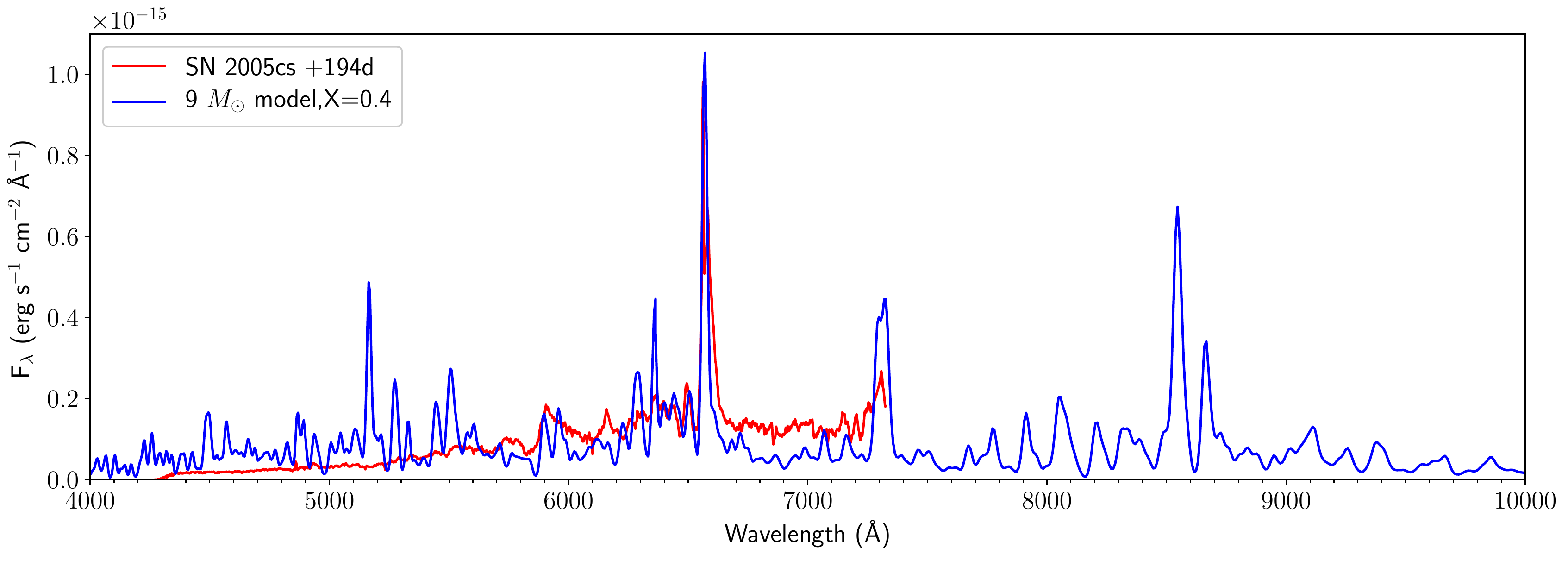}
\includegraphics[width=1\linewidth]{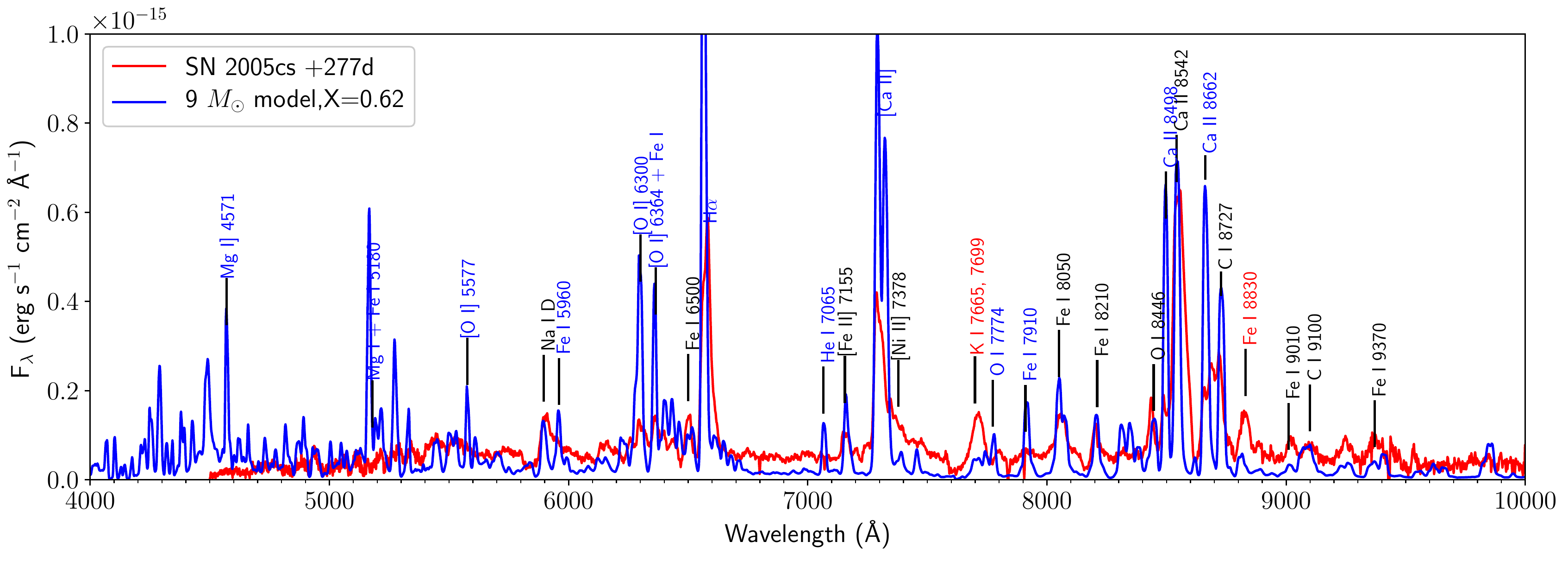}
\caption{\arj{Top: SN 2005cs at +194d (red) compared to the 9 \msun\ model (blue), at 200d rescaled with $\exp{\left(6/111\right)}$ and $X=0.4$. Bottom: SN 2005cs at +277d (red) compared to the 9 \msun\ model (blue), at 300d rescaled with $\exp{\left(23/111\right)}$ and $X=0.61$. The model spectra have been convolved with a Gaussian with FWHM=300 \kms~to match the telescope resolution. Well reproduced lines are labeled in black text, overproduced in blue, and underproduced in red.}}
\label{fig:2005cs}
\end{figure*}

\begin{figure*}
\includegraphics[width=1\linewidth]{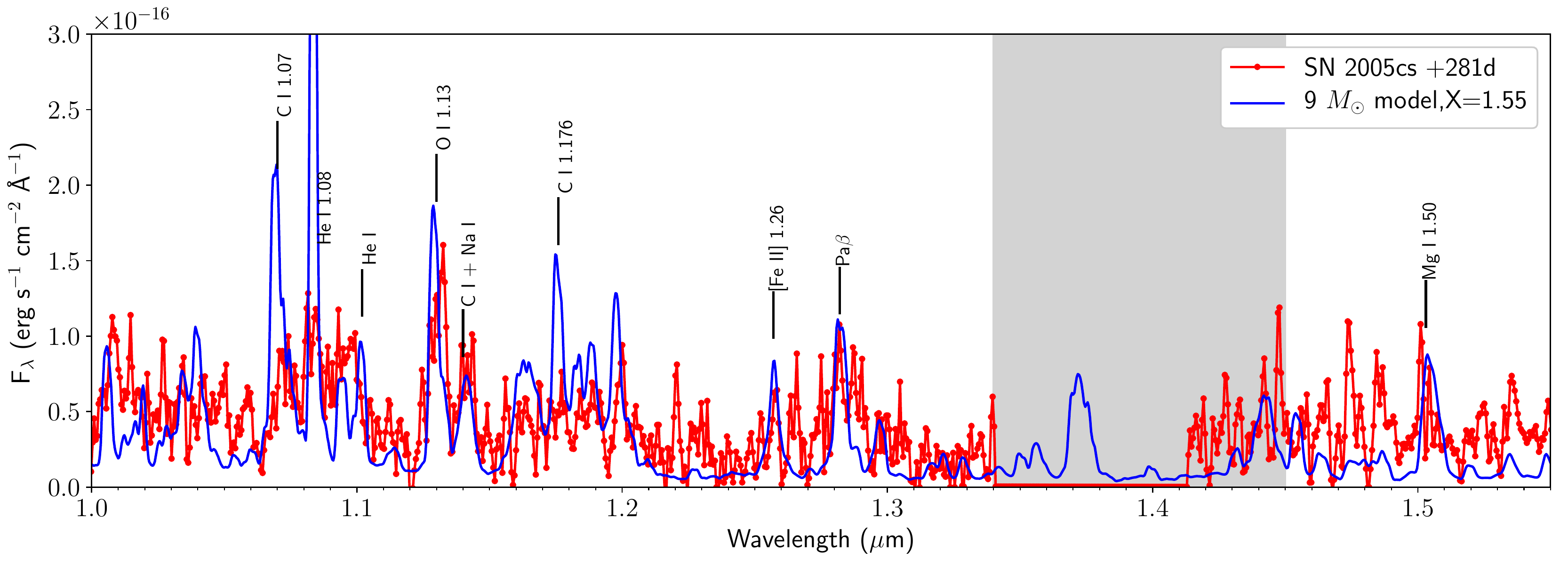}
\includegraphics[width=1\linewidth]{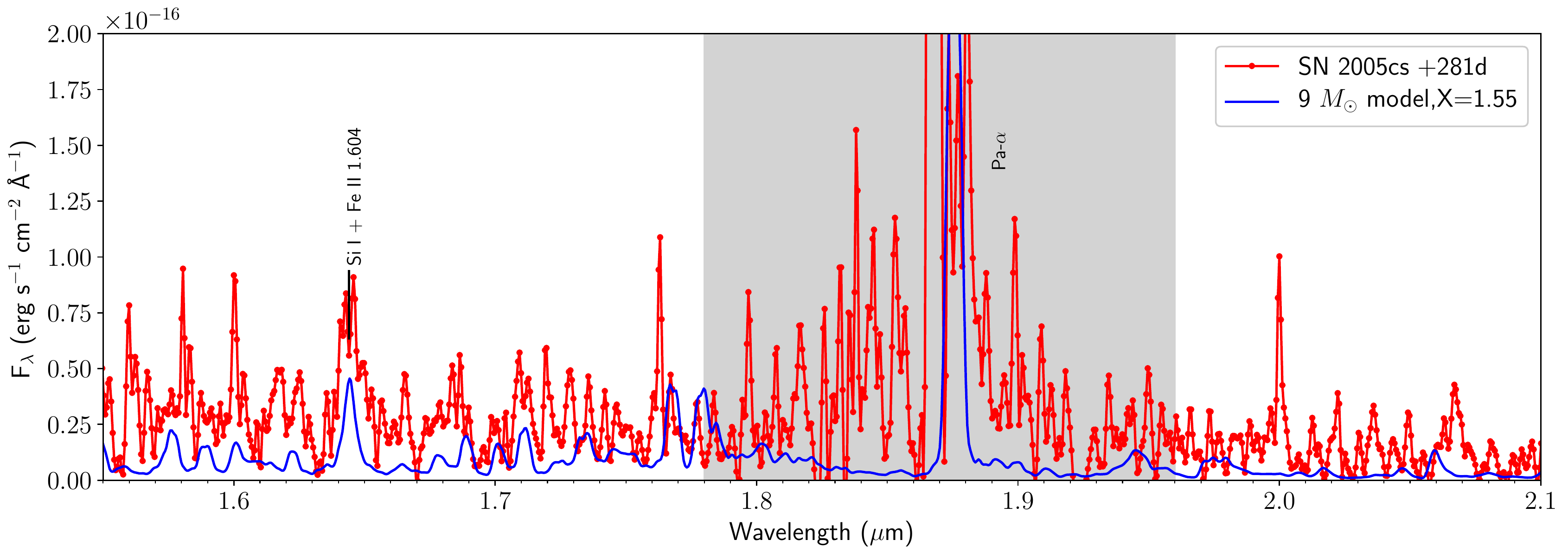}
\caption{\arj{SN 2005cs at +281 in the NIR (red) compared to the 9 \msun\ model (blue).} The model spectra have been convolved with a Gaussian with FWHM=300 \kms~to match the telescope resolution.}
\label{fig:2005csNIR}
\end{figure*}

\begin{figure*}
\includegraphics[width=0.32\linewidth]{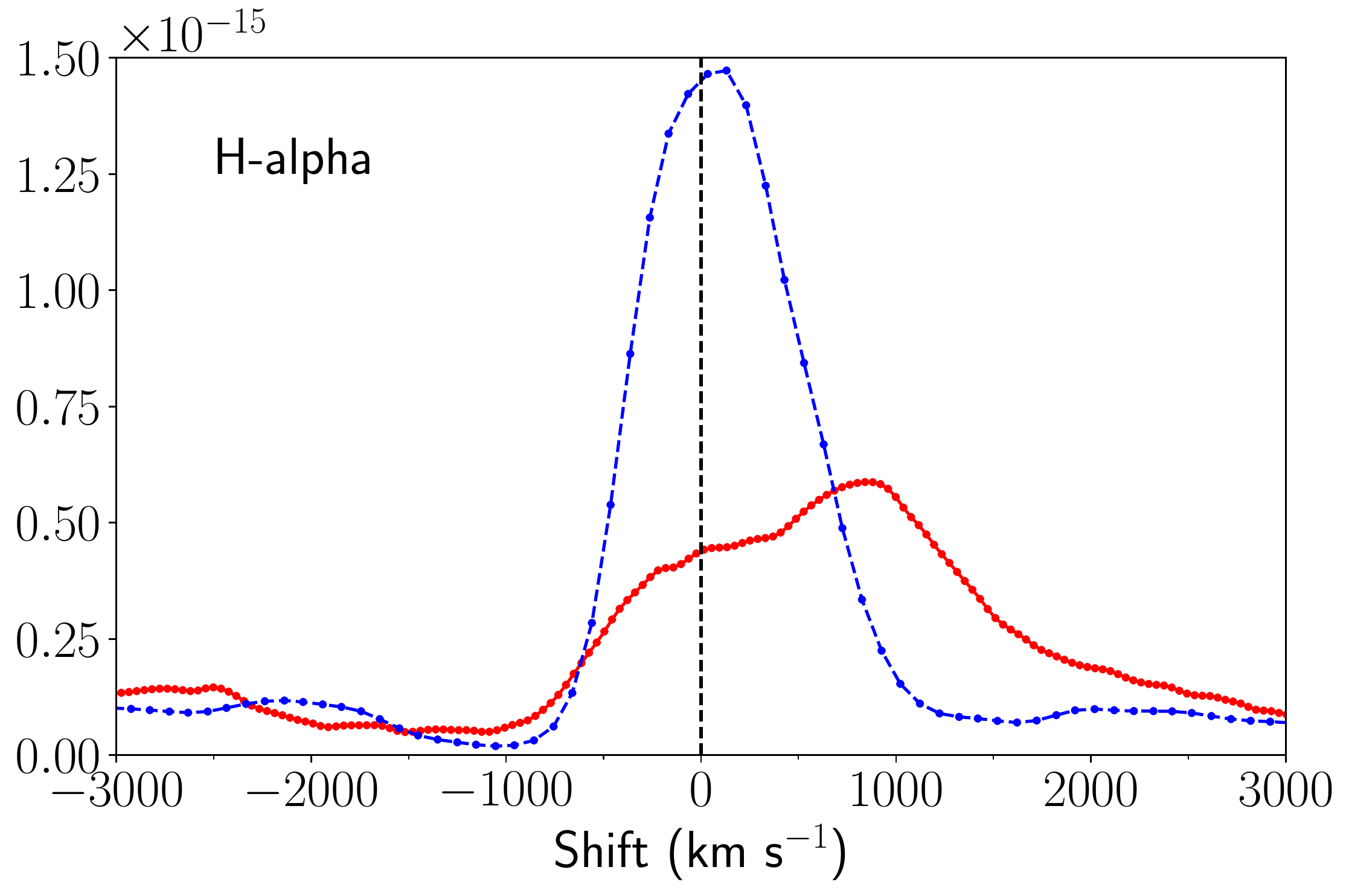} 
\includegraphics[width=0.32\linewidth]{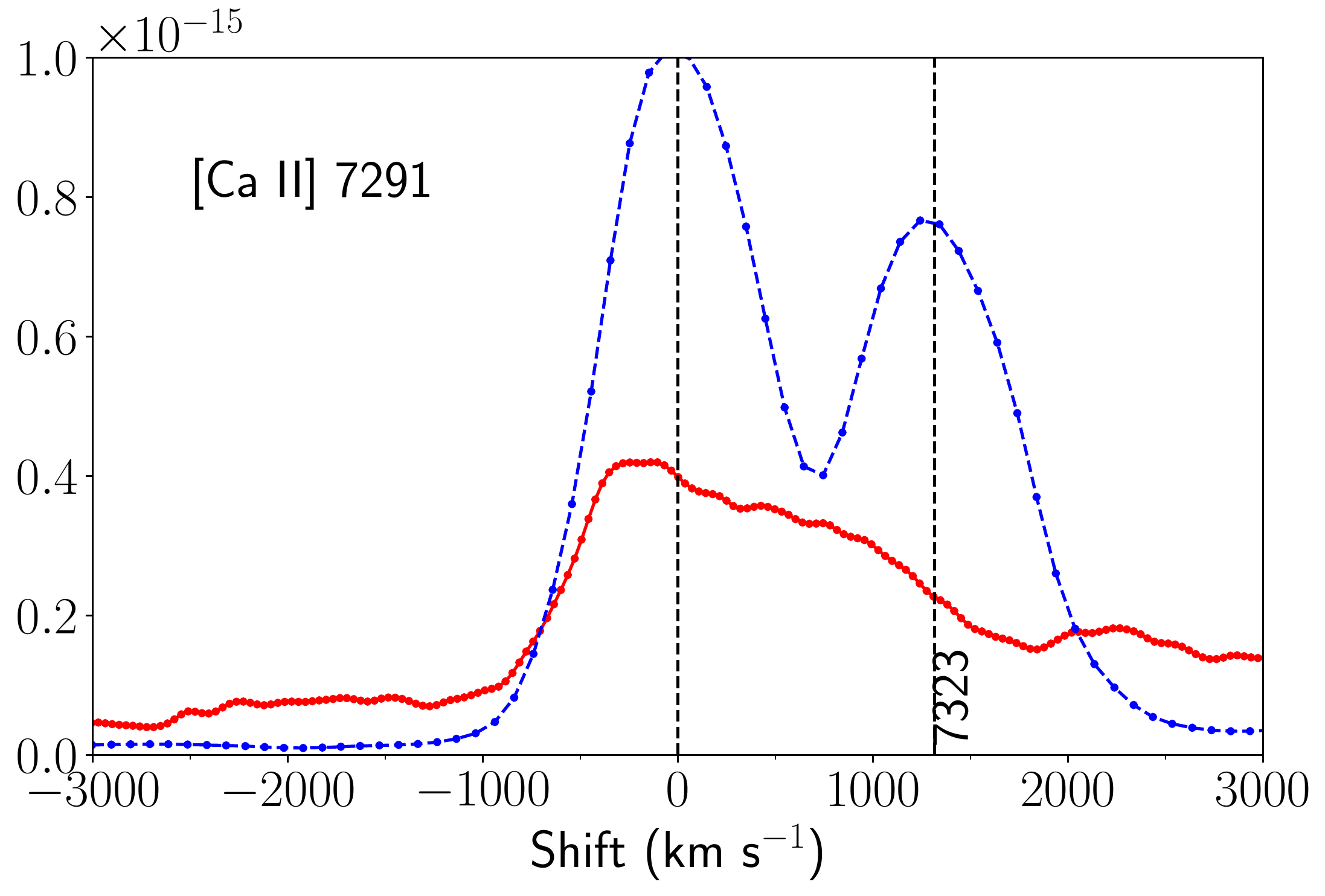} 
\includegraphics[width=0.32\linewidth]{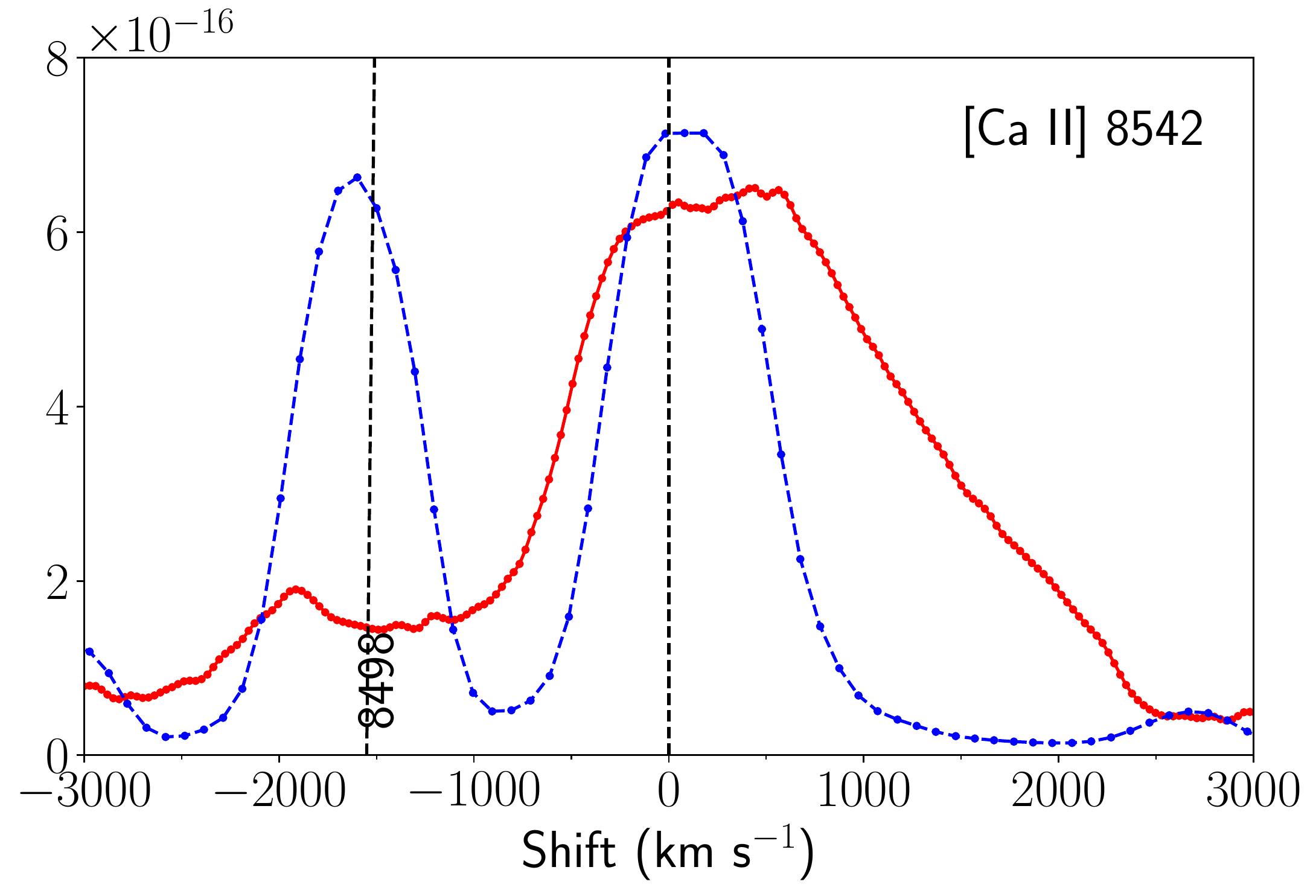} 
\includegraphics[width=0.32\linewidth]{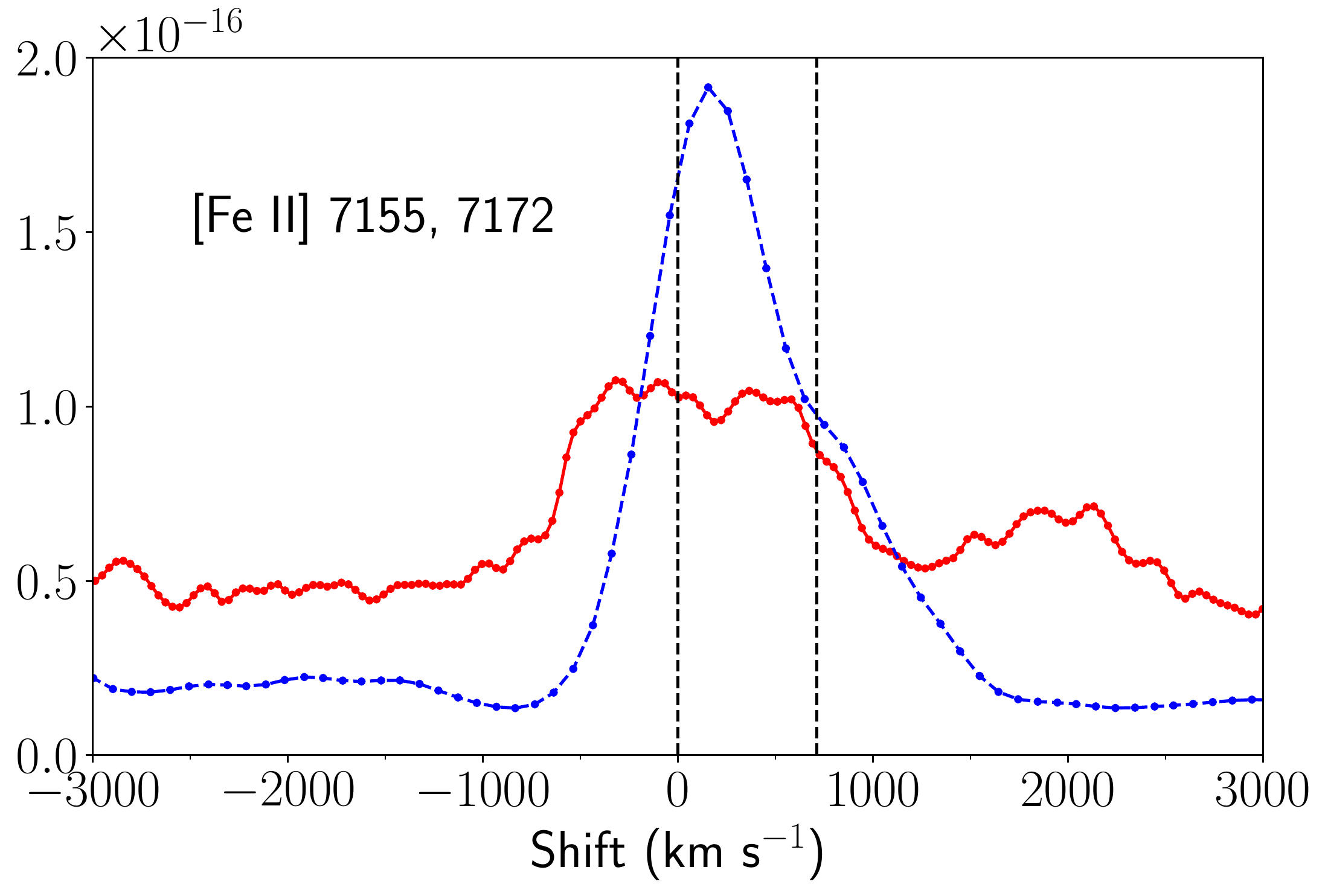} 
\includegraphics[width=0.32\linewidth]{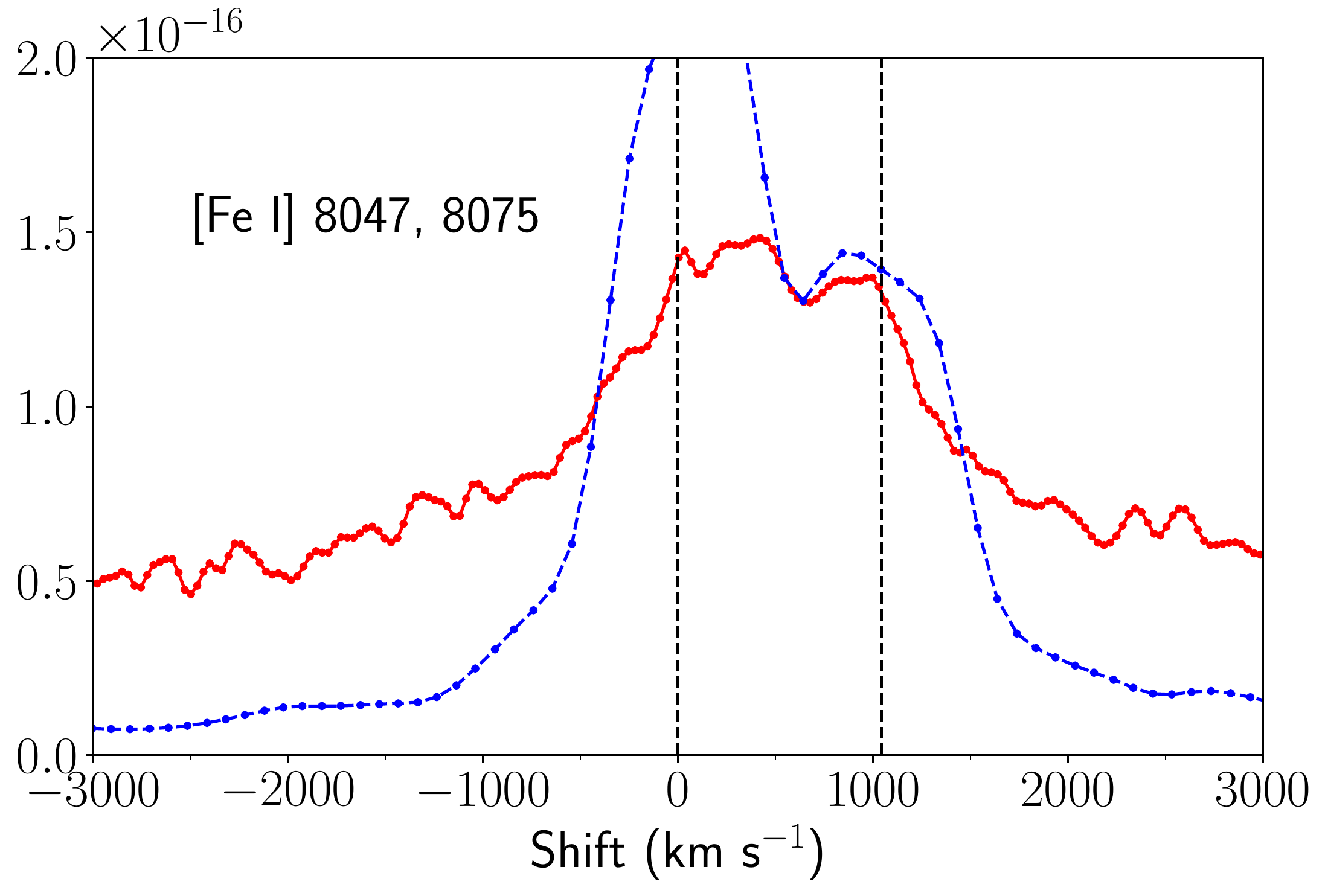} 
\includegraphics[width=0.32\linewidth]{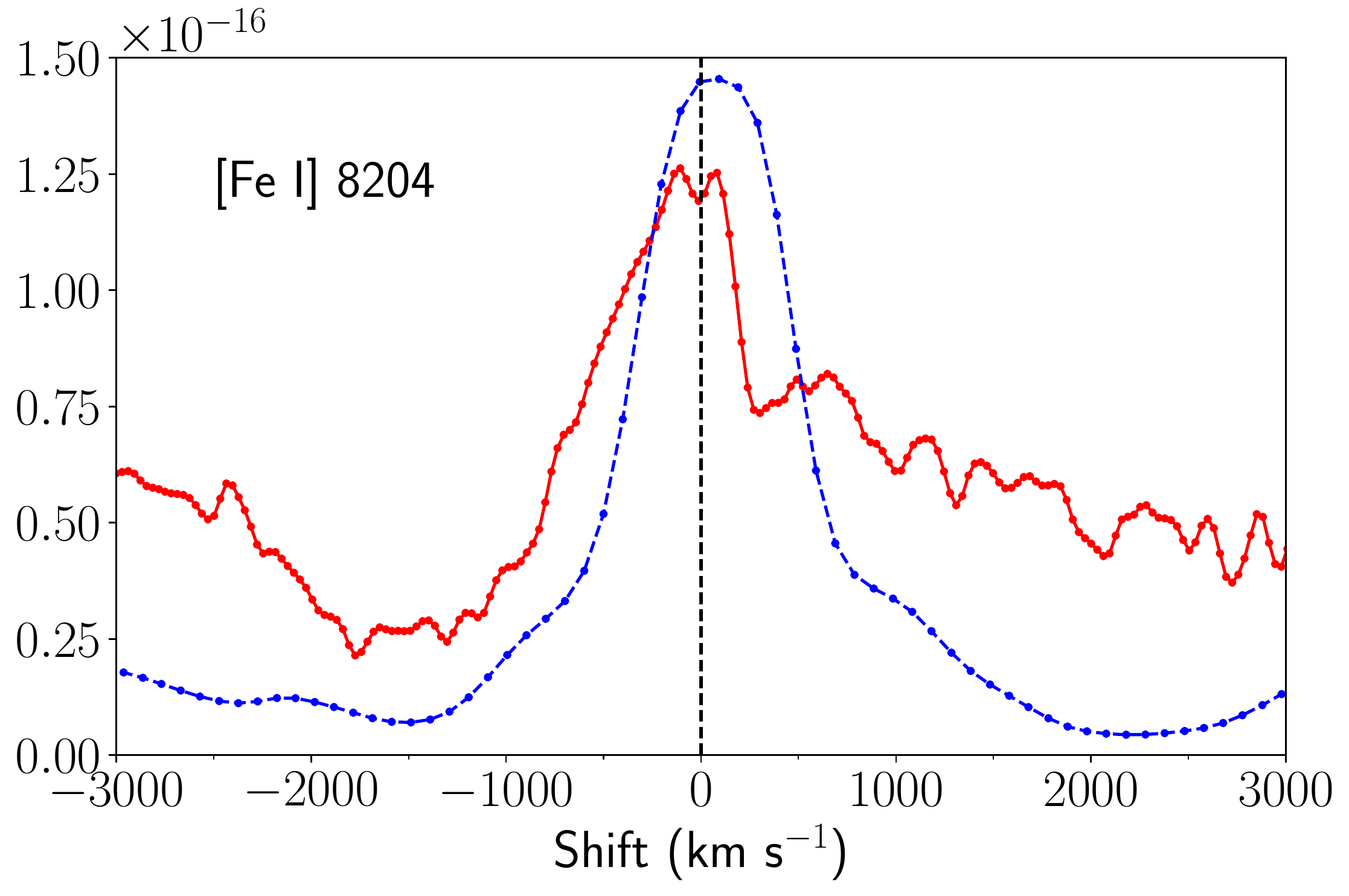} 
\caption{\arj{Observed (red, solid) and modelled (blue, dashed) H$\alpha$ in SN 2005cs.}}
\label{fig:2005cszoom}
\end{figure*}

\begin{figure}
\includegraphics[width=1\linewidth]{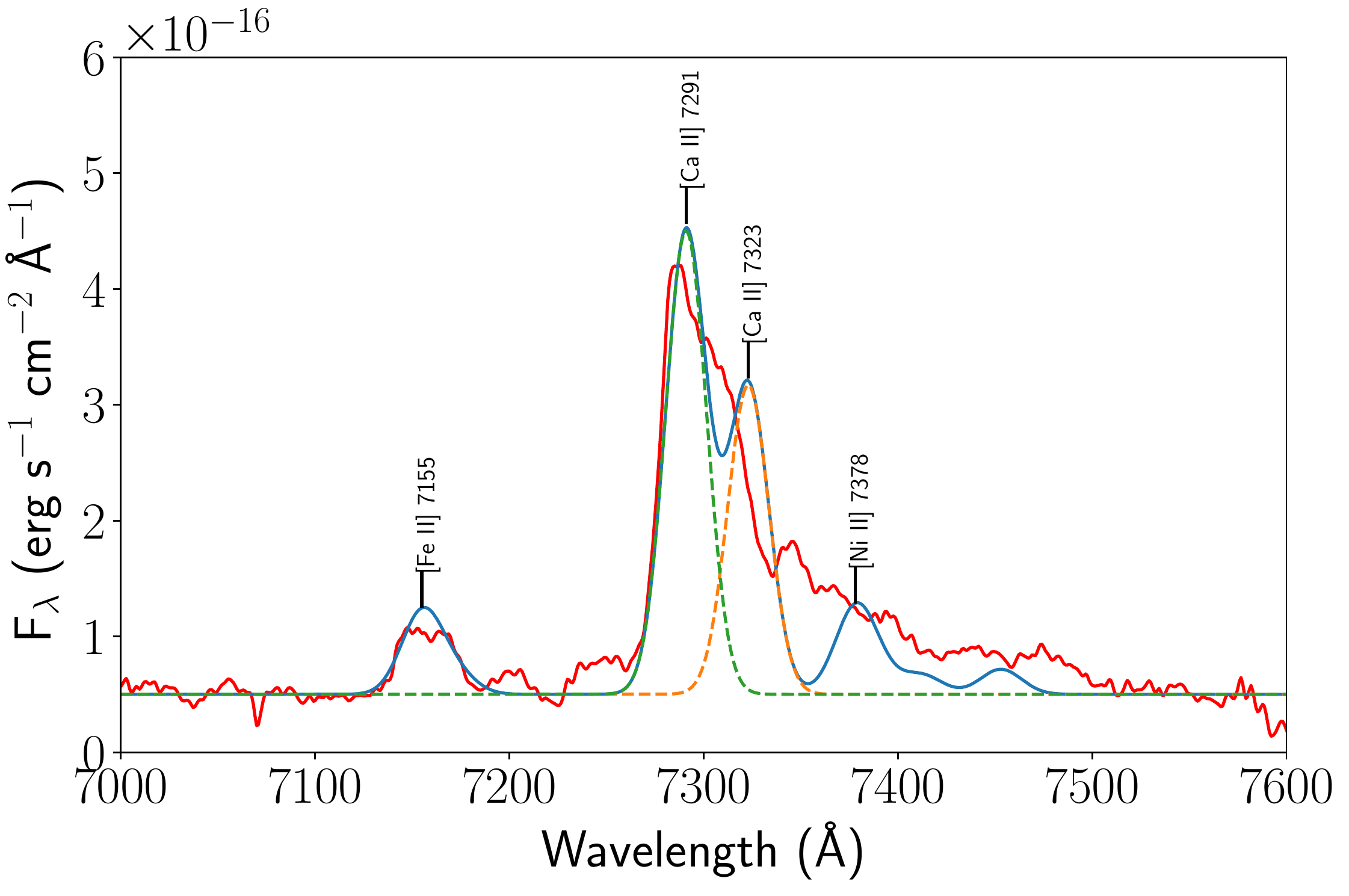} 
\caption{\arj{The 7000-7600 \AA~region in SN 2005cs (red) at +277d, and the Fe, Ca, Ni Gaussian fits described in \citet{J15a} (blue). From these fits a limit $L_{7378}/L_{7155} \leq 1$ can be put.}}
\label{fig:Cazoom}
\end{figure}

\subsubsection{+194d}
Figure \ref{fig:2005cs} (top) compares SN 2005cs at +194d with the 9 \msun\ model at 200d. A model scaling of $X=0.4$ was used to reach overall flux agreement, which would correspond to a $^{56}$Ni mass of $2.4\e{-3}$ \msun. This is close to the mass estimated by \citet{Pastorello2009} by comparison of the bolometric luminosity with SN 1987A ($3\e{-3}$ \msun).

But the spectral agreement is poor. The observed spectrum still appears photospheric in character, with few distinct emission lines apart from H$\alpha$. It is almost completely featureless below 5700 \AA, although as stated above the blue region involves significant calibration uncertainty. The model, on the other hand, produces many narrow lines in the blue. %The 11 model shows similarly poor agreement (not shown).  
This tendency was present also for SN 1997D at 200d, although not as extreme as here.
We note that the factor 2 lower \ni\ mass needed for SN 2005cs compared to the model would correspond to a lower explosion energy and yet lower velocities, which would (probably) worsen the problem of too narrow lines in the model.

The structure of the model at this epoch is a highly powered He core (expanding with $\sim$450 \kms), absorbing 90\% of the gamma rays. Powering appears to occur at higher velocities in SN 2005cs, but this would involve obtaining a lower \ni~mass at a higher explosion energy. Another, and perhaps more plausible situation, is that there is more mixing in this SN. 

\subsubsection{+277d}
% Line IDs in P09 is a bit random and not much to comment on, some of it way off
At +277d the observed spectrum shows somewhat more prominent emission lines and fine-structure (Figure \ref{fig:2005cs}, bottom panel), and general agreement with the model is somewhat improved. The model has still, however a too high ratio of narrow emission lines to ``quasi-continuum''. %It is tempting to ascribe this to residual background light. However, Figure \ref{fig:Z} shows that SN 2008bk (analauzed in next section) has a virtually identicial spectrum at this epoch.

Figure \ref{fig:2005csNIR} shows the NIR spectrum compared to the same model. 
The model requires an upscaling of $X=1.55$. The observed spectrum is of low S/N, and there are few certain line detections. O I 1.13 $\mu$m and [Fe II] + [Si I] 1.64 $\mu$m seem robustly detected. A few more look like plausible detections (He I + S I 1.08 $\mu$m, Na I 1.14 $\mu$m, Si I 1.20 $\mu$m, [Fe II] 1.26 $\mu$m, Pa$\beta$ 1.28 $\mu$m, ?1.47 $\mu$m, Mg I 1.50 $\mu$m, ?1.58 $\mu$m,  Br$\delta$ 1.79 $\mu$m, and ?2.0 $\mu$m), where question marks denote unknown line matches. There are potentially many more real lines, but what is noise and what is a line is hard to tell.

The observed optical spectrum shows strong, broad Na I 5890, 5896 and K I 7665, 7699 scattering lines. The blue absorption edges of these extend to about 3000 \kms, with the red emission reaching 2000-3500 \kms\ for Na I and 2000 \kms\ for K I. The model envelope ends at 2000 \kms~and thus does not fully capture these lines. These lines arise by scattering by primordial Na and K in the H envelope, which must have a density at 3000 \kms~high enough to makes these lines optically thick.
% Density limits not so useful as Na I and K I very low neutral fractions --> (weak) lower limits on density only

Some predicted lines of Fe, Ca, and C can now be clearly identified in the observed spectrum: [Fe II] 7155, [Ca II] 7291, 7323, Fe I 8047, Fe I 8204, Ca II NIR, [C I] 8727, Fe I 8830, Fe I 9010, Fe I 9370. The presence of Fe I lines in the observed spectrum is quite convincingly established, in particular through the strong Fe I 8047, 8075. However, as at 200d the model puts out stronger blue lines (e.g,. Mg I] 4571, Mg I + Fe I 5180, [O I] 6300, 6364) than observed. Some more detailed comments on emission lines from the major species are given below.

\textit{Hydrogen.} The model produces mainly H$\alpha$, a factor several too bright (the peak flux is $1.6\e{-16}$ erg s$^{-1}$ cm$^{-2}$ \AA$^{-1}$). The observed line (Fig. \ref{fig:2005cszoom}) is broad (FWHM=2000 \kms), quite flat-topped, and strongly asymmetric with emission on the blue side extending to 800 \kms, but to 2500 \kms~on the red side. The model line is too narrow (FWHM=1100 \kms), as mainly the inner part of the H envelope absorbs the gamma rays (Fig. \ref{fig:gammaaccum}). This suggests that the model core is too dense. %The model also makes a dim H$\beta$, which is not distinct in the observations}. 
% Nothing really to say about NIR lines

\textit{Helium}. He I 7065 is strong in the model but, in contrast to SN 1997D, is not observed in SN 2005cs. %He I 5876 and 6667 are dimmer, but also appear overproduced.  
Similarly, in the NIR He 1.08 is strongly overproduced (Fig. \ref{fig:2005csNIR}). %, unclear whether there is a line in data. 
Thus, there is no clear sign of He in the observations. It is tempting to draw a conclusion that there is no He shell present. However, carbon lines from this zone appears to be seen (more below).
% DC 2017-06-19 Also no He I 7065 at 330d.

\textit{Carbon.} [C I] 8727 appears to be observed in reasonable correspondance with the model, although it is blended with another (unidentified) line on the blue side. Of the two other reasonably strong C I lines predicted, C I 9100 also appears to be seen (although at low contrast), but [C I] 9850 not, however the last line lies at the noisy end of the spectrum. In the NIR, the model overproduces C I 1.07 and 1.176 $\mu$m, completing an overall picture that the model somewhat overproduces C lines.
% 1.07 consists of 6 lines from the 3p(3D) triplet (levels 15-17). E= 60,000, so probably recombination lines.
% 1.176 is a doublet (1.173+1.176, with last stronger) from 3p(1P) level 14.

\textit{Oxygen.} There is no sign of O I 7774, which is distinct in the model. Regarding [O I] 6300, 6364, there are two observed bumps, exactly peaked at 6300 and 6364 \AA, that likely are due to [O I] 6300, 6364. They have FWHM of 1500 \kms ([O I] 6300) and 1000 \kms ([O I] 6364), likely differing somewhat due to blending effects. The model makes much too strong [O I] lines, of which there is also significant Fe I contribution ($\sim$ 50\%) at 300d (6280 and 6360 \AA). 
At 300d there are contributions from all zones to this feature, although mainly from the O/He shell.
The model also predicts a distinct [O I] 5577 that is not observed. Thus, overall the model strongly overproduces emission from the O shell. The O I 8446 line is observed and reasonably well matched by the model, but this line is from the H zone.

\textit{Magnesium.} The model predicts a distinct Mg I] 4571 that is not observed, although the spectral coverage ends exactly around this wavelength. The strong model line at 5180 \AA~(also unobserved) has also a Mg contribution. Overproduction of Mg lines is consistent with the overproduction of O lines, as these arise from similar regions. There is no detection of Mg I] 4571 in any earlier nebular spectra of SN 2005cs that had extension further in the blue (Fig. 13 in \citet{Pastorello2009}, note the region was not observed in the last epoch of 333d), which is dim and featureless in the blue throughout the nebular phase. The reason may be that with the bright blue background and low \ni~mass, the SN had faded below detection limit.

The model also predicts a strong Mg I 1.50 $\mu$m (somewhat blended with Fe I 1.51 $\mu$m), but neither of this is there a clear detection (Fig. \ref{fig:2005csNIR}). 

\textit{Silicon.} The model produces no optical lines, and none is observed. The model produces 1.20 $\mu$m and 1.64 $\mu$m emission, but these are swamped by iron line emission. Thus synthesized silicon cannot be diagnosed. %In the NIR, Si I 1.20 and 1.64 are plausibly detected

\textit{Ca doublet.} The [Ca II] 7291, 7323 line has an odd observed shape (Fig. \ref{fig:2005cszoom}). The blue wing of [Ca II] 7291 indicates a maximal expansion of about 800 \kms, similar as H$\alpha$. 
%same as the [Fe II] 7155 line. 
A peak is seen just blueward of 7291 \AA\ (300 \kms\ blueshift), but no 7323 \AA\ peak is seen. Instead, a smooth red wing is present, not seen in the model, but again reminiscent of the H$\alpha$ profile. 

The 7291 and 7323 \AA\ lines are expected to emit in proportion 3:2 under optically thin circumstances; the upper levels have almost the same excitation energy and collisions between levels 2 and 3 will bring populations close to the 3:2 statistical weight ratio (A-values are the same). However, in the model at 300d, the lines are marginally optically thick ($\tau \sim 1$). Then the luminosity ratio changes from 3:2 towards 1:1.

In the model the two peaks are separated because the FWHM (800 \kms) is smaller than the line separation (1310 \kms). Precisely such a double-peaked feature is seen in SN 1997D at +351 (Fig. \ref{fig:1997D}), and in SN 2008bk (see next section). Clearly, this is not the situation in SN 2005cs (the line profile stays the same also in the last spectrum at +333d). A smooth structure with a clear weighting of the observed blend towards 7291 \AA\ suggests A) optical thinness, and B) that the lines are broad enough to blend. Both properties would follow naturally from a larger emitting region, as also suggested by the blue wing of [Ca II] 7291. The 7291 and 7323 \AA\ lines are separated by 1310 \kms, and blending will occur at about half this value for the expansion velocity, meaning one needs $V_{max} \gtrsim 600$\ \kms~to avoid it. Finally, some selective suppression of the red wing of 7323 \AA, and possibly also 7291 \AA, is needed.

Narrow [Ca II] is produced by mainly the inner \ni~and O shells, where the newly made Ca resides. Thus the discrepancy for [Ca II] means a discrepancy for this innermost region in the model.

\textit{Ca II triplet.} The triplet is strong both observationally and in the model (Fig. \ref{fig:2005cszoom} shows a zoom-in). In the observed spectrum the 8498 \AA\ line has scattered in the 8542 \AA\ line, whereas in the model this scattering has not occurred. The red wing of 8542 line indicates again asymmetry with a maximal emission velocity of 2500 \kms. The line shows a clear flat-topped over about  $\pm500$ \kms. The model line is too narrow, with FWHM of $\sim$ 900 \kms.
% 8542 in 028_300 : 910 kms FWHM. 8498 : 750 (probably partially eaten). 8662: 910.

\textit{Nickel.} There is no stable nickel in the model (except a solar abundance added in the H envelope). [Ni II 7378] is not clearly observed, although there is a red smeared tail of the [Ca II] doublet. Figure \ref{fig:Cazoom} shows a zoom-in on the 7000-7600 \AA~region, and a Gaussian fitting as described in \citet{Jerkstrand2015b}. A firm upper limit $L_{7378}/L_{7155} < 1$ can be put, which corresponds to a Ni/Fe ratio of solar or less. Thus, we see no indication of the high Ni/Fe production that occurs in the inner neutrino-heated ejecta in ECSNe and low-mass CCSNe \citep{Wanajo2011,Wanajo2017}. This conclusion is supported by non-detection of [Ni II] 1.94 $\mu$m line (Fig. \ref{fig:2005csNIR}). The nickel abundance is further adressed in more detail in Sect. \ref{sec:sensnucleo}.

\textit{Iron.} The observed [Fe II] 7155, 7172 line has a box-like shape (Fig. \ref{fig:2005cszoom}), with expansion velocity of about 600 \kms. With the telescope resolution limit of 300 \kms\ kept in mind, this emission may arise from a shell at $V\sim$500 \kms.

While detection of Fe I 8047, 8075 and Fe I 8204 lines give impetus to confirming presence of Fe in the ejecta, equally strong predicted lines of Fe I 7912, 8307/8310, and 8372/8382 \AA~are not clearly observed. The problem of identifying Fe is made worse by several overproduced lines at bluer wavelengths, and that the observed line at 8204 \AA~has a quite strong absorption trough, suggesting a P-Cygni formation. Looking into the model, Na I 8194 is optically thick out to about 1100 \kms~at 300d. 
%The next optically thick lines are Fe I 8047 on the blue side (which however has no absorption trough so should be a genuine emission line), and Ca II 8498 on the red. 
As the model underestimates the scattering in Na I D and Ca II 8542, it is possible that it underestimates the scattering in Na I 8194. Overall it remains ambigous whether the \ni~component of the model is in fact detected or not.
% After 8047 only Fe I 7912, then next one is K I 7700

Figure \ref{fig:2005cszoom} shows that Fe I 8204 and Fe I 8347, 8375 do not show the same flat-topped inner regions as lines from H and Ca. This gives some further weight to the hypothesis that this is indeed deep-lying synthesized iron.

\subsubsection{Discussion}
\textit{Signature of He core.}
% Skip He I 5876 because it scatters in broad Na I D. Skip also [C I] 9850 its weak
From Sect. \ref{sec:Hecore}, we know that the He core material imprints signatures of He I 7065, [C I] 8727, [C I] 9850, O I 7774, Fe I 5950 and Fe I lines at 7900-8500 \AA, with the Fe I and C I lines the most distinct. The same comparison at 300d (not shown) shows a similar pattern, although now O I 7774 is also made by the H-zone model.

In SN 2005cs at +277d there is no observation of He I 7065, O I 7774, or Fe I 5960. There is, however, observation of most of the Fe I lines in the 7900-8500 \AA~range (see above), and [C I] 8727. [C I] 9850 is not clearly seen but this lies at the end of the spectrum. Some other weaker Fe I lines in the 8800-9500 \AA~range are also reasonably well reproduced. Thus, there appears to be signatures of some of the He core material (Fe and C), but not all (O and He). The presence of a He/C zone is ambiguous; [C I] 8727 appears to be seen, but He I 7065 is not seen at all.  

The last spectrum at +333 (not shown here) shows no particular differences to the +277d spectrum; no new lines have emerged and the line profiles stay roughly the same.

The most important conclusions we draw from the analysis above is that 

\begin{itemize} 

\item{\arja{The spectral fit of the 9 \msun~model is not particularly good, and further explosion models need to be considered for SN 2005cs. The spectral lines of H and Ca show evidence for significant asymmetry (enhanced receding side emission), something not seen in either SN 1997D or SN 2008bk (next section), and the emission seems to come from a more extended velocity range.}}

\item{\arja{Attemps to determine whether certain composition zones are present or not are inconclusive. There is no clear evidence for a O shell in the SN, still at +333d after explosion. A He shell is indicated by detected by [C I] lines, but contraindicated by absence of He lines.}}

\item \arja{The Ni/Fe production ratio seems low, which is inconsistent with the theoretical nucleosynthesis in ECSNe and lowest-mass Fe CCSNe}.

\end{itemize}

\arja{The model fits are not quite good enough to claim a strong association of SN 2005cs with a low-mass Fe CCSN. The blue regions have too much line structure and are too bright (although the large uncertainty in data should be kept in mind), and the metal lines from the O and He shells are not in very good agreement. SN 2005cs maintains a rather smooth spectrum even 300d post-explosion, with few hints to its ejecta composition}.

\subsection{SN 2008bk} % ==============================================================
\label{sec:2008bk}

\arja{SN 2008bk was discovered by \citet{Monard2008} on March 25 2008 (MJD 54550), in NGC 7793. Photometric data for the first 400d, and spectroscopy up to day 190, were presented in \citet{vanDyk2012}, and photometry at later epochs in \citet{vanDyk2013}. %The explosion epoch is estimated as MJD 54548 by Pignata (in prep), just 2 days before discovery. 
Here we study a nebular spectrum taken on 2009-09-05 (MJD 55079 = +531d post explosion, assuming MJD 54548 for explosion epoch) presented in \citet{Maguire2012} (available on WiseRep)}. 

\arja{\textit{Extinction.} No Na I D absorption lines were seen, which suggests low host extinction \citep{vanDyk2012}.
This is also supported by low average extinction of the host galaxy, and dim MIR emission at site of SN 2008bk.
\citet{vanDyk2013} infer also a low average extinction $A_V=0.05$ towards the \textit{echo} of the SN.
Higher extinctions were preferred in modelling of the progenitor photometry.
\citet{Mattila2008} infer $E(B-V)=0.3$ mag towards the progenitor, but it is unclear how much of this
is due to circumstellar dust that would be destroyed in the explosion.
The Milky Way extinction is small, $E(B-V)=0.02$ mag. Given the tension in results for the host extinction, we deredden here with only this foreground extinction $E(B-V)=0.02$ mag.}% With this value, the observed nebular spectrum tend to be somewhat too bright in the blue regions compared to the models, which may be taken as indication that $E(B-V)$ is in fact low ($\lesssim 0.1$).

\arja{\textit{Distance and recession velocity.} We use the median distance reported at NED of 3.7 Mpc. The standard deviation is  0.7 Mpc.
%We adopt a distance of 3.5 Mpc \citep{Maund2014}, which is relatively well known.
For recession velocity we adopt 226 \kms (NED)}.
%For a direct comparison one may rescale the model with the relative difference in \ni\ mass.  However, no estimate of this has been presented in the literature. We therefore allow a scaling factor $X$ to allow for this.

\arja{\textit{Calibration to photometry.} The raw spectrum has synthetic $V=20.03$ and $I=19.69$. This compares to $V=20.10$ and $I=19.32$ photometry in \citet{vanDyk2013} at a very similar epoch. $I$-band is difficult for emission line spectra with strong Ca lines, and $V$-band is more relevant for calibration. As the agreement in $V$-band is within the error bar of the photometry, we do not perform any scaling but retain knowledge that there should not be any significant slit losses or other calibration effects.}

\arja{\textit{$^{56}$Ni mass estimate.} There is no refereed publication with a \ni~mass estimate of SN 2008bk. In a conference poster, \citet{Pignata2013} (who also assumed $E(B-V) = 0.02$ mag) estimated $M(^{56}\mbox{Ni}) = 0.009$ \msun}.

\arja{\textit{Convolution to telescope resolution.} The VLT spectrum is stated to have a resolution of 10 \AA\ in \citet{Maguire2012}, corresponding to 600 \kms\ at 5000 \AA. The model spectra were therefore convolved with a Gaussian with FWHM of 10 \AA\ throughout. Note that this gives somewhat broader model lines than in the SN 2005cs comparison, where data had a higher resolution}.
% DC 2017-06-12 Data is binned at 3.32 A throughout --> convolution in fixed number of pixels actually in fixed number of Angstrom.

\subsubsection{Dust formation}
\arja{The optical light curve of SN 2008bk started declining significantly faster than $^{56}$Co after $\sim$400d (Fig 2 in \citet{vanDyk2013}, suggesting the possibility of dust formation. After $\sim$600d the light curves flattened again, suggesting that the bulk of dust formation was over. The dust formation then occurred on similar time scales as in SN 1987A \citep{Lucy1989}. \citet{Pignata2013} similarly infer beginning of dust production at 490d from the spectroscopic evolution. We note that the similarity between SN 2008bk and SN 1997D and SN 2005cs probably means that dust is not a factor in the interpretation of the $<$350d spectra fort these previous section.}

%\arj{The evolution of line profiles (Fig. \ref{fig:dust}) give strong additional evidence that dust formation occurs in the SN, although strong blueshifts seen at +389d suggests this is the latest phase for initiation. At 265d the lines are centred at their rest wavelengths, but at 389d they have all begun to blueshift ($\sim 300$ \kms), and 
\arja{At 531d the blueshifts are of order $300$ \kms. The blueshift is consistently seen in Mg I] 4571, H$\alpha$, [O I] 6300, 6364, [Fe II] 7155, and [Ca II] 7291, 7323, and in several more. The similar effect seen in all lines suggest A) Emitting and absorping material is distributed in a similar manner for Mg, O, H, Fe, Ca, suggesting large scale mixing B) The dust behaves as a gray absorber, which is most easily explained by a distribution of optically thick clumps \citep{Lucy1989}.}

\arja{A peak blueshift of 300 \kms~on lines expanding with 1000 \kms~corresponds to $\tau_d = 1$ in a uniform sphere model. For a 700 \kms\ sphere (ratio becomes 0.45) it corresponds to $\tau_d=2$. Thus, the indication is extinction with something of this order ($\tau_d=1-2$). We here compare the data to a model with $\tau_d=2$. Dust likely explains why the optical 531d spectrum is significantly weaker than expected for a \ni~mass of 0.009 \msun; some 3/4 of the luminosity is probably emitted by dust in the IR}.

\subsubsection{+531d}
\begin{figure*}
\includegraphics[width=1\linewidth]{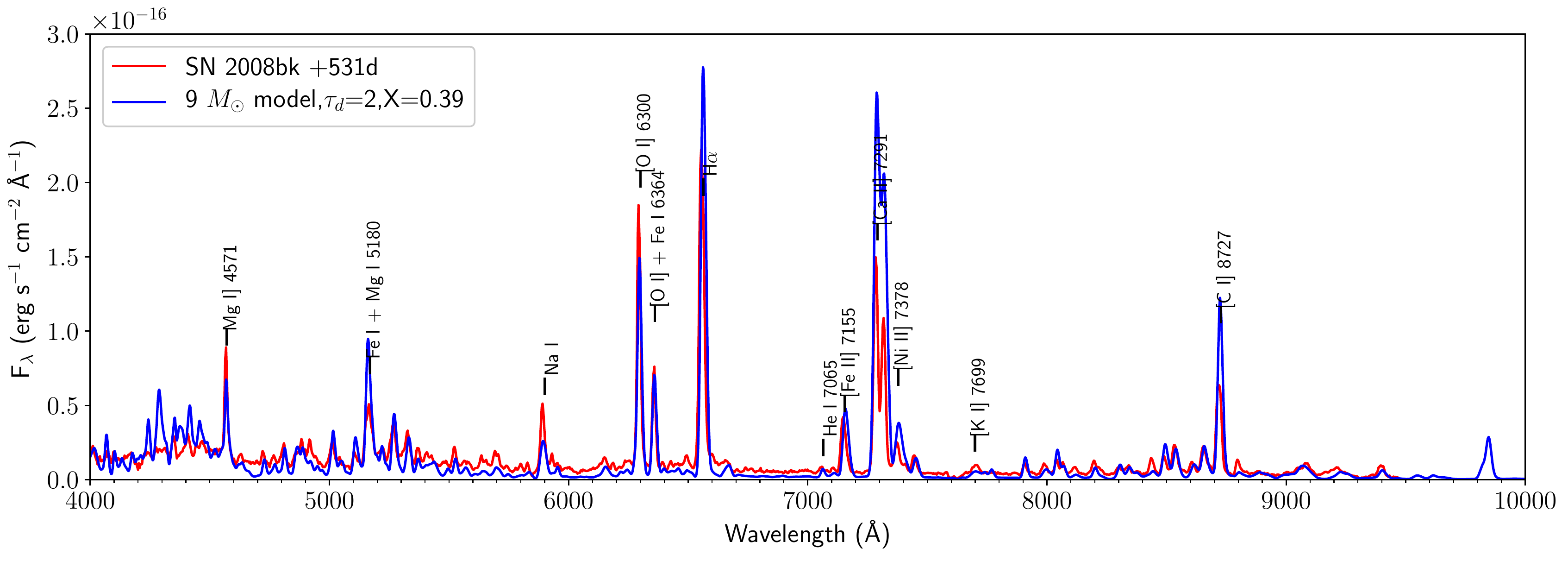}
\caption{\arja{SN 2008bk at +513d (red) compared to the 9 \msun\ model (blue).} The model spectra have been convolved with a Gaussian with FWHM=10 \AA ($\sim$ 600 \kms)~to match the telescope resolution.}
\label{fig:2008bk}
\end{figure*}

\begin{figure*}
\includegraphics[width=0.32\linewidth]{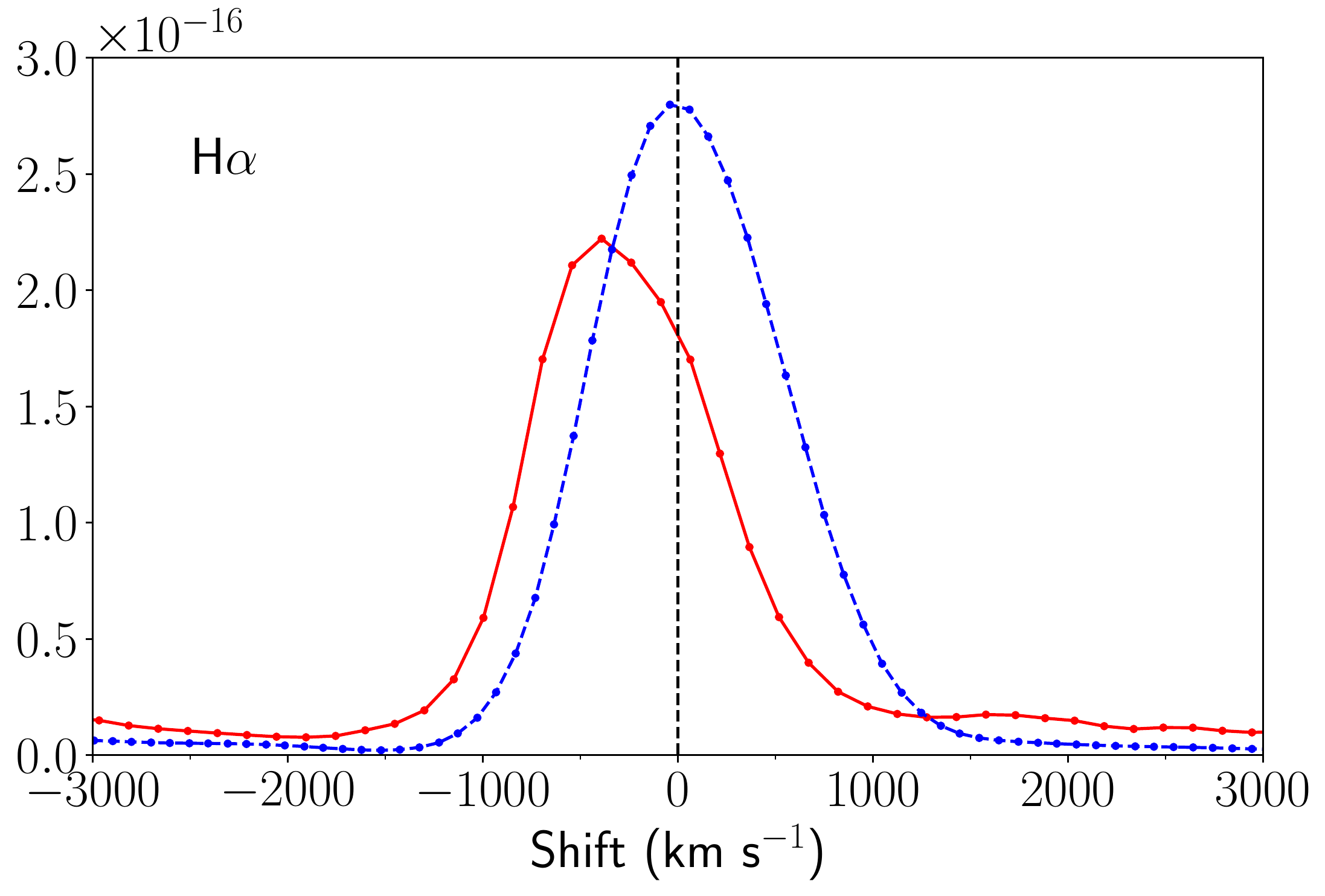}
\includegraphics[width=0.32\linewidth]{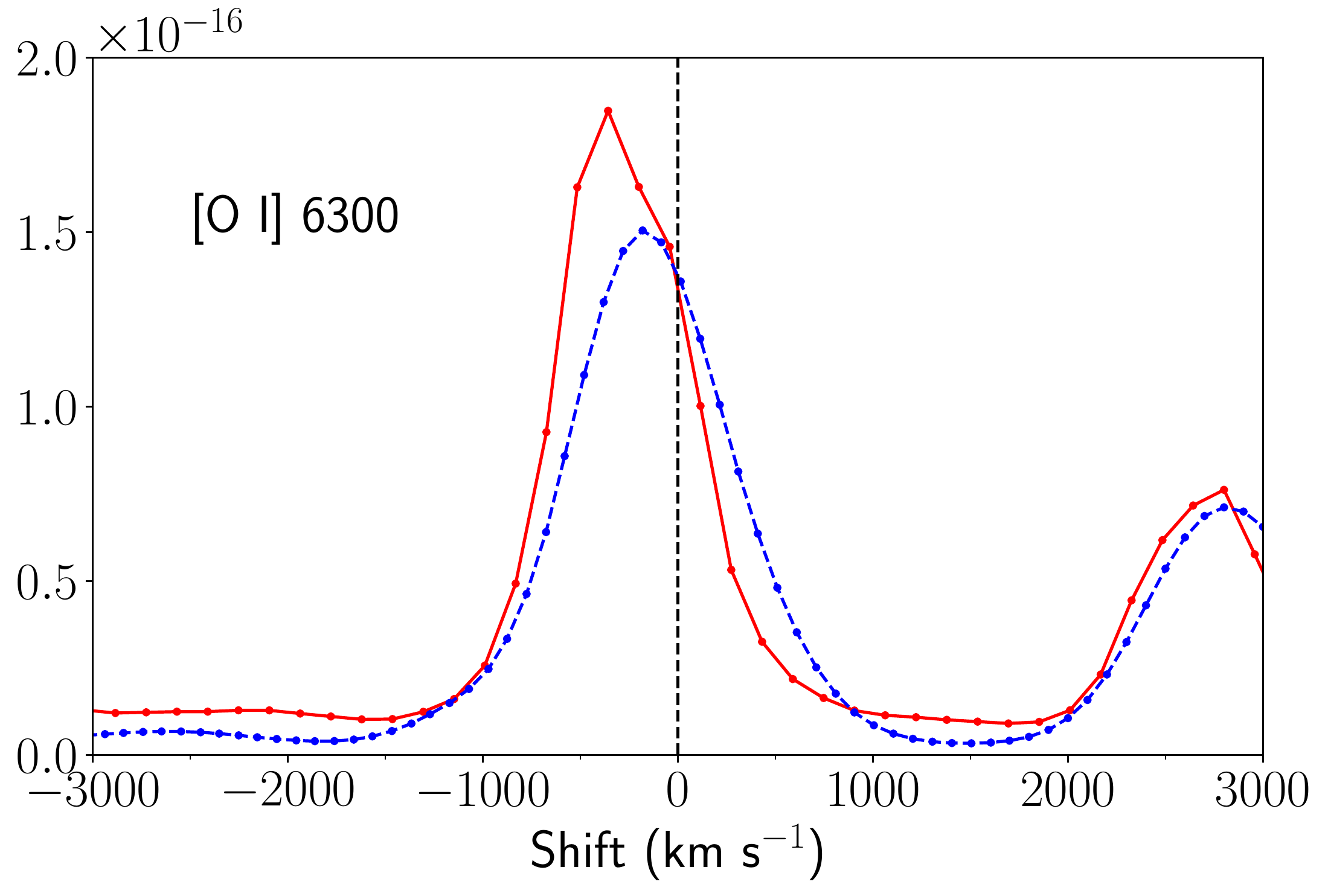}
\includegraphics[width=0.32\linewidth]{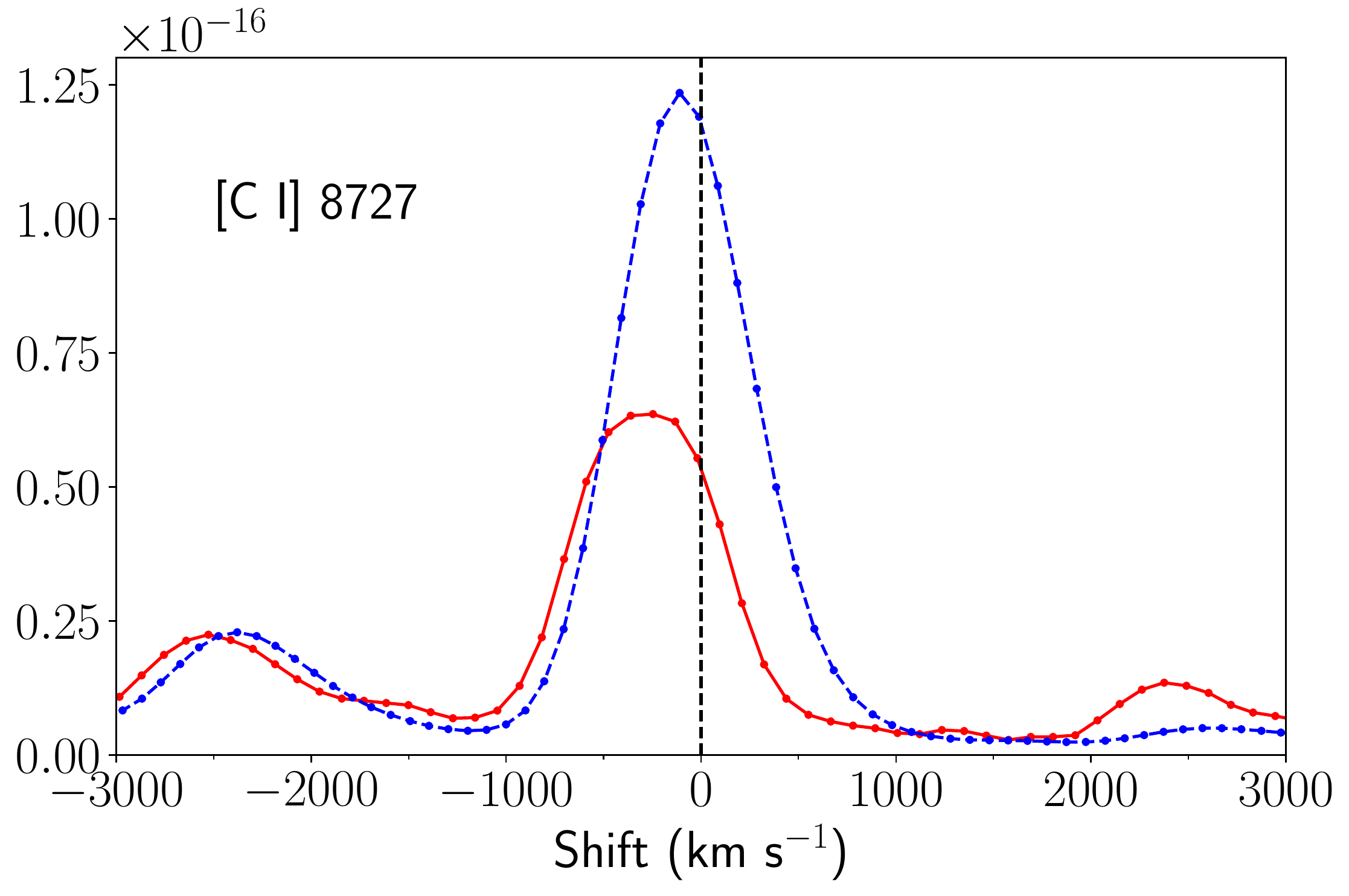}
\includegraphics[width=0.32\linewidth]{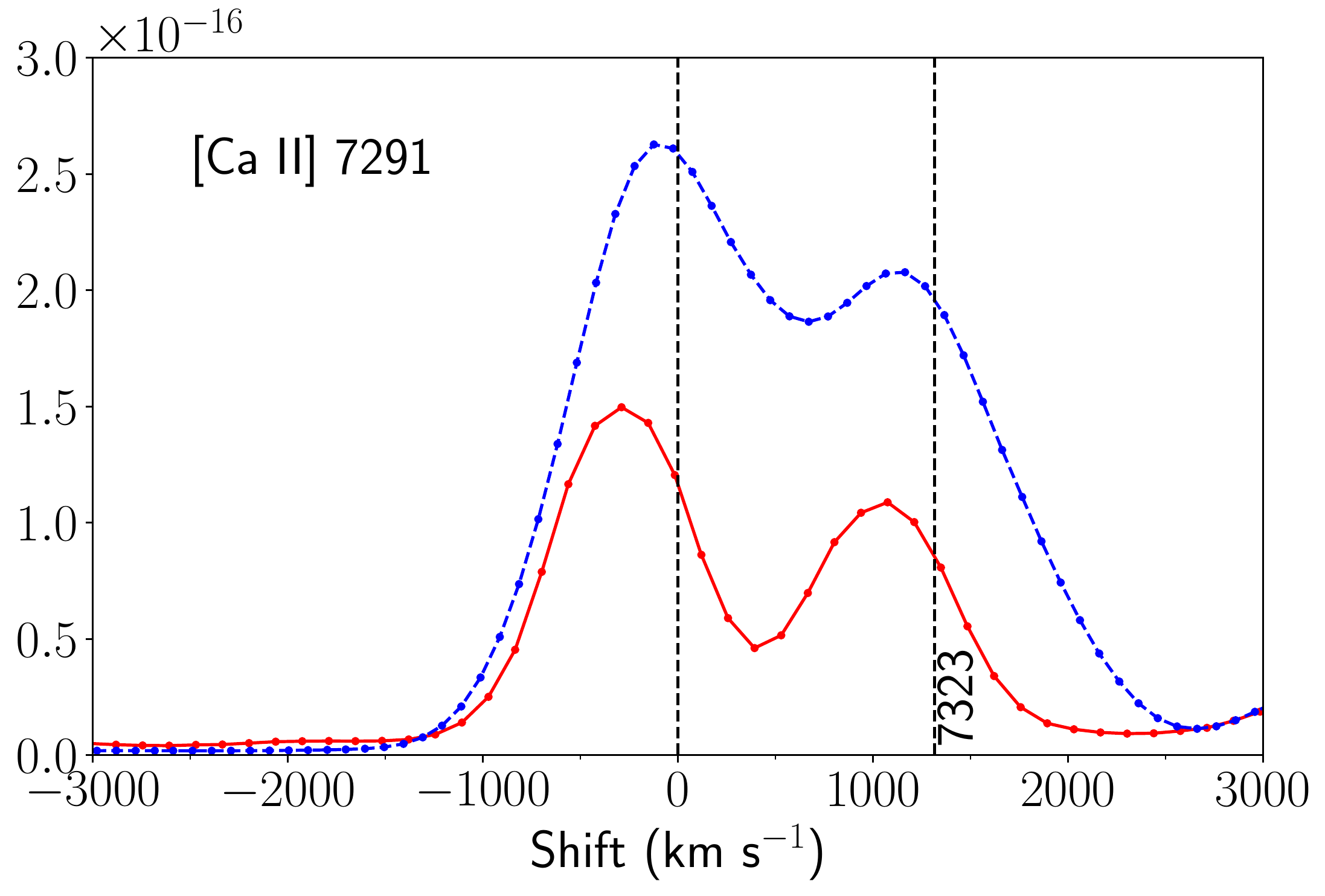}
\includegraphics[width=0.32\linewidth]{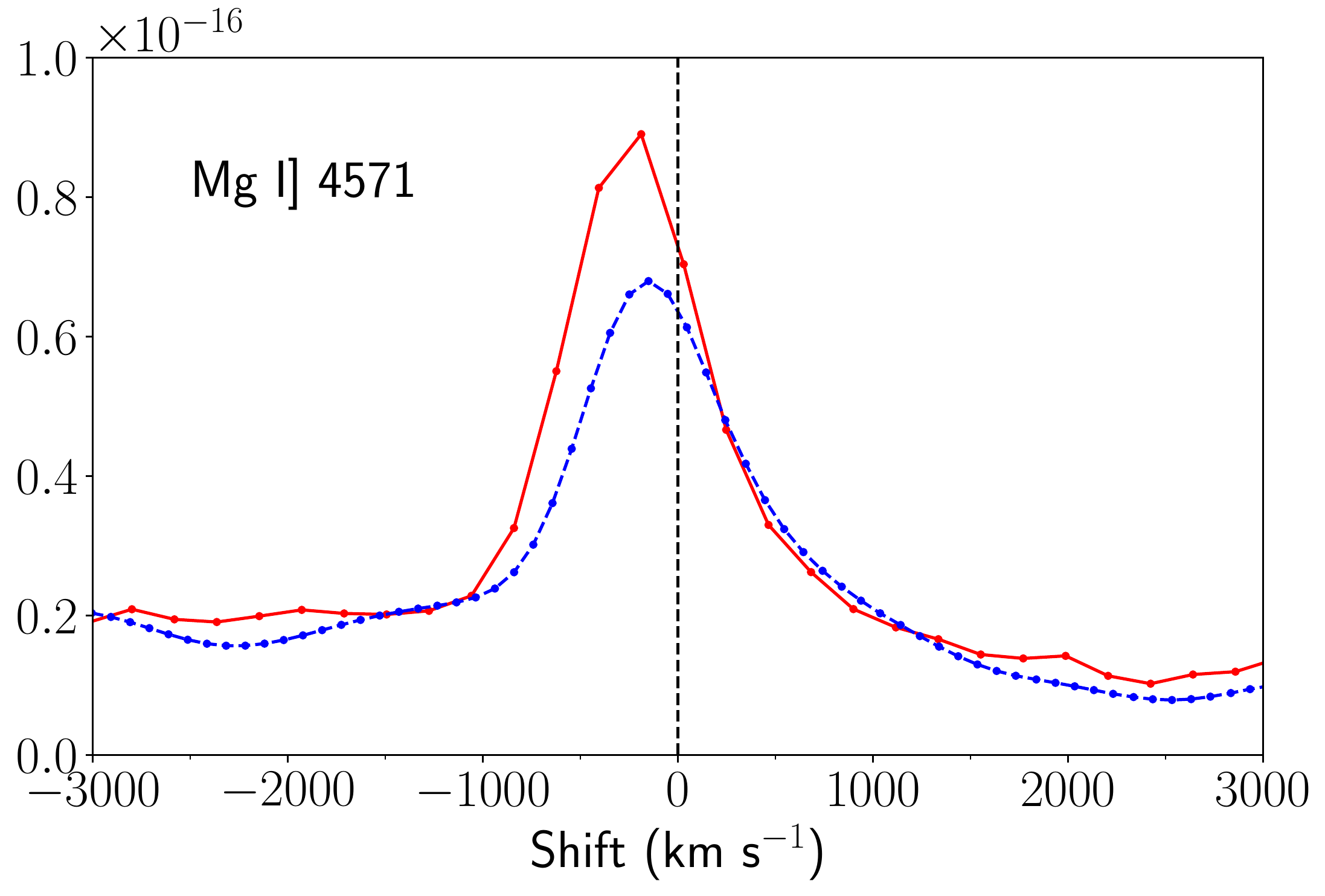}
\includegraphics[width=0.32\linewidth]{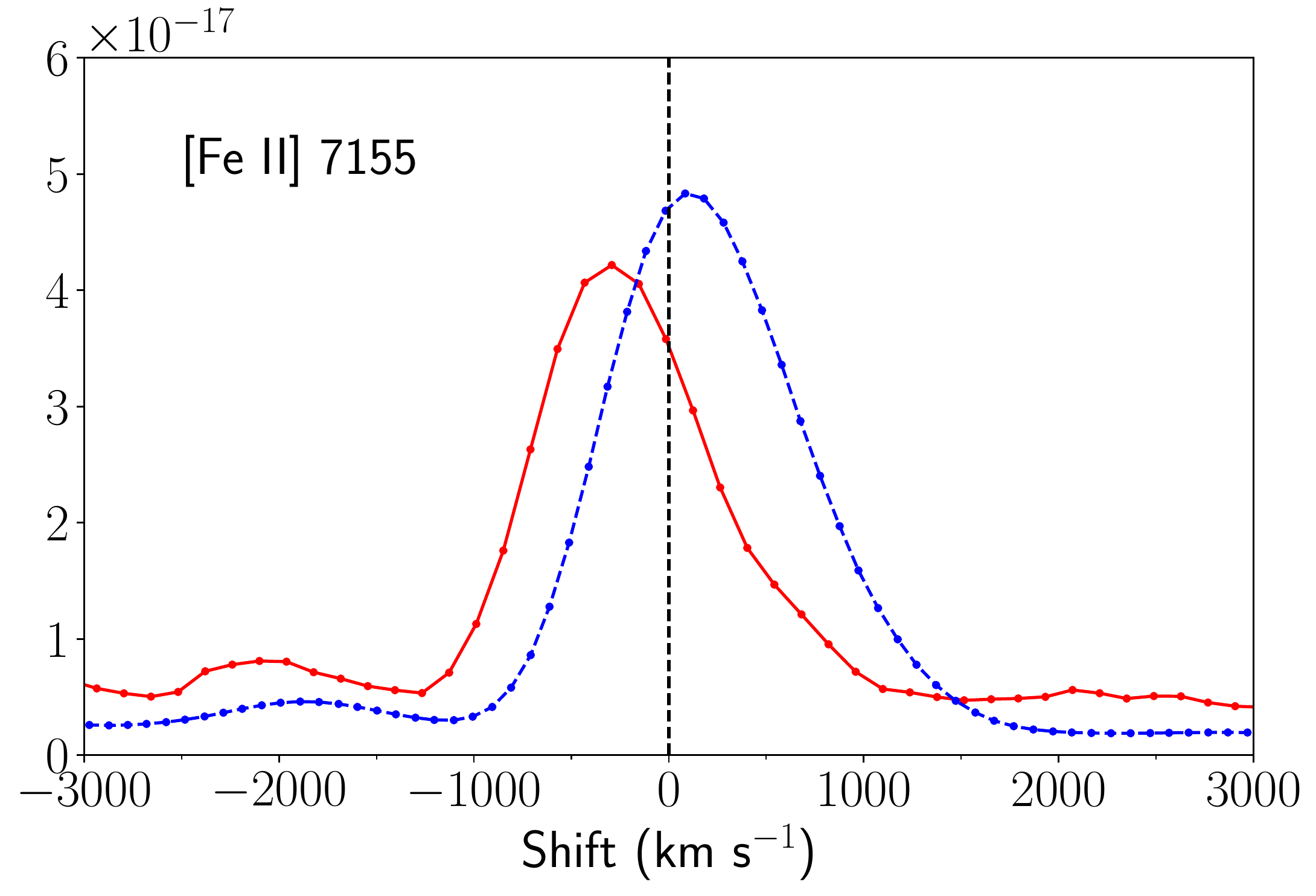}
\caption{\arja{Line profiles in SN 2008bk at +531d. Red solid is the data, blue dashed is the model.}}
\label{fig:dust}
\end{figure*}

\arja{In Figure \ref{fig:2008bk} the observed spectrum is compared to model 9 at 500d, with $\tau_d=2$, and with a further factor $X=0.39$ to reach overall optical flux agreement. %This last scaling suggests at face value a \ni\ mass of about $2\e{-3}$ \msun.
We make the following observations}

\begin{enumerate}
\item \arja{\textit{The [C I] 8727 line is distinctly observed, and within a factor 2 of the model prediction.} The [C I] 8727 line is clearly seen in the observations, which
confirms the presence of carbon in the ejecta. %The model luminosity is factor 2 high high, which suggests that A) The He zone is too massive (0.16 \msun), B) The He zone obtains too much gamma energy (25\%), or C) The He zone has too high carbon content (13\%). 
The factor 2 overproduction is similar to SN 2005cs (Fig. \ref{fig:2005cs}). %As molecules are not expected to form efficiently in a He-rich environment, a fourth option of CO cooling is not an attractive solution. 
There is also a (weak) He 7065 present, which further strengthens the presence of a He/C shell.}

%This line comes mainly from the He zone (shells 8-12,or 250-400 kms).
\item \arja{\textit{The [Ca II] 7291,7323 lines are seen with clear separation, indicating low velocity emitting material as in the model}. The model luminosities are too high by a factor of 2. These lines have contributions from many parts of the ejecta, but there is in particular a strong broad contribution from primordial Ca in the H envelope (Fig. \ref{fig:zones9}, this contribution also grows from 400d to 500d) that is overproduced. It therefore appears unlikely that the energy deposition in the H-rich gas is too low in the model (e.g. we cannot solve the overproduction of [C I] 8727 by transferring gamma energy to the H zone). We cannot reduce it too much either because then H$\alpha$ would become too weak}. % No good scenarios to explain how both H-alpha and Ca II can improve.
% DC 2017-06-18 H envelope in beta model strong based for [Ca II], accounting for 2/3 of total lum.

\item \arja{\textit{Synthesized Fe (from \ni) is clearly detected in the observed spectrum, from the Fe I 7912, Fe I 8047, 8075, Fe I 8204, and Fe I 8307, 8310 lines}. These lines are in excellent agreement with the model prediction, and completely lacking in the corresponding pure H nebula model at this epoch, confirming them as unique signatures of the \ni~region. These Fe I lines show maximum expansion velocities of $\sim$ 800 \kms, and FWHM $\sim$ 1000 \kms, which, given the telescope broadening means that the bulk of the \ni~resides inside $\sim$500 \kms. This provides important direct constraints on the distribution of \ni.}
% Fe 7912 : blue wing -700 km/s all pochs. At 290d -540 to 340 FWHM = 880. Line show same blue shift as other lines.

\item \arja{\textit{[O I] 6300, 6364 is well reproduced}. At this epoch these lines are almost entirely due to O, with little Fe contamination. About 1/3 comes from the H zone, and 2/3 from synthesized oxygen. This success means that this low-mass model, with just 0.07 \msun~of oxygen in the ejecta (mostly primordial), passes one critical test for matching low-velocity Type II SNe. As this unmixed 1D setup gives efficient gamma deposition in the thin O shell compared to mixed models, an underproduction would have posed serious problems to account for the [O I] lines (mixing can reduce the line strengths but not easily increase them). We cannot yet say if this success survives in 3D, or what level of agreement more massive ejecta models would give, but for now can state that the model passes the test at this point}. 

\arja{Some further arguments can however be derived from analytic considerations of the oxygen doublet. From the observed 3:1 ratio, the [O I] lines are optically thin at 500d. We may use this result to put an upper limit of the number density using the Sobolev optical depth formula
\begin{equation}
n_{OI} < \tau 8\pi \frac{g_l}{g_u} A^{-1} \lambda^{-3} t^{-1}
\end{equation}
where $n_{OI}$ is the number density of O I, $g_l$ and $g_u$ are the statistical weights to lower and upper levels, $A$ is the Einstein A value, $\lambda$ is the wavelength and $t$ is time.
For $\tau<0.1$ and $t=500$ we get $n_{OI}<4\e{7}$ cm$^{-3}$. The mass is, letting $f$ denote the filling factor,
\begin{equation}
M_{OI} = n_{OI} \frac{4\pi}{3} V^3 t^3 f < 0.18 f\ M_\odot
\end{equation}
where we have used $V<1000$ \kms.
As $f<1$ the maximum O I mass responsible for the observed [O I] lines is $<$0.18 \msun. Allowing for molecules to have extinguished the optical thermal emission from O/C and O/Si compositions, 0.18 \msun~corresponds to the O/Mg zone in a progenitor of maximum $M_{ZAMS}= 12$ \msun \citep{WH07}. This makes it unlikely that high-mass models would be able to reproduce the O lines in SN 2008bk; such  stars would give 1:1 ratios for the doublet lines for these low expansion velocities. The same conclusion can be drawn for SN 1997D, which had 3:1 ratio at +480d.}

\item \arja{\textit{Mg I] 4571 is well reproduced}. As for the [O I] lines, the suitability of the properties of the O/Mg shell in the model is confirmed. While much of the [O I] line luminosity may be recovered in the H-nebula model, this is not true for the Mg I] 4571 line, providing evidence that there is synthesised material present. Note that the Mg I/Fe I/Fe II 5180 blend is mainly due to Fe II 5160 in the envelope at this epoch. In this spectrum of SN 2008bk, this line is in fact the strongest in the window between Mg I] 4571 and Na I D, providing confirmation that there should be a strong line at this wavelength, although comparisons to SN 1997D and SN 2005cs suggests it emerges too early in the model.}

\item \arja{\textit{[Ni II] 7378 is distinctly seen in the observed spectrum}. It is quite weak, and the model, which has only primordial nickel in the H envelope, overproduces the line, suggesting small amounts of newly made nickel. This issue is addressed in more detail in Section \ref{sec:sensnucleo}. In the model, almost all of [Fe II] 7155 is from primordial iron at this epoch, and a similar situation would be expected to hold for [Ni II] 7378.}

\end{enumerate}

\subsubsection{Discussion}
\arja{\textit{Signature of He core.} There are clear signatures of the He core from Mg I] 4571, [O I] 6300, 6364,[C I] 8727, and the Fe I lines at 7900-8500 \AA. The O, Mg and C lines rule out an ECSN model, and SN 2008bk is again best explained as a Fe CCSN.}

\arja{Figure \ref{fig:dust} shows zoom-in on selected line profiles. Our simplified treatment of dust does not give the right deformation of line profiles. Some blueshift is obtained for He core lines ([O I] 6300, 6364, [C I] 8727, Mg I] 4571), although too little ($\sim$200 \kms) compared to the observed shifts (300-400 \kms). For lines mainly coming from the H envelope (H$\alpha$, [Fe II] 7155) there is no shifts at all in the model, as the dust lies interior to the emitting shell and this does not give blueshifts. The observed shifts in these lines are as strong as for the other lines, and this indicates that the dust distribution should overlap also with H-zone material. We hope to return to more detailed modelling of these line profiles in a future publication where 3D effects and more detailed dust treatment will have to be considered.}

\section{Sensitivity to explosive nucleosynthesis} 
\label{sec:sensnucleo}
\arja{The detailed composition of the inner explosively burnt region is not very accurate in the 13-element alpha network in P-HOTB. Here, we consider sensitivity to this composition by analysis of models with hand-made composition changes guided by more detailed simulations of the burning in this region.}

\arja{\citet{Wanajo2009} presented detailed post-processing nucleosynthesis calculations of the 1D ECSN simulation of \citet{Kitaura2006}. The ejecta obtain an electron fraction in the range $Y_e=0.47-0.53$ (see \citet{Wanajo2011} for figure), with an average value below 0.5 and significant production of neutron-rich isotopes such as $^{58}$Ni and $^{60}$Zn. In each mass shell of the \citet{Wanajo2009} model, the dominant isotope is either $^{56}$Ni,$^{58}$Ni, or $^{60}$Ni}. 

\arja{The total yields in this model of the most common elements are He ($5.12\e{-3}$ \msun), $^{58,60,62}$Ni ($4.39\e{-3}$ \msun), $^{56}$Ni/$^{56}$Fe ($2.50\e{-3}$ \msun), and $^{64,66}$Zn ($1.09\e{-3}$ \msun), together making up $1.31\e{-2}$ \msun\  or 94\% of the mass of the explosively burnt and neutrino-processed region, in (normalized) mass fractions 0.46 (He), 0.40 (Ni), 0.22 (Fe), 0.09(Zn). In our standard 13-element 9 \msun\ model these mass fractions are instead 0.72 (He), 0 (Ni), 0.25 (Fe), and 0 (Zn). Only a negligble fraction ($<10^{-15}$) of the \ni~remains undecayed at nebular times, but a few percent of the $^{56}$Co remains undecayed (7\% at 300d and 1\% at 500d), also giving a cobalt mass fraction of about 1\%.} The Ni/Fe and Zn/Fe ratios in the \citet{Wanajo2009} model are 32 and 310 times their solar values,
and these elements may therefore be unique signatures of ECSN nucleosynthesis.
% solar Zn/Fe = 1.3e-3 by mass

\arja{\citet{Wanajo2011} followed up by computing the nucleosynthesis for the same progenitor exploded in 2D. The 2D model showed roughly the same $Y_e$ distribution, except that a low-$Y_e$ tail was created. There were only minor ($<$ factor 2) differences in yields for elements with $Z<34$ (see their Fig. 3). For some heavier elements ($Z=34-40$) larger variations were seen, but these elements still remain too rare to have any visible emission in the nebular spectra. Thus, we conclude the the 1D yields for He, Fe, Ni and Zn from \citet{Wanajo2009} can be used to good accuracy}.

\arja{\citet{Wanajo2017} expanded these 2D simulations to iron cores. It was found that the lowest mass iron cores ($M_{ZAMS} = 8.1, 9.6$ \msun) produce nucleosynthesis very similar to the ECSNe of \citet{Wanajo2009} and \citet{Wanajo2011}, implying that ECSNe cannot be uniquely identified based on signatures from their explosive nucleosynthesis. For more massive iron cores ($M_{ZAMS}=11,15,27$ \msun), on the other hand, slower shock expansion led to significantly higher $Y_e$ values and lower production of neutron-rich isotopes. Therefore, high Ni/Fe and Zn/Fe ratios appear to be at least diagnostics of the lowest mass range ($\sim$8-10 \msun), giving ECSNe or ECSN-like explosions}.

\arja{None of these specific models are available to us to use here as they were not evolved into
the homologous phase. Instead, to mimic such a structure, we created a variant of the 9 \msun\ model in which the inner region, taken as where the \ni\ fraction exceeded 10\%, had some of the helium in the explosive region removed and replaced by $^{58}$Ni giving $^{58}$Ni/$^{56}$Ni=3 and 20 times solar (this was done in each zone where the \ni~mass fraction was over 1\%). Note that beyond this added nickel, there is also primordial nickel in the hydrogen zone.
SUMO does not currently treat zinc, and we therefore have to defer study of zinc abundance.
%and a Zn abundance given by Zn/Fe = 300 times solar. 
The He abundance was correspondingly reduced to ensure mass fractions added to unity. %If this was insufficient to accomodate the added Ni and Zn, all mass fractions were scaled down to add to unity. 
These model were computed at 300d (for comparison with SN 2005cs) and at 500d, with dust optical depth $\tau_d=2$ (Sec. \ref{sec:dust}) for comparison with SN 2008bk}.

\begin{figure*}
\includegraphics[width=0.48\linewidth]{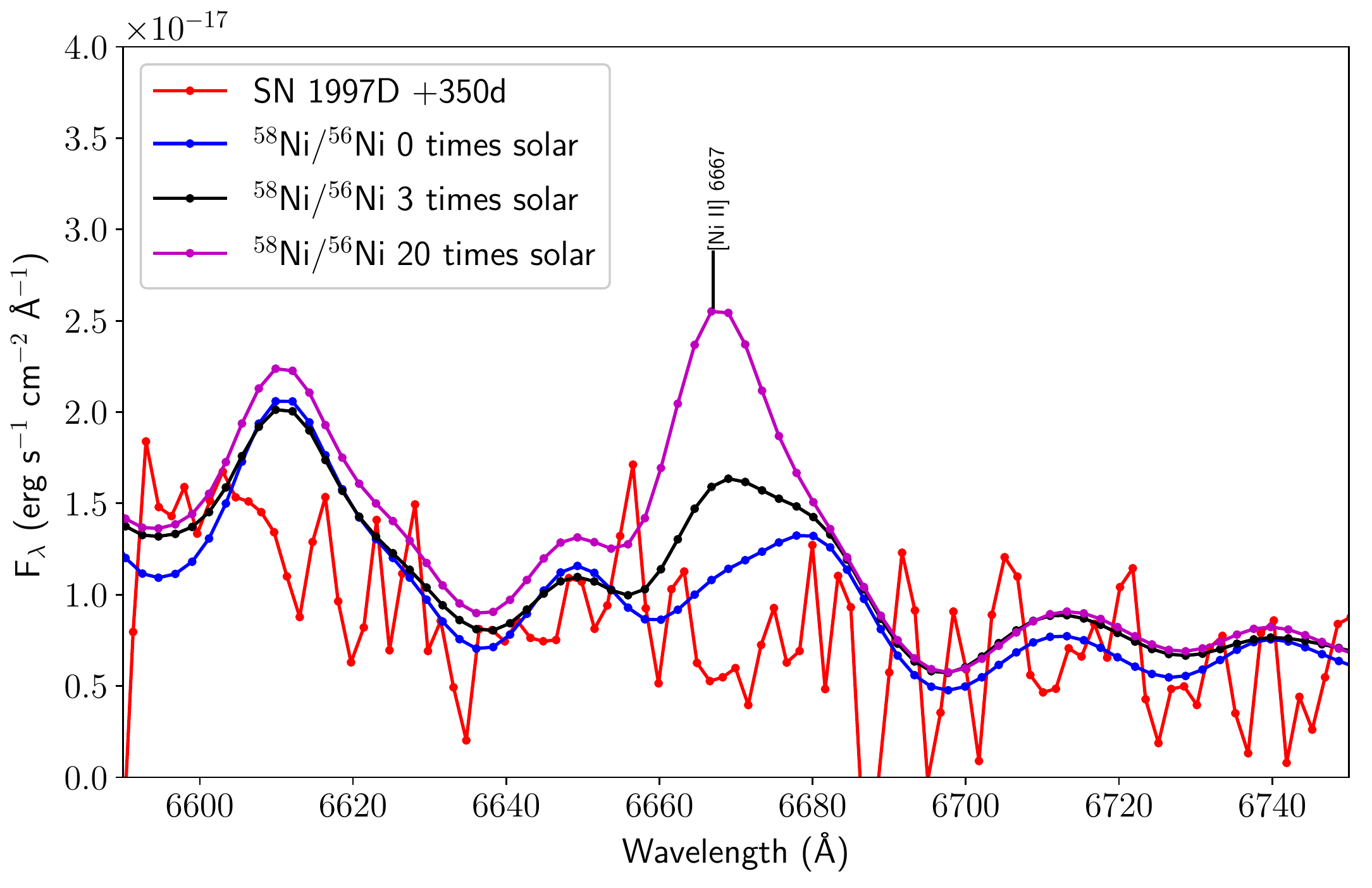} 
\includegraphics[width=0.48\linewidth]{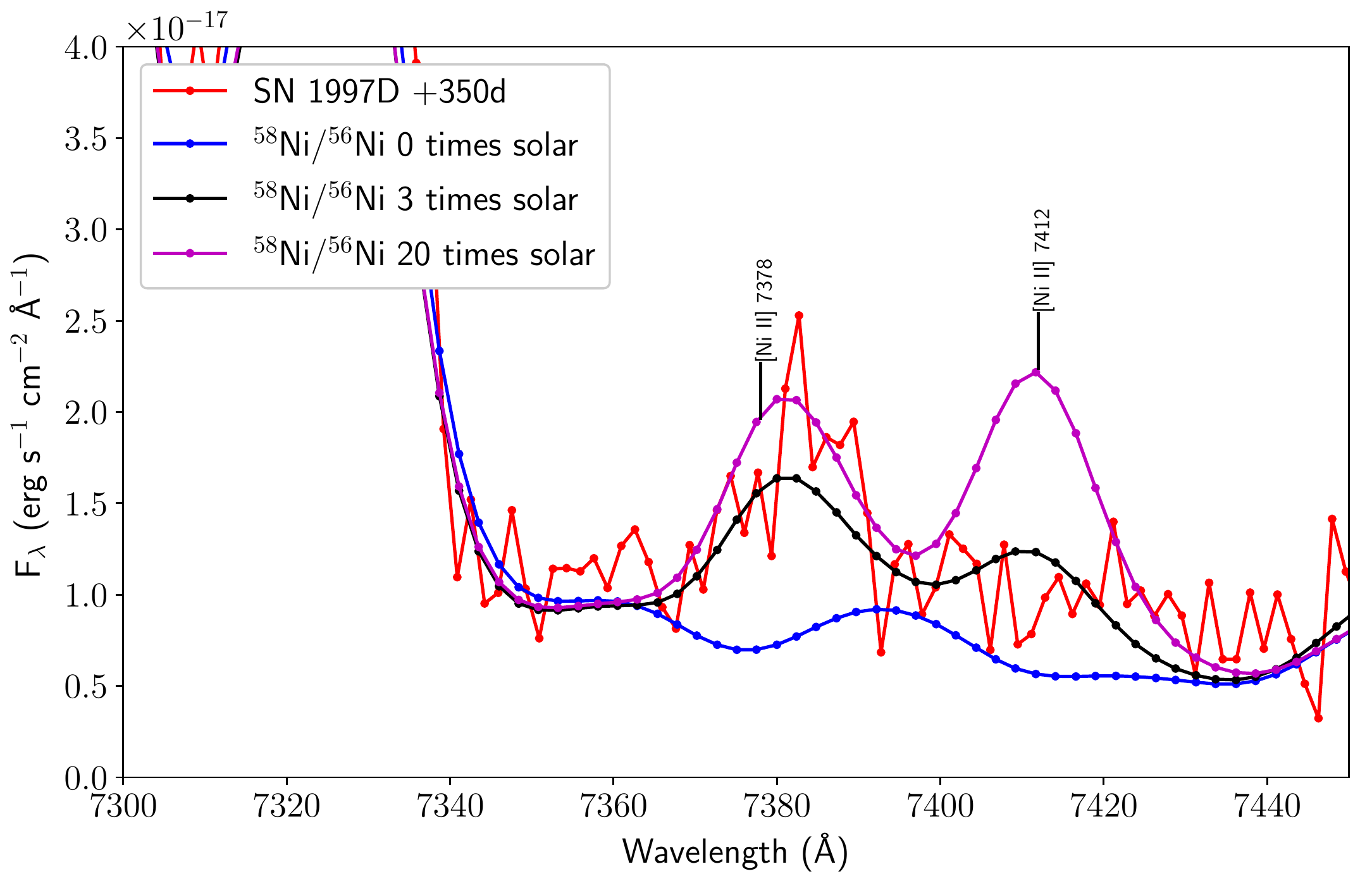} 
\includegraphics[width=0.48\linewidth]{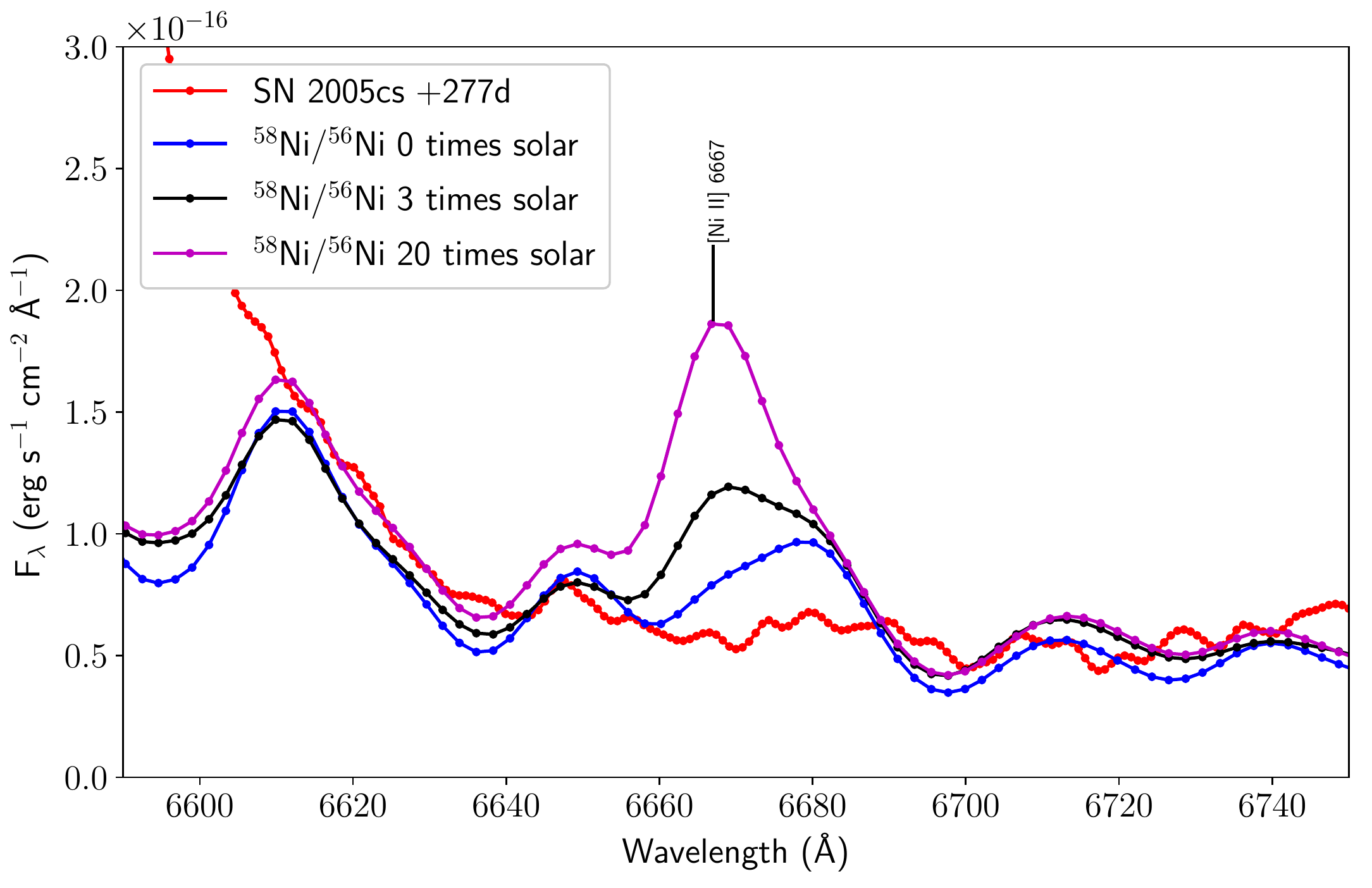} 
\includegraphics[width=0.48\linewidth]{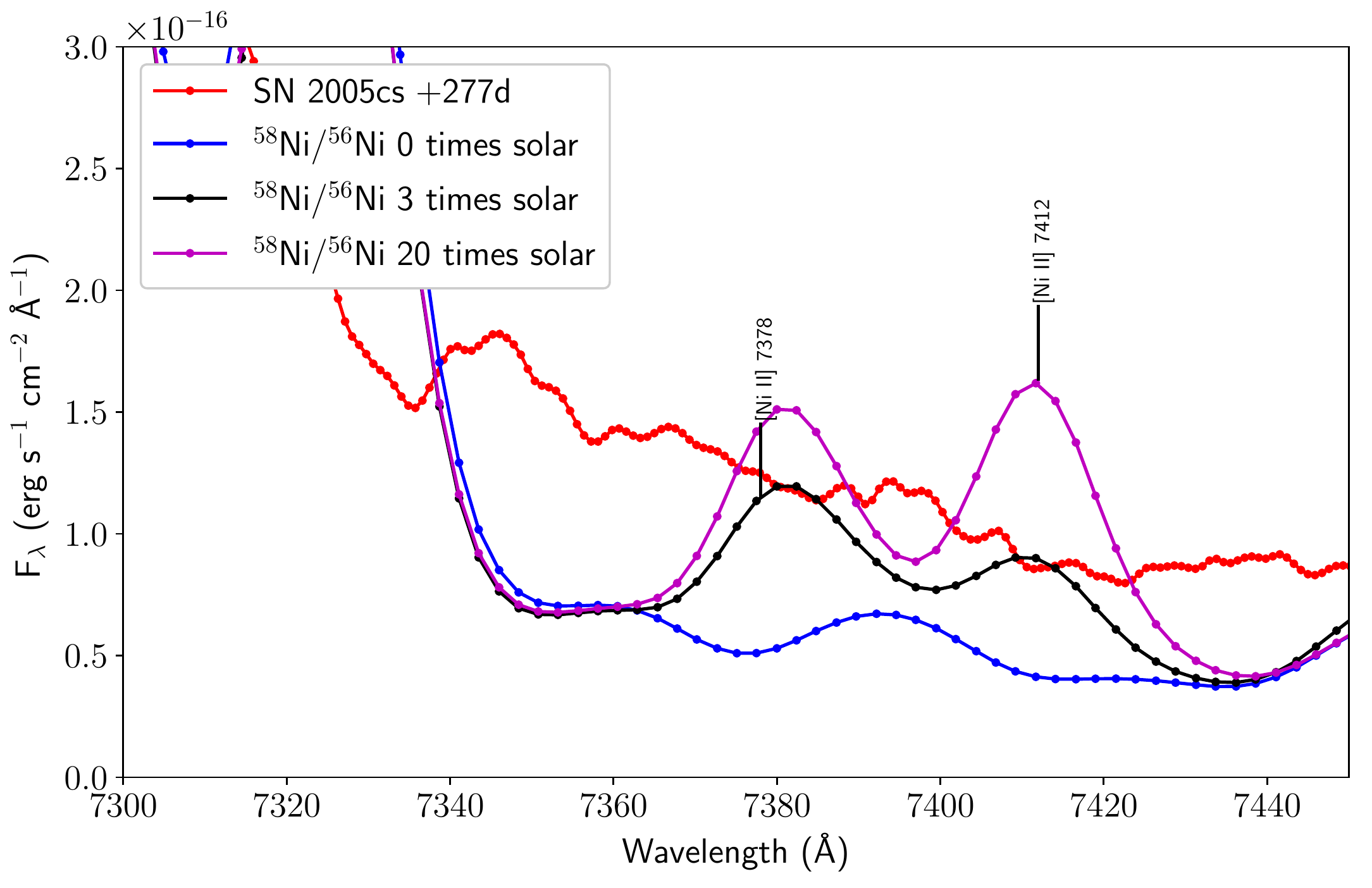} 
\includegraphics[width=0.48\linewidth]{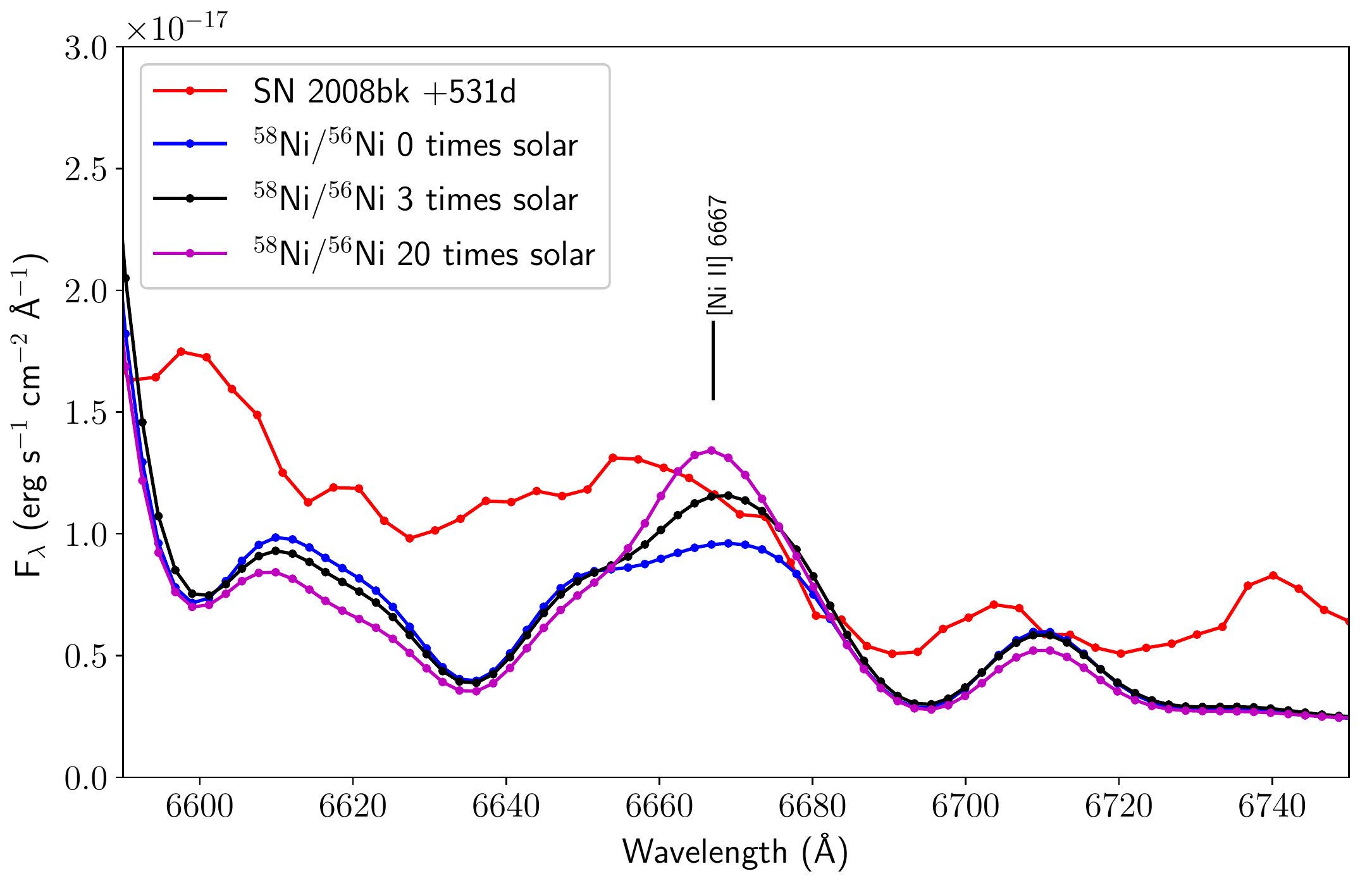} 
\includegraphics[width=0.48\linewidth]{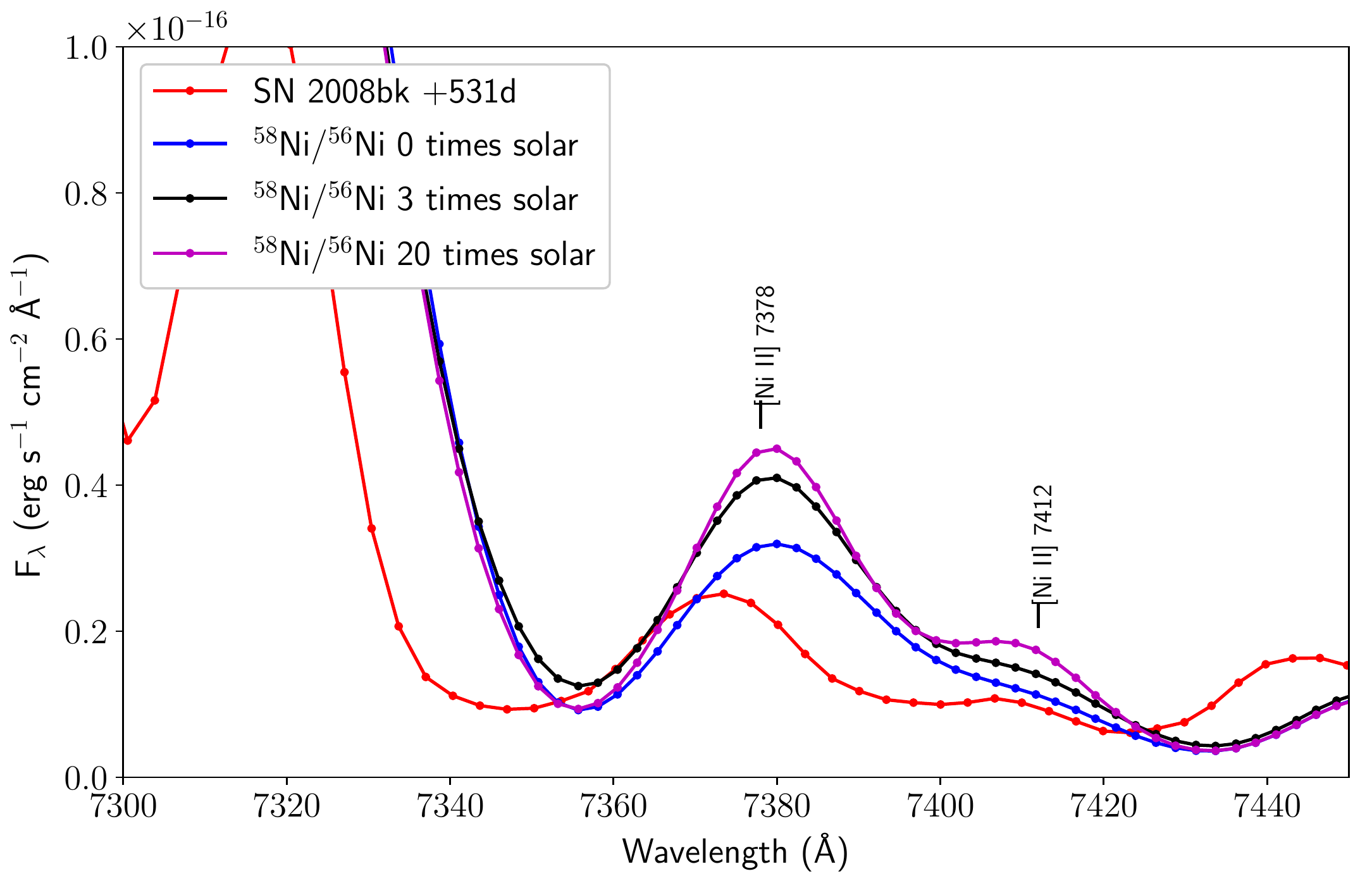} 
\caption{\arja{Comparison of models with varying amounts of synthesized stable nickel added in the explosive burning region (as by the stated $^{58}$Ni/$^{56}$Ni value), and SN 1997D (top), SN 2005cs (middle) and SN 2008bk (bottom). No high production of stable nickel is suggested in either}. Ni/Fe=0 refers to the original model where nickel only exists as a primordial abundance in the H envelope.}
% Redo w beta decay models
\label{fig:wanajocomp}
\end{figure*}

\subsection{SN 1997D}
\arja{Figure \ref{fig:wanajocomp} (top) shows a comparison between the model at 300d and SN 1997D at 350d, zoomed in at the two regions containing the only major changes to the optical spectrum; [Ni II] 6667 and [Ni II] 7378, 7412. While [Ni II] 7378 initially shows promise for a high Ni/Fe ratio, the observed [Ni II] 7412 and [Ni II] 6667 lines are clearly too dim for such a high ratio. In the model at 400d, the [Ni II] 7378 line has emerged more prominently in the Ni/Fe=0 model (from primordial nickel in the envelope), and the overall analysis therefore strongly points to Ni/Fe between 0-3 times solar.} 

\arja{The figure illustrates that while the Ni lines grow in strength with increased abundance in the explosive region, the growth is quite slow. This is because the lines become optically thick. As an illustration of the line formation, in the model with 20 times Ni/Fe, the Ni II fraction is 0.7 at 200 \kms (very similar to Fe value), and Ni II contributes 6\% to the cooling (compare 70\% in Fe I and 5 \% in Fe II). The optical depths are 7.1 in [Ni II] 7378, 2.4 in [Ni II] 7412, and 1.7 in [Ni II] 6667. This optical depth means the difference between the model with Ni/Fe=3 and Ni/Fe=20 is much smaller than 20/3}.

\arja{The very strong [Ni II] 7378 in the Crab nebula has been taken as an argument for a very high Ni/Fe ratio and an origin in the 8-10 \msun~range. At the much higher age of the Crab this line has become optically thin and is therefore a more direct diagnotic of the stable nickel content. Nevertheless, the diagnostic is still powerful enough here to rule out a high Ni/Fe ratio in SN 1997D}.

\subsection{SN 2005cs}
\arja{Figure \ref{fig:wanajocomp} (middle) shows a comparison between the models and SN 2005cs at 300d. %Here the models have been artifically broadened by a factor 3 to approach the observed line widths. 
While there is a broad emission region between 7350-7600 \AA~seen in the data, the distinct bump expected for [Ni II] 7378, 7412 is not present. The discrepancy for [Ni II] 6667 is even clearer, and again a Ni/Fe ratio as high as 20 is disfavoured.}

\subsection{SN 2008bk}
\arja{Figure \ref{fig:wanajocomp} (bottom) shows a comparison with SN 2008bk at 531d. The analysis here is complicated by the dust formation, which eats away at line profiles and is not modelled in detail with our simplistic dust treatment}.

\arja{At these later epochs, there is less deposition in the inner region (20\%), the gas is more neutral ($x_e=0.04$) and colder ($T=3000$ K), and so the nickel lines from synthesized nickel are weaker compared to 300d. There is therefore a smaller difference for the three models. [Ni II] 6667 is predicted too weak to observe even at Ni/Fe=20}.
% Ni II cools 5.0% in model 025, but mostly IR lines. Cooling dominated by Fe I (75%).

\arja{The emission from primordial nickel in the envelope dominates emission from synthesized nickel, so the models with synthesized nickel are only somewhat brighter in the nickel lines than the model with no synthesized nickel. Note that the temperature and ionization decrease slower, and sometimes even increase, in the envelope compared to the \ni\ core (Fig. \ref{fig:temps}, \ref{fig:xe}), so the ratio of emission from primordial to syntesized iron-group elements increase with time. The observed [Ni II] 7378 line is weaker than even the Ni/Fe=0 model, and thus there is no sign of strong nickel production either in SN 2008bk, although the caveat with dust formation should be kept in mind. The [Ni II] 1.94 $\mu$m line shows a similar behavior as the [Ni II] 7378 line.}
% Very small differences in NIR between model with no Ni and model with 3 and 10 times nickel

\subsection{Discussion}
\arja{A caveat to these results is that the fraction of energy absorbed by the \ni\ clumps may vary in a 2D/3D model. Thus, we cannot completely rule out that a high Ni/Fe ratio could be present and still give weak [Ni II] lines, if mixing significantly reduces the local gamma trapping. However, we note that the Fe I lines at 7900-8500 \AA, whose strength depend on this self-absorption, are relatively well reproduced with the current setup. Also, in the model at 500d, the fraction of energy absorbed in these clumps is just 8\%, and half of this is positron energy which definitely cannot escape. Thus, there should be maximum a factor 2 overproduction of the synthesized component. We therefore think our result here, that none of these three SNe show Ni/Fe $\gg$1, and thereby do not originate from ECSN or ECSN-like explosions, is robust.} %Still, the \emph{ratio} between the nickel lines and say [Fe II] 7155 is too high in the model, and this should be less sensitive to the trapping fraction}.

\begin{figure}
\includegraphics[width=1\linewidth]{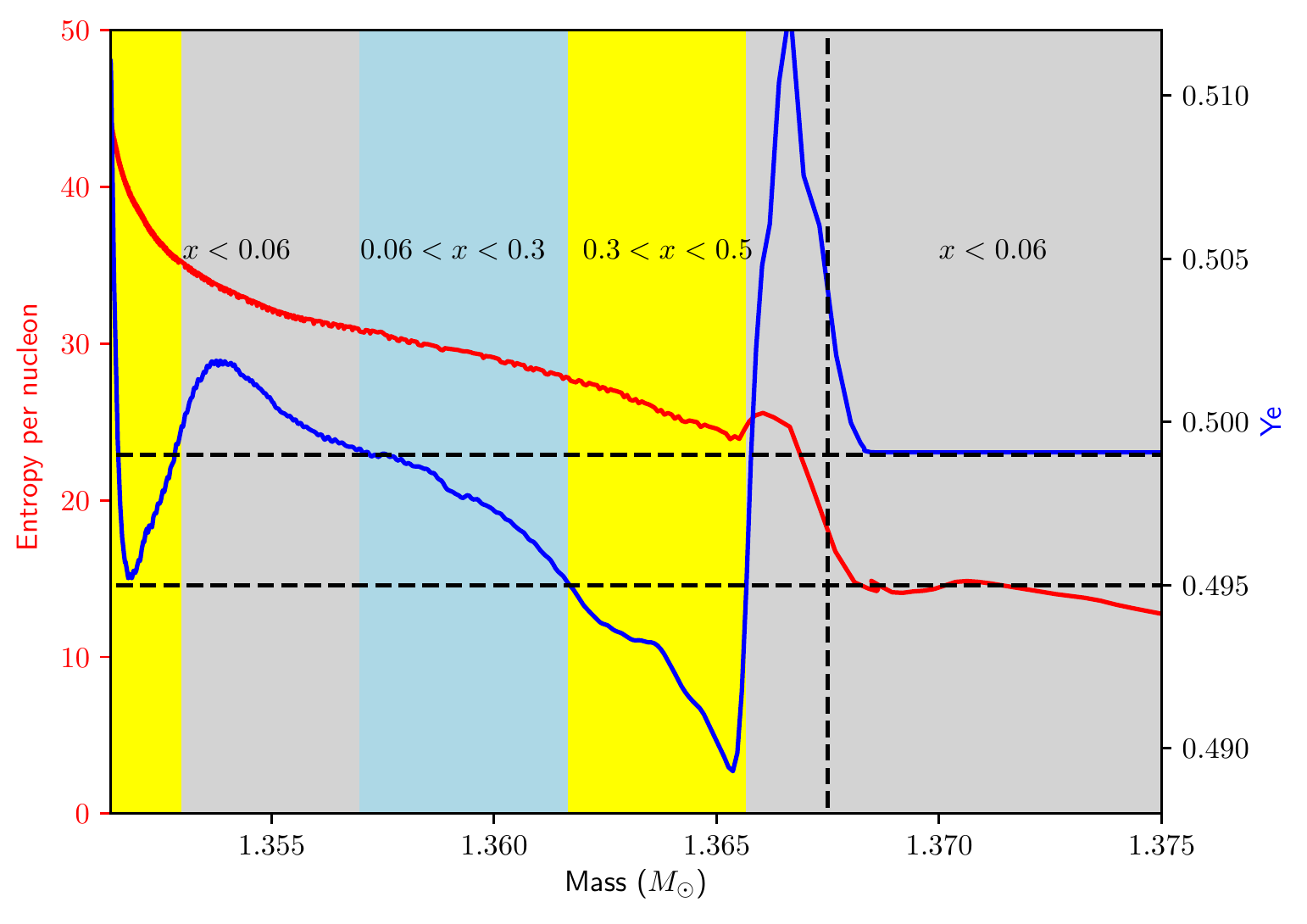}
\caption{Profiles of $Y_e$ (blue line, right axis) and entropy per baryon (red line, left axis) for the innermost ejecta, at the radial location where $T_9 = 7$ and NSE freezes out. Regions of intervals mapped to the $^{58}$Ni mass fraction (shown as $x$), using the results of \citet{Jerkstrand2015c} are indicated. The vertical dashed line at $M=1.3675\ M_\odot$ marks the outer edge of the neutrino processed region. The horizontal dashed lines mark $Y_e=0.499$ and $Y_e=0.495$ that were used for mapping. Iron-group production occurs throughout the plotted domain. The left edge of the region is the final mass cut at 1.35138 \msun.}
\label{fig:Ye}
\end{figure}

Neither our P-HOTB nor KEPLER models can make an accurate prediction for $Y_e$ and $^{58}$Ni nucleosynthesis, as this requires both a large network and neutrino processed ejecta. Without neutrinos, the KEPLER model gives a Ni/Fe ratio close to solar. Inspection of the $Y_e$ profile in P-HOTB gives some indication of what production of $^{58}$Ni can be expected. Figure \ref{fig:Ye} shows that $Y_e$ varies significantly within the explosive nucleosynthesis region. Using the charts in \citet{Jerkstrand2015c} we can estimate for the relevant entropies ($S_\gamma/R=20-30$) what the $^{58}$Ni mass fraction roughly would be (for KEPLER, the Ye is close to the one of the progenitor shown in Fig A1 as it does not include neutrino processes).

Adding the regions together (with the average $^{58}$Ni fraction in each zone), the total expected $^{58}$Ni production is about $3\e{-3}$ \msun, about half the \ni~production and an Ni/Fe ratio of 8 times solar. This is between the solar production occurring in massive progenitors and the 20-30 times solar value occurring in ECSNe-like explosions.

The simplified neutrino calculation and 1D geometry prevents us from firmly saying whether this would hold true in a more advanced model. A Ni/Fe ratio of 8 would from our comparisons above not fit these 3 observed candidates, and a more massive model (with lower Ni/Fe) would be preferable ($M_{ZAMS} \gtrsim 10-11$ \msun). However, it is also possible that this 9.0 \msun~model obtains a lower $^{58}$Ni production in a multi-D simulation with full neutrino physics (work in progress).

\section{Discussion} % =========================================================
\label{sec:discussion}

\subsection{Multi-D effects}
\arja{As our model is 1D, and it has been previously shown that multi-D hydrodynamic effects often are strong in Type IIP SNe, an assessment of the possible impact of mixing effects is warranted}.
% Janka 2008 : A 400 ms 2D ECSN model but little on "mixing", which would occur on longer time-scales
% Mueller 2012,2013 2D models for some 8-12 Msun iron CCSN prog, but for short time --> no dynamics
\arja{Regarding the explosion itself, a 2D ECSN explosion was simulated by \citet{Janka2008}, who demonstrated a weak sensitivity of shock propagation and explosion energy to dimensionality \citep[see also][for other published examples]{Muller2012, Muller2013}. The only 3D simulation done for the low-mass range is by \citet{Melson2015}, who discuss the impact on explosion energy in a 400 ms simulation of a 9.6 \msun\ progenitor. In 3D a somewhat more rapid explosion occurs compared to 2D, with a 10\% higher explosion energy.}

\arja{These runs were however too short ($<$1s) to reach conclusions on large scale mixing occurring as reverse shocks are created at later times. If these produce as strong stirring of the inner ejecta as at higher $M_{ZAMS}$ \citep[e.g.][]{Hammer2010}, the resulting spectrum, both line profiles and luminosities, can change significantly as the radioactive gamma deposition changes between the different layers. With models using strong mixing, \citet{Maguire2012} and \citet{Lisakov2017} found reasonable fits to SN 2008bk for a 12 \msun~star. Thus, the precise best-fitting mass likely depends on the amount of mixing.}

However, for two reasons we expect less mixing of \ni\ into the He and H layers than in more massive progenitors. First, the explosion energy of $\sim 10^{50}$ erg is considerably lower than the more canonical $10^{51}$ erg of 15-20 \msun\ stars. Second, the relative amount of \ni\ and ejecta is also lower ($\sim 0.006/7.5$ compared to $0.07/(13-18)$ for massive progenitors). Both facts favor less entrainment of high-velocity \ni\ into the He and H shells. Since the detailed structure of the progenitor star, however, plays a crucial role for the mixing \citep{Wongwathanarat2015}, 3D long-time explosion simulations are needed for a final answer.

We note also that strong mixing would lead to brightening of lines such as H$\alpha$. In our unmixed setup the gamma deposition in the He envelope is about 20\% at 200d, and this gives a well reproduced H$\alpha$ compared to SN 1997D and SN 2005cs (no data on SN 2008bk). This line, and also other lines such as [Ca II] 7291, 7323 would become too strong in a mixed scenario.

\arja{The impact on nucleosynthesis going from 1D to 2D for low-mass SNe was studied by \citet{Wanajo2011} (ECSNe) and \citet{Wanajo2017} (low-mass CCSNe). In both cases, the authors find significant differences for some high-A isotopes \citep[see also][]{Yoshida2017}, but for the dominant species that produce the nebular spectra (e.g. Fe, Ni) the differences were small compared to 1D models \citep{Wanajo2009}}.

\subsection{The link between subluminous IIP SNe and 8-12 \msun~stars.}
\arja{When subluminous Type II SNe were discovered in 1997, one hypothesis put forth was that this were massive stars exploding with a low energy and fallback of \ni\ \citep{Turatto1998}}. From several arguments presented below we consider it more plausible that this observational class is linked to low-mass progenitors. 
\begin{enumerate}
\item \arja{The observationally inferred explosion energies and amounts of \ni~are naturally obtained in current ab-initio explosion simulations of low-mass progenitors \citep{Kitaura2006,Melson2015,Muller2016,Wanajo2017}. If subluminous IIP SNe do not correspond to these progenitors, what observational class would}?
\item \arja{While a single object or two could perhaps be explained by fall-back, extreme fine-tuning would be needed to consistently produce a \ni~mass of $\sim$$5\e{-3}$ \msun, as observed now in over a dozen objects \citep{Spiro2014}, making up a good fraction of closely studied Type IIP SNe. It has, in addition, not been demonstrated that such fallback is dynamically possible}.
% Spiro2014 table 13 shows 14 objects with M(56Ni) < 0.01 Msun and 2 0.01-0.02.
% Limongi2003 has a few ejecta with M(56)Ni <~ 0.01 Msun.
\item \arja{The observed fraction of these events relative to all Type IIP SNe is of order 1/3 \citep[using Fig. 16 in][]{Spiro2014}. This roughly correspond to the range of $8-11$ \msun~if we assume $M_{max} = 25$ \msun\ for Type IIP SNe (this gives 44\% with a standard Salpeter IMF)}.
\item \arja{In three of three cases have direct progenitor detections given low masses, $M_{ZAMS}=8-15$ \msun}.
\item The nebular model of a low-mass progenitor studied here fits line profiles and luminosities in 2 out of 3 events, without any tuning. Even without a model, the evolution of the [O I] 6300, 6364 ratio is not consistent with large amounts of oxygen as originally pointed out by \citet{Chugai2000}.
\end{enumerate}
%This leaves the question what arguments do we have for massive progenitors for this class? 

\arja{Tension still comes from hydrodynamic light curve modelling, which occasionally points to higher masses. Two main results are in particular giving substantially higher values; those of \citet{Zampieri2003} and \citet{Utrobin2008}. The result of \citet{Utrobin2008} is for SN 2005cs, which is the object we could not draw any strong conclusions on regarding nucleosynthesis. It has somewhat different line profiles than SN 1997D and SN 2008bk, and may originate from a different kind of progenitor, or have exploded in a different way, perhaps with more mixing}.

\arja{For very few CCSNe do we have strong constraints on the progenitor star that exploded. Subluminous IIP SNe hold promise to become the observational class where we have a firm idea of the progenitor mass range to within 1-2 \msun. If this is achieved, we can move on to constrain the stellar evolution and explosion physics by comparing detailed models starting from a narrow range of progenitor masses. Of particular importance here is SN 2008bk, which exploded just 3.7 Mpc away and is firmly indicated to be a $\sim$10 \msun~star from both progenitor detection, hydrodynamic modelling, and nucleosynthesis analysis. With a large available dataset (Pignata, in prep.) this will become an important testbed for low-mass explosion models}.

\arja{Another topic we can make progress on now is to attempt to answer whether ECSNe occur in Nature. Recent stellar evolution modelling has indicated that the progenitor mass range is narrow and may not exist at all. Here, we note that zero out of 3 subluminous IIP SNe show nucleosynthesis of stable nickel that would be expected from an ECSN \citep{Wanajo2009,Wanajo2011}. They also show evidence for He core lines, which would be lacking in ECSNe. While there remains a possibility that some of the He shell emission (He and C lines) may be moved to the H envelope in a progenitor with dredge-up, emission by O shell material should be truly absent in an ECSN.}

In a toy scenario where ECSNe would come from the range 8 to $8+\Delta M$ \msun~stars and subluminous IIP from $8-12$ \msun~stars, we can already make a rough constraint on $\Delta M$ by taking $\left(\left(1-\Delta M)/4\right)\right)^3 < p$ (probability $p$ to have zero of three events not belong to the ECSN category). This gives $\Delta M < 1.2$ \msun~on a 1-sigma level (ignoring IMF weighting). A problem here, though, is that recent 2D models have shown that also the lowest-mass Fe CCSNe could produce much stable Ni. \citet{Wanajo2017} found similar Ni/Fe ratios of $\sim$20-30 times solar in e8.8 (ECSN), u8.1(Fe CCSN) and z9.6 (Fe CCSN) models.  Note that the (unpublished) u8.1 and z9.6 models have extreme metallicities ($Z=10^{-4}$ and 0, respectively), and behave different from other models in this mass range. For example, the density profile of z9.6 model is much steeper than for the 9.0 \msun~model we have considered here, despite its higher mass (see e.g. Fig. 1 in \citet{Radice2017}). 

At $M_{\rm ZAMS}=11.2$ \msun~the Ni/Fe ratio was restored to close to solar in \citet{Wanajo2017}. It is of importance to investigate what happens in solar metallicity Fe CCSN models at 9-11 \msun, which have density gradients between the z9.6 and 11.2 models. If these were also produce high Ni/Fe ratios, it would become an issue that we do not observe strong $^{58}$Ni lines in any of these objects. We note here that several uncertainties in the neutrino treatment, e.g. neutrino oscillations, also leave some uncertainty for the calculations of these yields.

\subsection{The existence of faint supernovae from massive stars}
The high masses inferred for some subluminous Type IIP SN by hydrodynamic modelling led to the hypothesis of two branches of SNe in the $M_{ZAMS}-E$ plane; the bright and faint branches \citep{Nomoto2013}. One important observational puzzle that the existence of a faint branch could solve is the abundances of CEMP stars, which are enriched in intermediate elements like C and O, but lack iron-group elements. Only a fallback SN from a massive star could easily explain these.

As discussed above we think there is overall scant evidence for the existence of such a branch in the local Universe. In particular, nebular spectra of SN 1997D do strongly suggest a low-mass star, and there is almost no data on the photospheric phase to constrain light curve models. 

One attractive scenario would be that faint supernovae occurred in the early Universe, but not today. Indeed low metallicity makes it harder to explode stars, and fallback cases might become more likely \citep{Pejcha2015}.

\section{Conclusions and summary}
\label{sec:conclusions}
\arja{We have calculated the spectral appearance of a 9 \msun~CCSN in the interval 200-600d post explosion, using a neutrino-driven 1D model. We analyzed the spectral formation processes in this model, and compared the model spectra with observed spectra of subluminous Type IIP SNe. Our main conclusions are}

\begin{itemize}
\item \arja{After radioactive decays, the He core in a low-mass Type IIP SN producing $\sim 5\e{-3}$ \msun~of \ni~expands with $\sim$500 \kms. Inclusion of the ``Ni bubble'' effect broadens lines by a factor $\sim$2 and changes some line luminosities}.
\item \arja{Gamma rays are fully trapped for about 3 years, giving a light curve following the $^{56}$Co decay for this amount of time. The gamma rays are absorbed in the innermost 500-1000 \kms~of the ejecta, which leads to emission lines with FWHM$\sim$900 \kms~for He core lines (when convolved with typical telescope resolution of 300 \kms, this becomes 950 \kms), and somewhat larger for H lines ($\sim$1100 \kms)}.
\item \arja{Despite the low mass of \ni, O and He zones compared to the H zone, these shells absorb significant amounts of gamma ray energy and emit characteristic, detectable signatures}:
\begin{enumerate}
%\item 
\item \arja{The \ni~shell displays unique Fe I lines in the 7900-8500 \AA~range}.
\item \arja{The He shell displays unique He I 7065 and [C I] 8727.}
\item \arja{The O shell displays unique O I 7774, and enhances Mg I] 4571 and [O I] 6300, 6364.}
\end{enumerate}
\item \arja{The 9 \msun\ explosion model, without any free parameters except a rescaling factor of order 2 to allow for somewhat different \ni~masses, gives good reproduction of nebular spectra of subluminous Type IIP SN 1997D and SN 2008bk, associating this class with the lowest-mass end of the progenitor spectrum. Both line profiles and luminosities naturally emerge at their observed values when an explosion delivers $\sim$ 0.1 B to a low-mass progenitor. While high-mass models have not yet been calculated, one can find arguments against them from certain line ratios. In addition, even if such suitable models can be constructed, they would require fine-tuning to get the right amount of fall-back and core structure}.

\item \arja{For SN 2005cs, the link cannot be as clearly established, as by the end of the observational coverage at +330d, key diagnostic lines of e.g. Mg I] 4571, [O I] 6300, 6364 had still not emerged (this happened sometime between +207 and +350d in SN 1997D and between +265d and +416d in SN 2008bk). However, spectral similarities to SN 2008bk around 265d are high, which makes a common scenario plausible}.

\item \arja{None of the observed events show the high Ni/Fe ratio ($\sim$30 times solar) predicted for ECSN and ECSN-like explosions. SN 1997D and SN 2008bk also show clear evidence of He core material, which consistently rules out ECSN models for them. Thus, we assess that ECSNe remain to be discovered from spectroscopic analysis, and we note here that the expected signature is \textit{lack} of certain lines (He I 7065, [C I] 8727, Mg I] 4571) at $t\gtrsim$350d, combined with strong [Ni II] 6667 and [Ni II] 7378. With zero of three subluminous IIP objects matched to ECSN nucleosynthesis, their progenitor mass range is constrained to $\Delta M \lesssim$1 \msun~at 1 sigma}.

\item \arja{With very good consistency with spectral modelling performed here, progenitor detection, and light curve modelling, SN 2008bk is today our perhaps most robustly inferred explosion of a low mass Fe CCSN, with $M_{ZAMS} = 9-12$ \msun. This gives strong empirical support for the developing theoretical picture that low-mass stars explode with a low energy ($\sim$ 0.1 B) and a low \ni~mass ($\sim 5\e{-3}$) \msun. The close distance (3.7 Mpc) and a comprehensive data-set will allow this SN to become a test-bed for detailed stellar evolution/explosion models for the 9-12 \msun~range, similar to what SN 1987A has been for ~$\sim$20 \msun~progenitors}.
\end{itemize}

\appendix % =================================================================

\section{The progenitor structure}
\begin{figure*}
\includegraphics[width=1.0\linewidth]{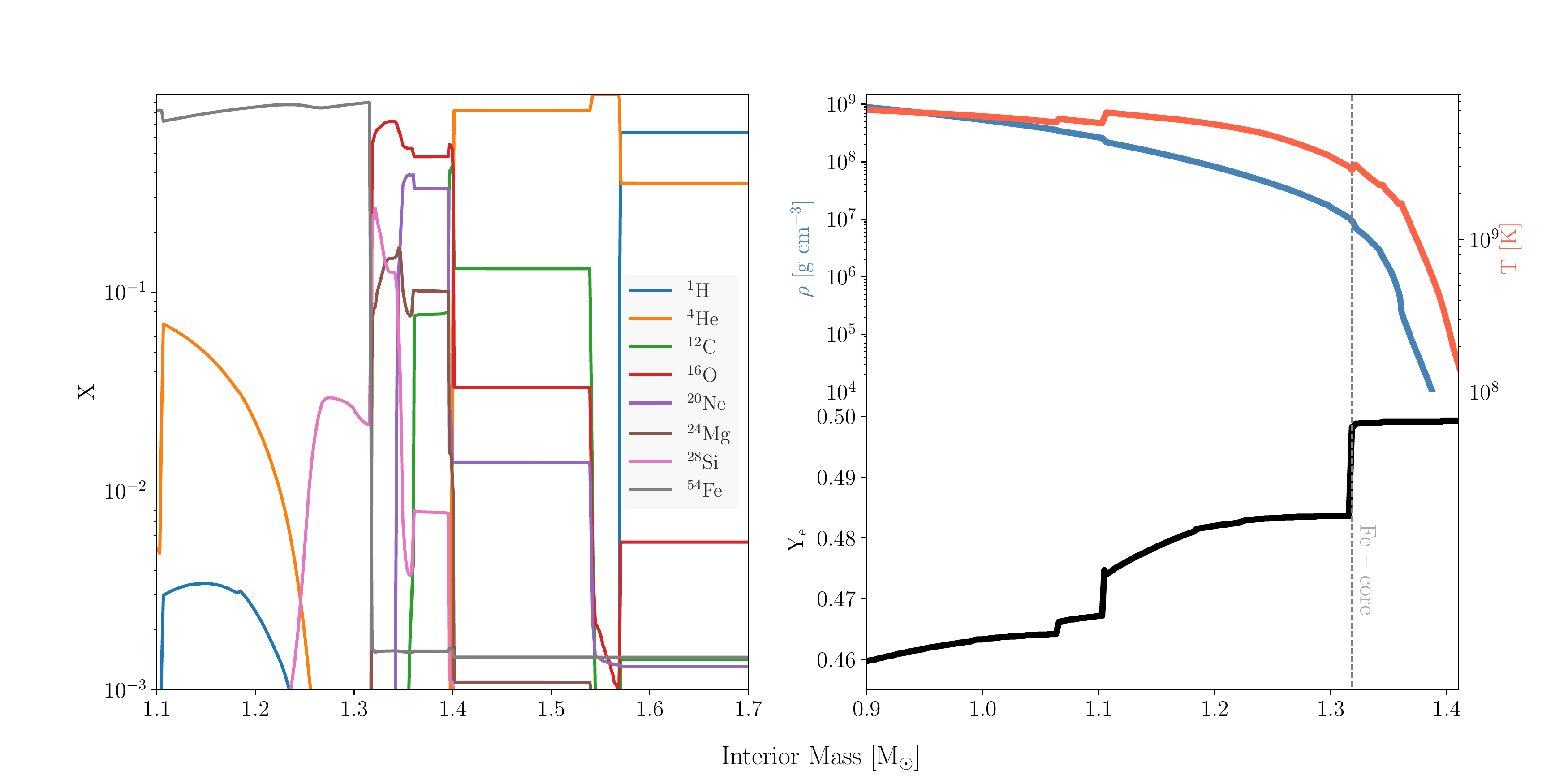}
\caption{Progenitor model. The left panel shows composition, the upper right density and temperature, and the bottom right $Y_e$.}
\label{fig:progenitor}
\end{figure*}

Figure \ref{fig:progenitor} shows the composition, temperature, density and electron fraction in the progenitor. While the progenitor was not calculated here but in \citet{Sukhbold2016}, we add this information here for completeness. 

\section{Comparison with KEPLER model}
\label{sec:KEPLER}
\begin{figure}
\includegraphics[width=1\linewidth]{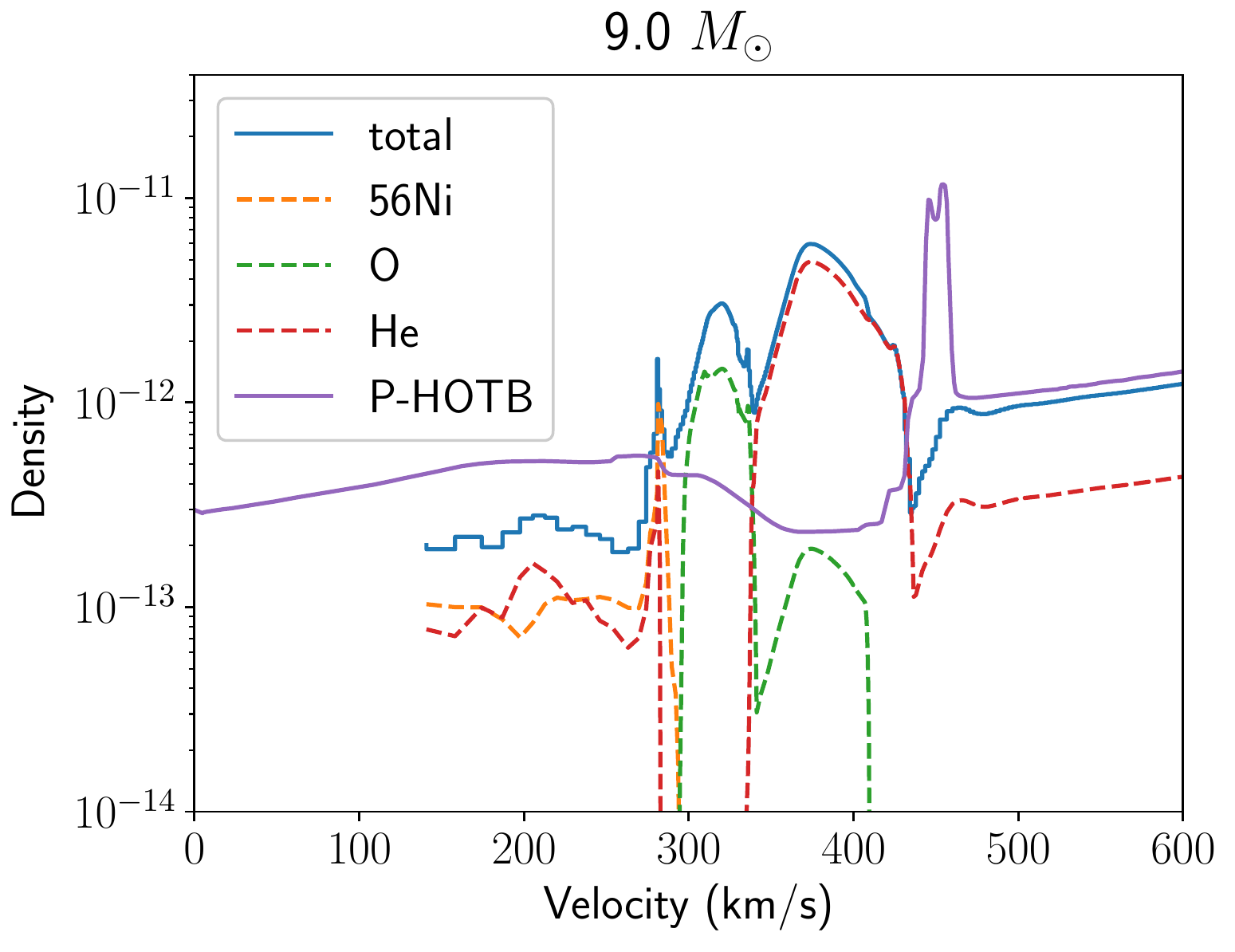} 
\caption{\arja{Density profile of the KEPLER model (blue), compared to P-HOTB (purple, see also Fig. 1). The density profiles of He, O, and \ni~are also shown.}}
\label{fig:KEPLERdens}
\end{figure}

Spectra of the KEPLER explosion simulation of the 9 \msun~model were also calculated. The density profile of this model compared to the P-HOTB model is plotted in Fig. \ref{fig:KEPLERdens}. The KEPLER model shows less dramatic shell compression from $\beta$ decay than the P-HOTB model. Part of the reason for this is the lower \ni~mass in KEPLER, $3.6\e{-3}$ \msun~versus $6.2\e{-3}$ \msun\ in P-HOTB (see Table 7 in \citet{Sukhbold2016}). The cause of this difference is that the KEPLER simulations lack the neutrino-processed matter ejected from the vicinity of the neutron star after the start of the explosion. The evolution of the reverse shock as it approaches the inner region of the domain also differs between the codes.

\begin{figure*}
\includegraphics[width=1\linewidth]{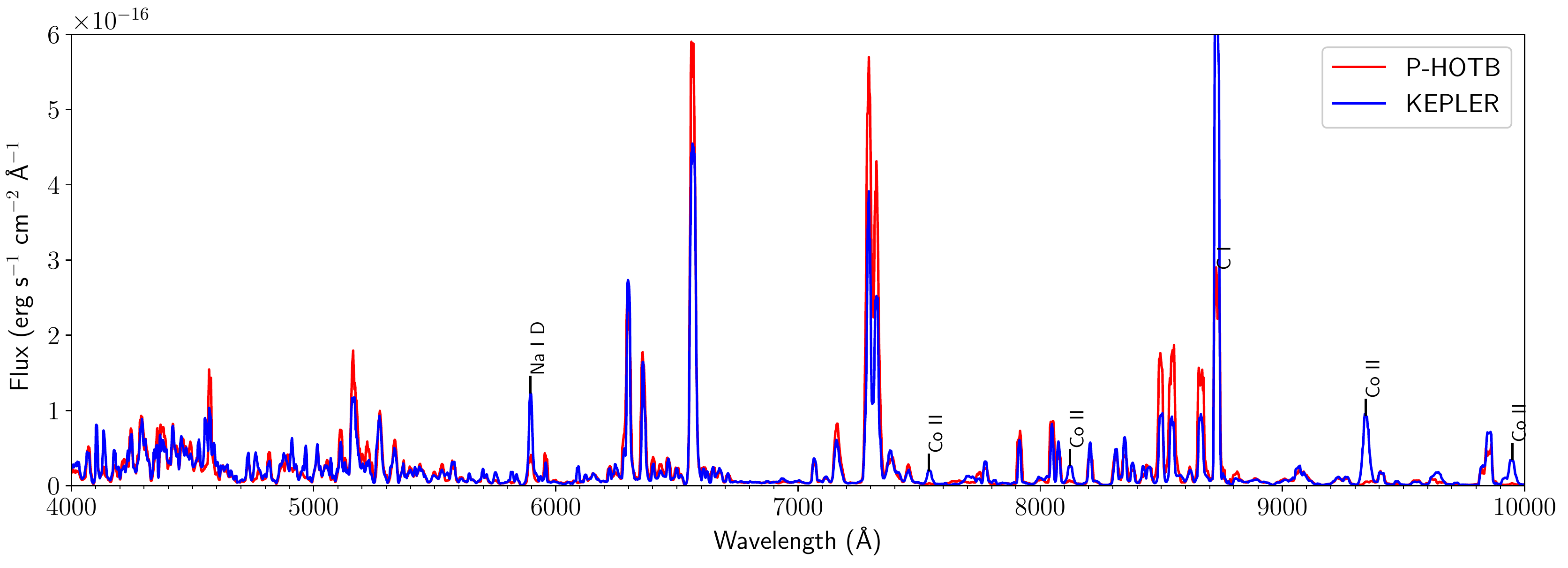} 
\caption{Comparison of P-HOTB (red) and KEPLER (blue) model spectra at 400d. The models have been scaled to have the same \ni~mass, and convolved with a Gaussian of FWHM=150 \kms. The lines showing the biggest difference are labelled.}
\label{fig:KEPLERcomp}
\end{figure*}

The spectra show relatively small differences (Fig. \ref{fig:KEPLERcomp}). The KEPLER model has a few distinct lines from cobalt emerging at 7550, 9350 and 9950 \AA, originating in the core. This is radioactive $^{56}$Co, as the abundance of stable cobalt is very low. It also gives a stronger [C I] 8727, Na I D, and weaker calcium lines. The factor $\sim$2 variation in these suggest conclusions based on smaller discrepancies with observations than factor 2 should be treated with care. The hydrogen, oxygen, magnesium, and iron lines show smaller differences.

\begin{figure*}
\includegraphics[width=0.48\linewidth]{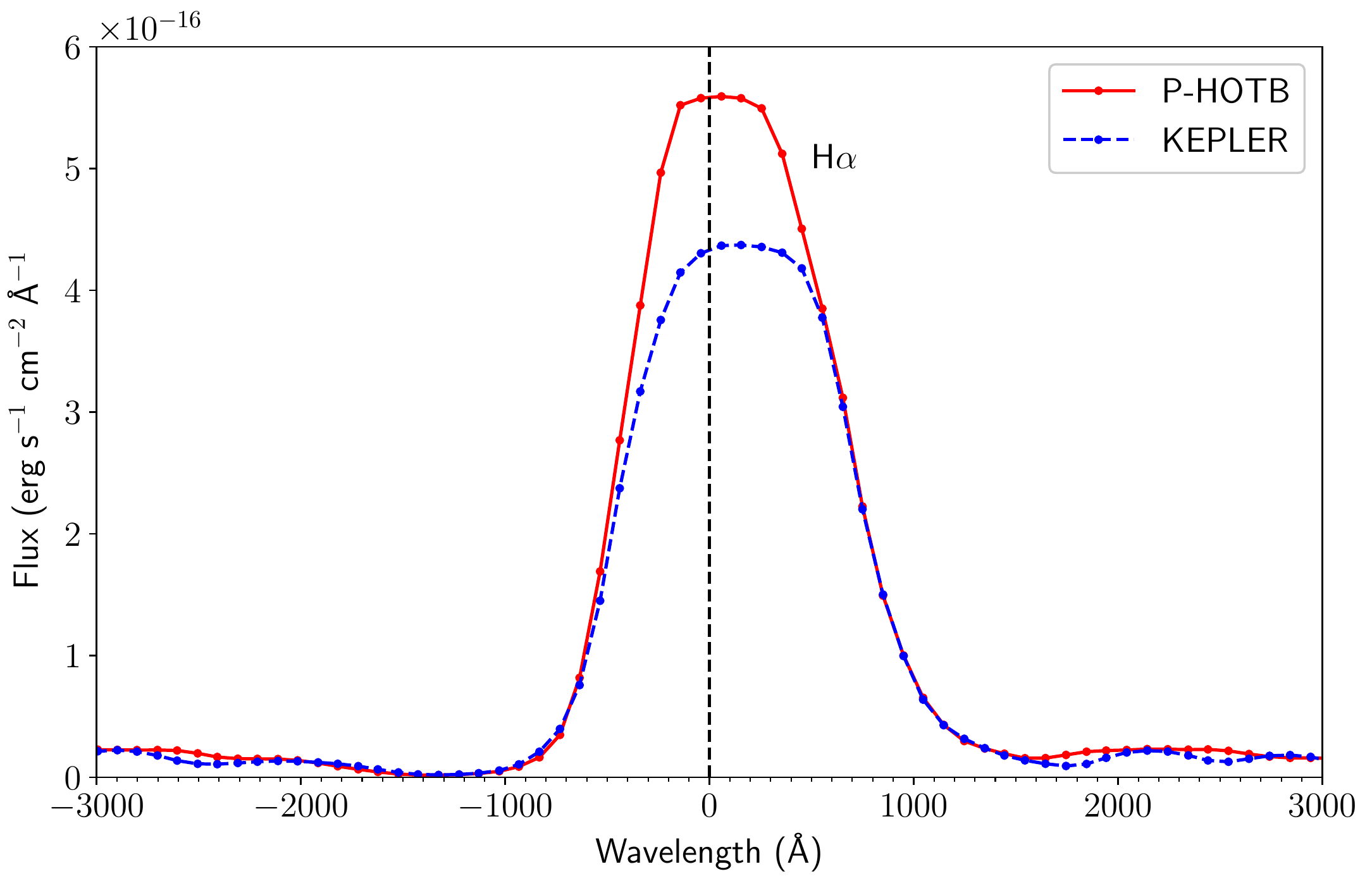} 
\includegraphics[width=0.48\linewidth]{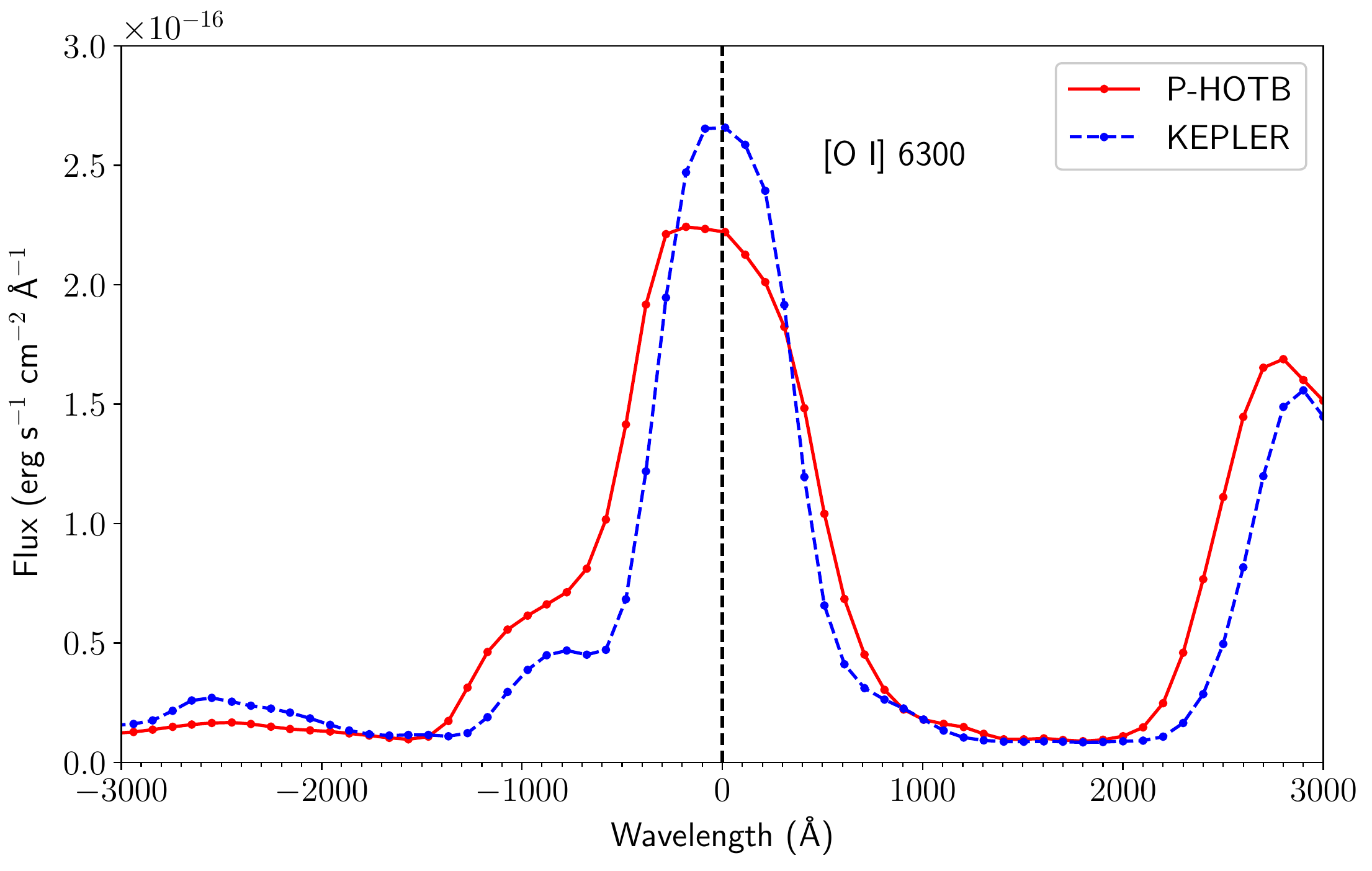} 
\includegraphics[width=0.48\linewidth]{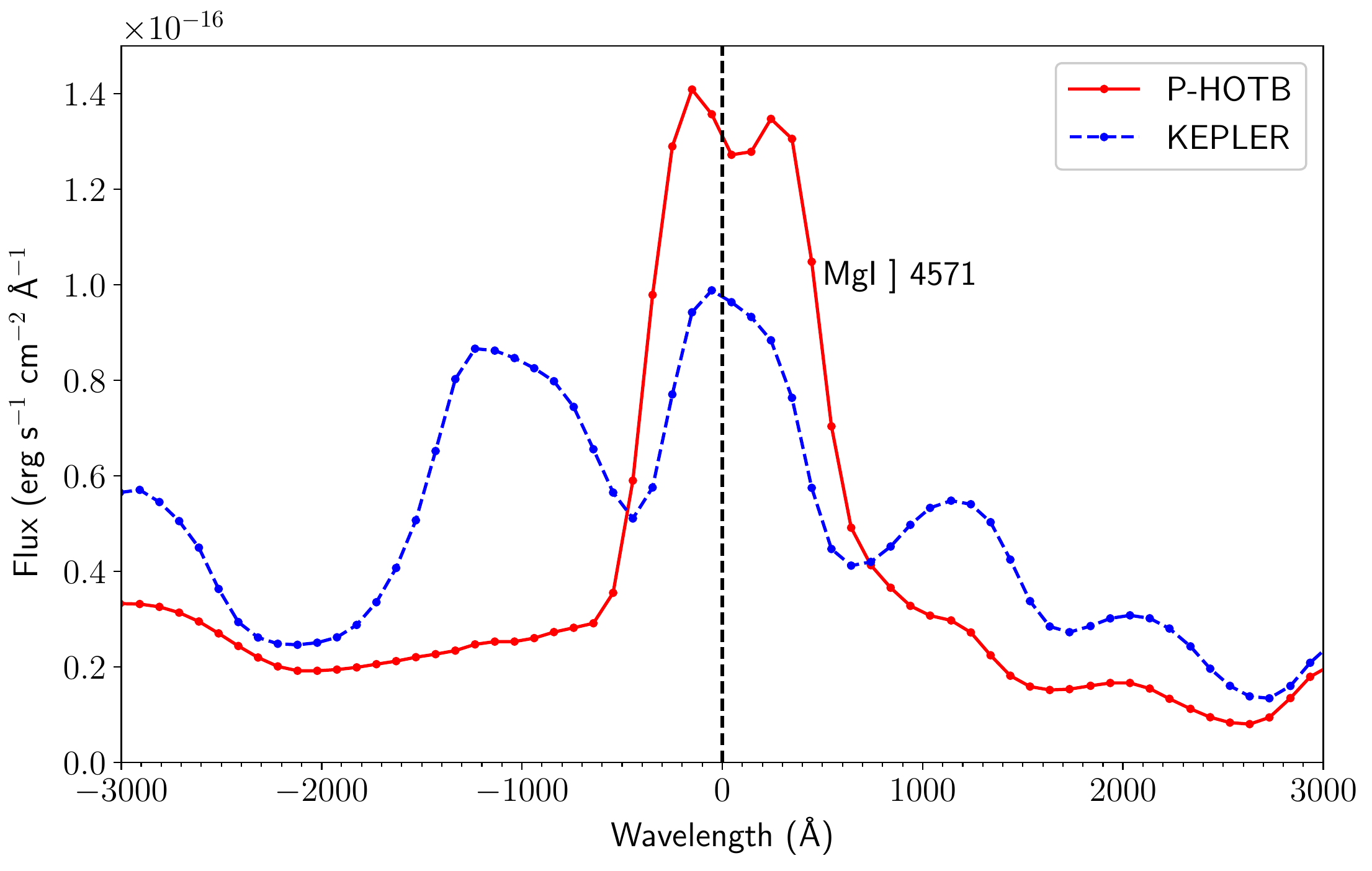} 
\includegraphics[width=0.49\linewidth]{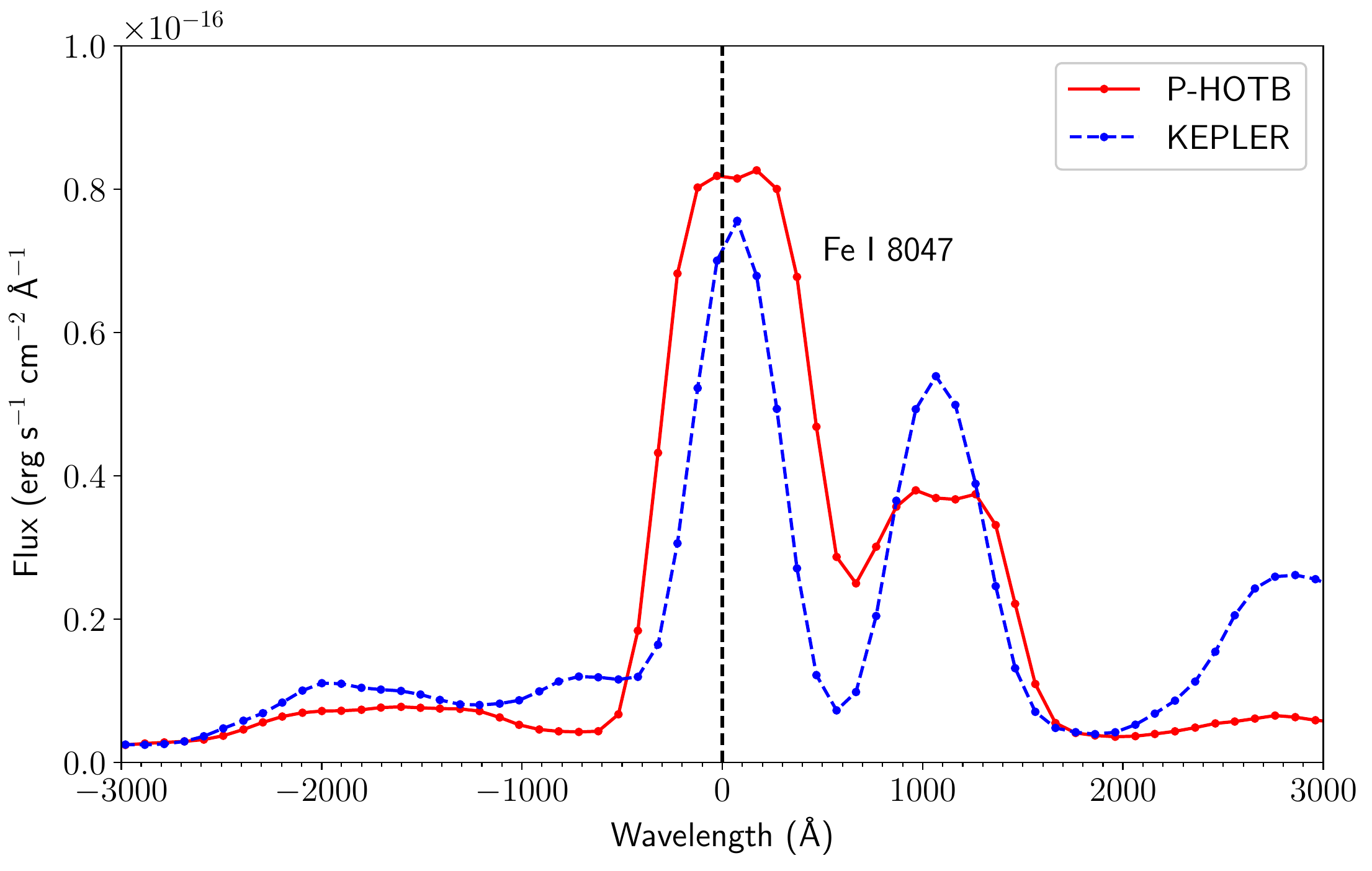} 
\caption{Comparison of selected line profiles in P-HOTB (red, solid) and KEPLER (blue, dashed) at 400d.}
\label{fig:KEPLERzoom}
\end{figure*}

Figure \ref{fig:KEPLERzoom} shows a zoom-in on some lines. Whereas lines from the envelope such as H$\alpha$ show little differences in their line profiles, the slower core in the KEPLER model is clearly seen in oxygen, magnesium, and in particular in iron lines.

Nevertheless, it is encouraging for the robustness of the modelling that the differences due to the different explosion methods and nucleosynthesis networks in the two codes is not bigger.

The mass of stable nickel ($^{58}$Ni and $^{60}$Ni) in the KEPLER model (which uses a large nucleosynthesis network) is $1.6\e{-4}$ \msun, so $\mbox{Ni}/\mbox{Fe}=0.043$ (after decay of \ni), about the solar value of 0.056 \citep{Lodders2003}. Thus, this explosion simulation shows no continuation of the growing Ni/Fe ratio with declining $M_{ZAMS}$ obtained in the \citet{Woosley1995} grid, as illustrated in Fig. 8 of \citet{Jerkstrand2015c}. %The Si shell of the progenitor extends to 1.32 \msun, whereas 
The mass cut is at 1.35 \msun, which is in the oxygen shell, where $Y_e \sim 0.499$ and where solar Ni/Fe production occurs. The KEPLER simulations also lack the neutrino processing that gives rise to neutron-rich production in the inner layers.

\section{Composition mapping}
\begin{figure}
\includegraphics[width=1\linewidth]{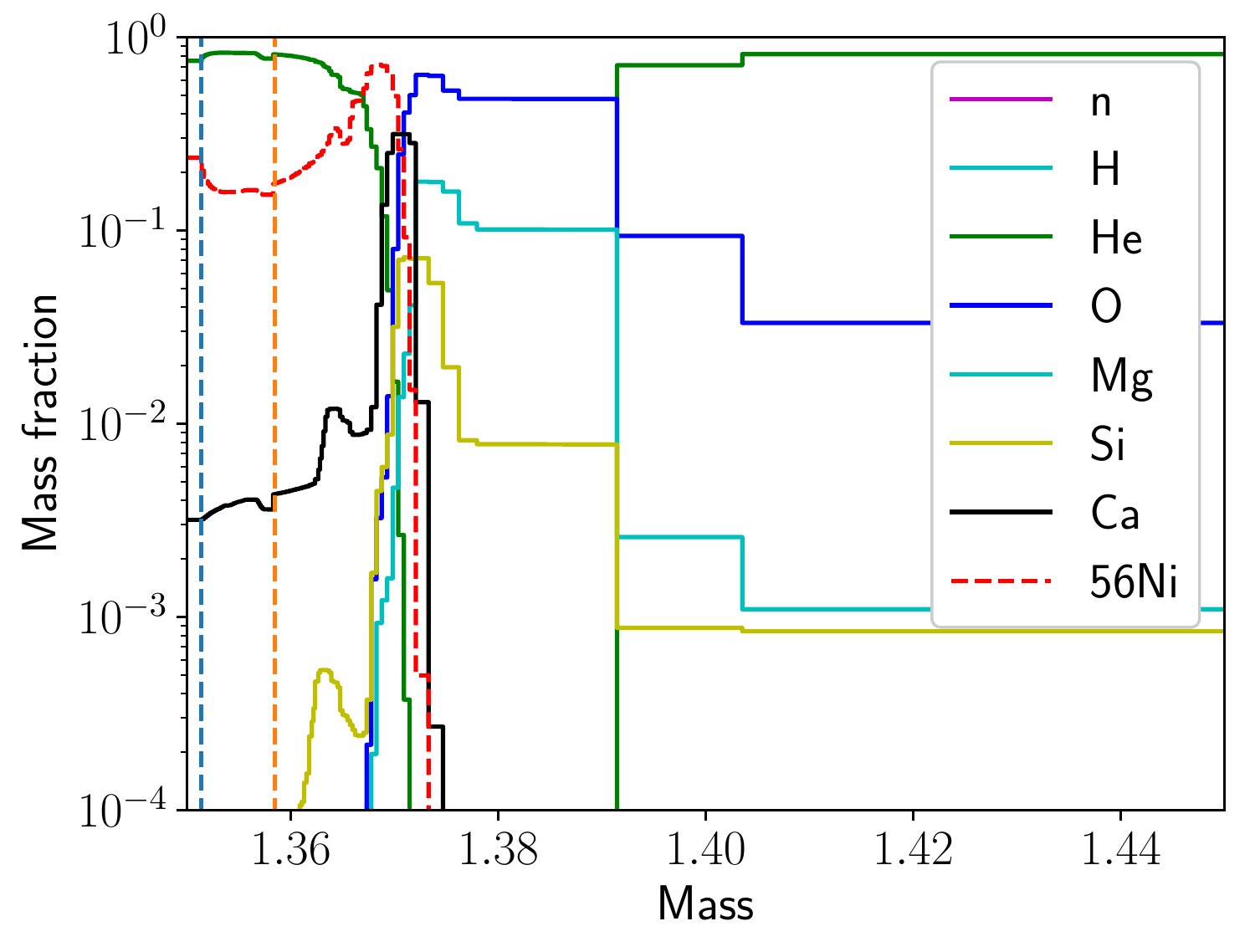}\\ % plotspectra.py 
\includegraphics[width=1\linewidth]{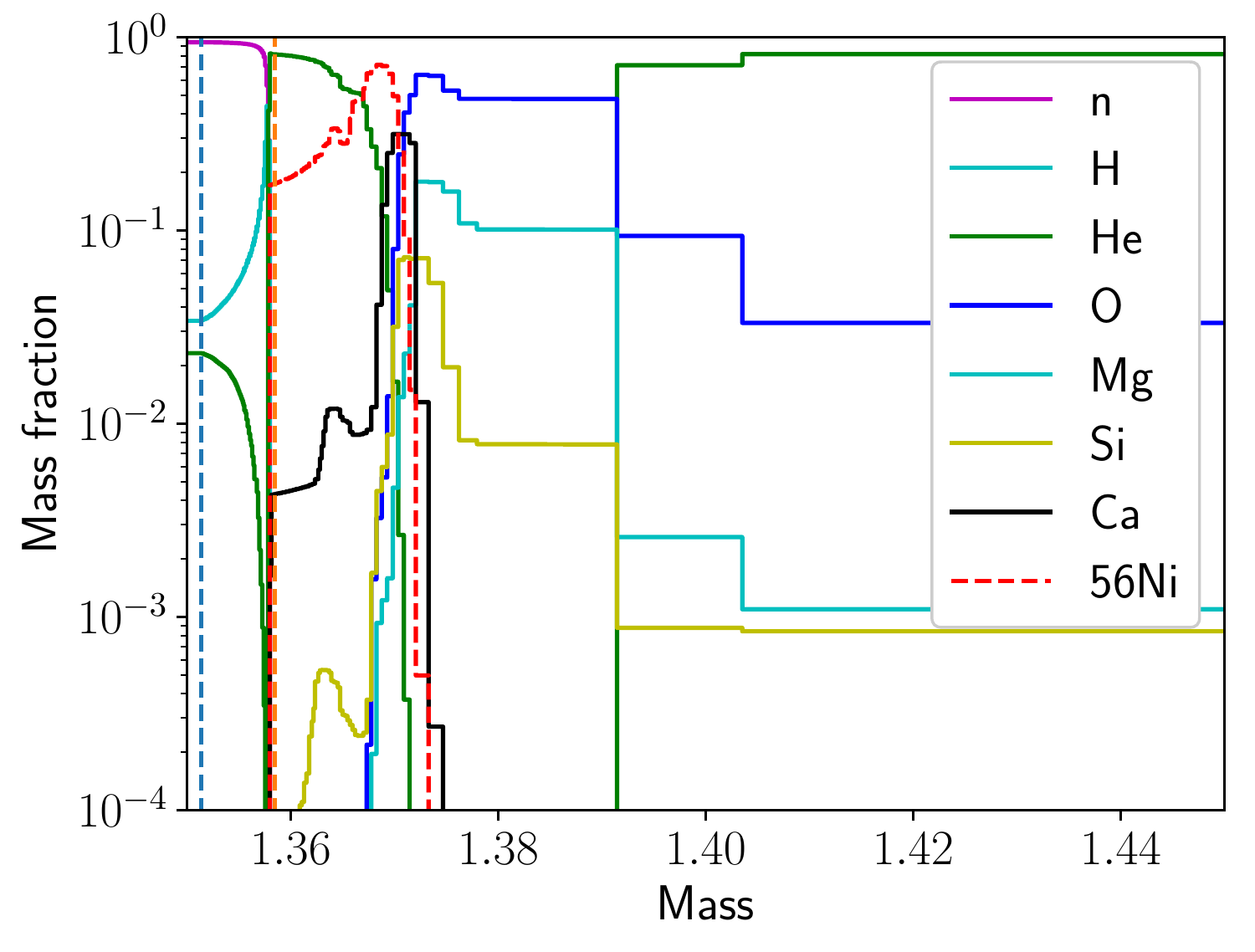} 
\caption{\arja{Illustration of numeric diffusion of composition. The top panel shows composition versus mass coordinate in a simulation run to 100d. The bottom shows a run to 2s. The left-most vertical line at 1.35138 \msun~denotes the ``mass cut'' (taken as $V>$50 \kms~at 100d). The second vertical line at 1.35803 \msun~marks the mass coordinate inside which the composition is n and p at 2s (not yet assembled to \ni).}}
\label{fig:compcompare}
\end{figure}

\arja{One limitation of Eulerian explosion simulations like P-HOTB is that compositionally distinct
layers get artificially mixed as they are 'cut' by the Eulerian spatial gridding at each time step.
The final composition profiles of a model at late time are significantly different than at early time.
Figure \ref{fig:compcompare} compares composition at 2s and 100d, showing the effect}.

\arja{For the computed models, the composition versus mass coordinate was taken at 2s. Material inside 1.35803 \msun\ in the 9 \msun\ model is still evolving at 2s (see Fig. \ref{fig:compcompare}, bottom panel), processing of another 0.0067 \msun~will increase the total \ni~yield from $5.0\e{-3}$ \msun~at 2s to a final $6.2\e{-3}$ \msun. To include this, the composition in this innermost region ($1.35138-1.35803$ \msun) was taken from the 100d model rather than the 2s model}. %textcolor{red}{Explanation for proton-rich layer?}
%It may be better to use a model at say 5s or 10s.

\arja{In the 2s model, \ni~extends only into a small fraction of the O shell ($\sim$10\% by mass). Nevertheless, the O shell is moving at only a slightly higher velocity ($\sim$25 \kms) than the \ni~rich shell at 1.36-1.37 \msun. Therefore, with limited velocity resolution (10 \kms), \ni~gets mixed into about half of the O-rich region in our grid (Fig. \ref{fig:comp}). Binning has to be finer than 5 or maximum 10 \kms~to resolve the individual composition shells. Beyond some limit, increased velocity resolution will become meaningless as several physical processes treated in on-the-spot approximations in the model occur in reality over some finite volume/velocity dispersion. What exactly these volume scales are would require a detailed separate investigation.}

\section{Sensitivity to radioactive decay} % ======================================================
We study here the impact of the ``56Ni bubble'' for the spectral formation. This long-term dynamic evolution of the ejecta is often ignored in explosion
simulations and spectral models, so it is of interest to improve understanding of its significance.

\begin{figure*}
\includegraphics[width=1\linewidth]{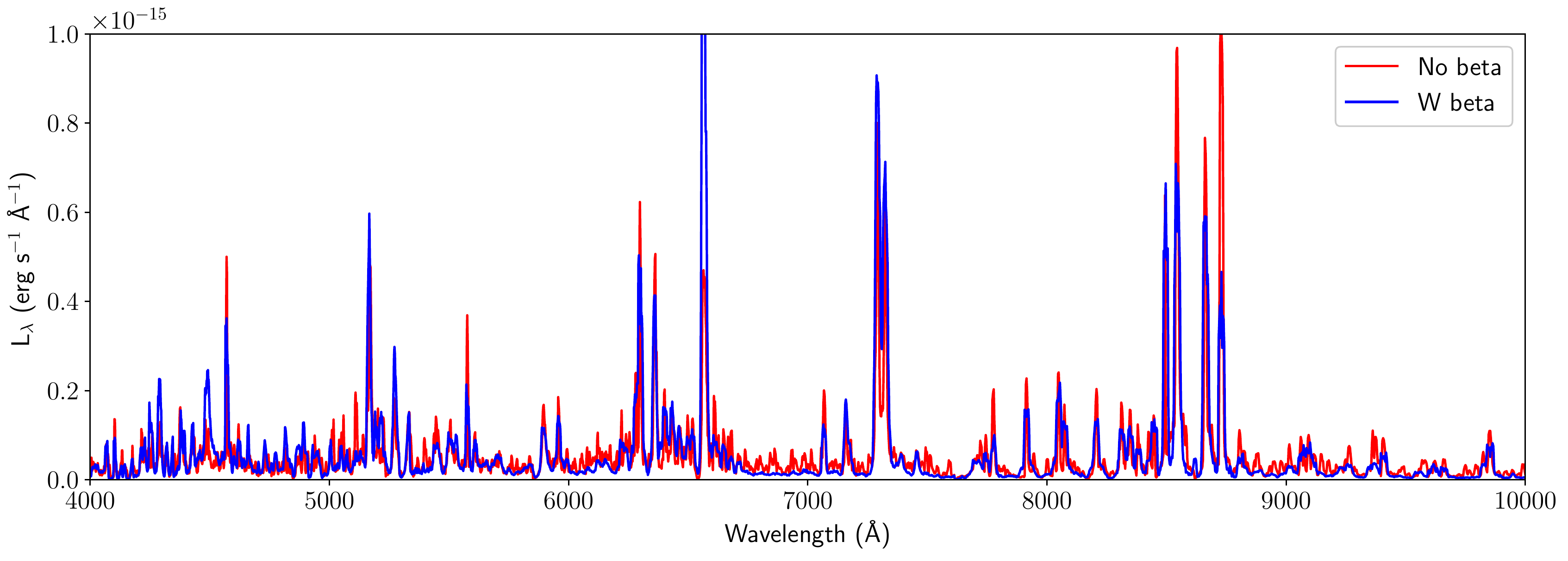}\\ 
\caption{\arja{Comparison for the 9 \msun\ model without (red) and with (blue) dynamic effects of beta decay, at 300d. The beta decays broadens the lines and changes some luminosities significantly, e.g. in H$\alpha$ and [C I] 8727.}}
\label{fig:comp4}
% DC 2017-06-12 Same edep for model 022 and 028
\end{figure*}

\arja{Figure \ref{fig:comp4} compares models computed with and without the dynamic effect of radioactivity on the ejecta, at 300d. 
Figure \ref{fig:density} compares the density profiles for the two cases. The factor $\sim$ 2 larger core velocities in the model with decay leads to broader lines and less fine-structure in the spectrum. The overall spectral properties are not much changes, but some lines show sensitivity.}

\begin{figure}
\includegraphics[width=1\linewidth]{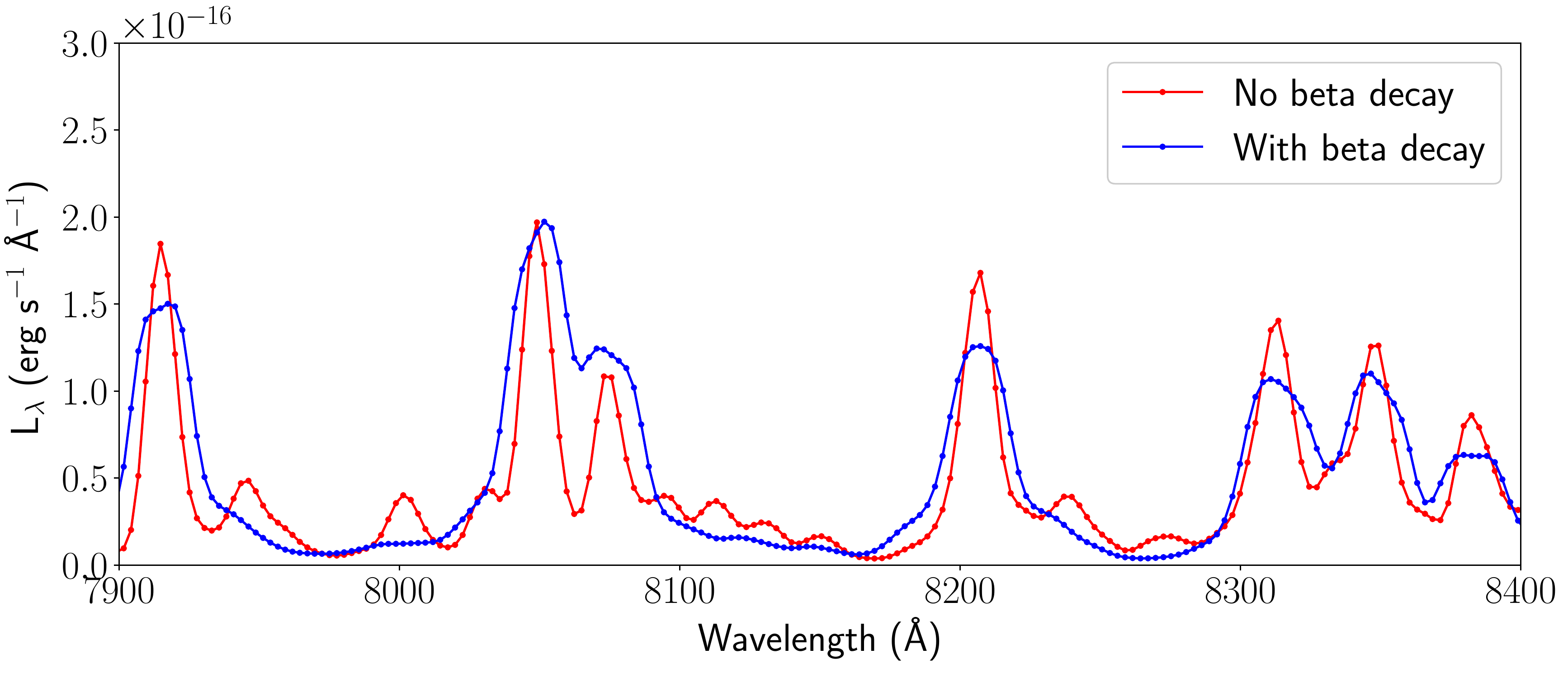}\\ 
\caption{\arj{Zoom-in of Figure \ref{fig:comp4} on the 7900-8500 \AA~region. Here a telescope convolution of 300 \kms~ has also been added.}}
\label{fig:comp4b}
%Gaussian smoothing (checked process.f90). In paper Im stating FWHM=1000 km/s. 
%
\end{figure}

% Line profiles
\arja{Figure \ref{fig:comp4b} shows a zoom-in on the 7900-8500 \AA~region, where also convolution to 300 \kms~has been applied to simulate an actual observation. The difference in width of Fe lines can be seen also after this convolution}.

\arja{Regarding line luminosities, the beta decay model has in particular increased the brightness in H$\alpha$, and reduced the brightness in [C I] 8727 (factor 2). The reduced column density of the He core leads to more gamma rays escaping into the H envelope, producing the boosted H$\alpha$ effect. At 300d the deposition in the 4 main zones is 35-35-25-5\% in the model without decay, and 20-15-40-30\% in the model with decay. The marginally optically thin Ni and O shells show a factor $\sim$ 2 drop in absorption (expected as $V^{-2}$ in the optically thin limit, and here their $V$ increase by factor $\sim$1.5), and this energy instead goes into the He shell and optically thick H envelope. In that sense, the decay mimics the effect of macroscopic mixing of H into the He core}.

\arja{However, not all lines show a simple behaviour. The Fe I lines, and also [O I] 6300, 6364 remain mostly unchanged, despite a factor 2 lower energy deposition in the core zones. This illustrates the complexity of the spectral formation, where radiative cross talk, optical depth effects, and changing ionization balance all can play a role. The extreme compression of the O/He shells in the beta decay model also leads to some compositional mixing}.

\section{Sensitivity to zoning/velocity resolution} % -------------------------------------------------------------------------------------------
\arj{The composition of the core changes on relatively small velocity scales, of order 10 \kms\ in the O-layer (Fig \ref{fig:comp}). As the emission spectrum can change quite dramatically depending on details of composition, it is important to ensure that the simulations are converged with regard to velocity resolution (zoning)}.

\begin{figure*}
\includegraphics[width=1\linewidth]{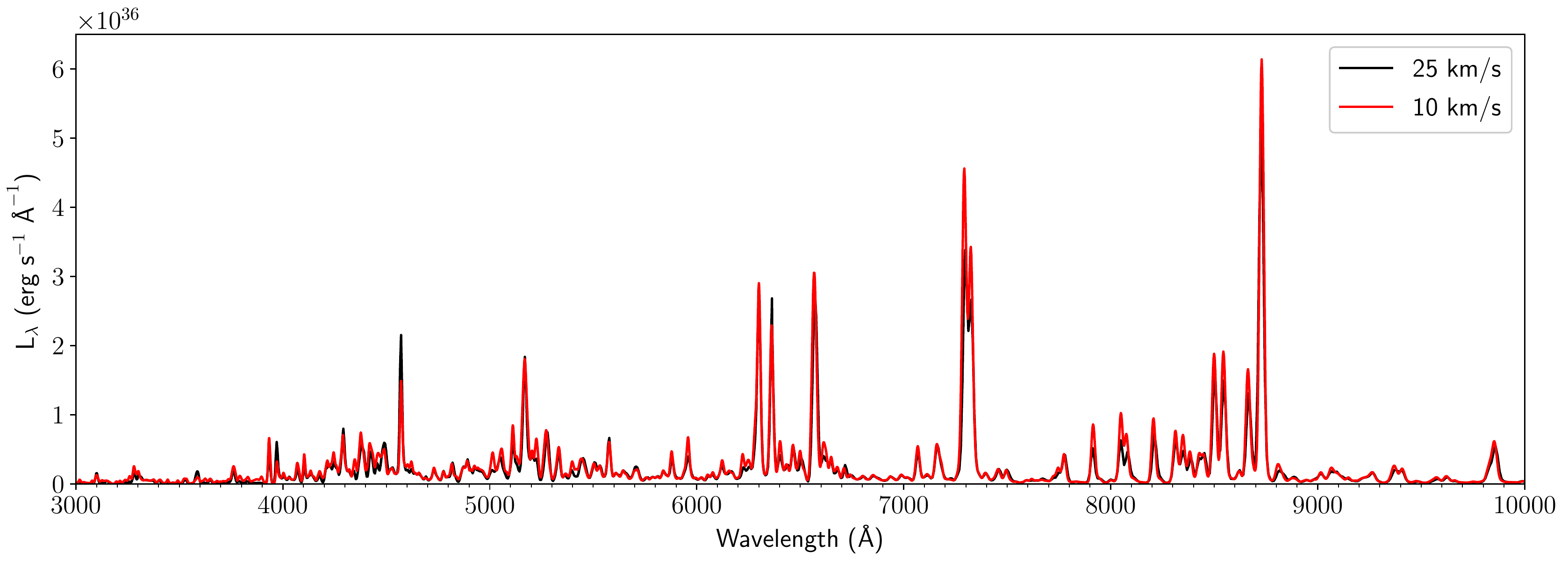} 
\caption{\arj{Comparison between using 10 (red) and 25 (black) \kms\ zoning (9 \msun\ models at 400d, exluding $^{56}$Ni dynamics). A 25 \kms~zoning is fully converged.}}
\label{fig:spec1}

\end{figure*}

\arj{Figure \ref{fig:spec1} compares spectra computed with 10 and 25 \kms~in the core (models with no beta decay). There are almost no differences, which demonstrates that zoning is sufficient}. %Only one line shows distinct sensitivty to this; [C I] 8727, which is significantly stronger in the finely zoned model.

\section{Sensitivity to frequency resolution}
\begin{figure*}
\includegraphics[width=1\linewidth]{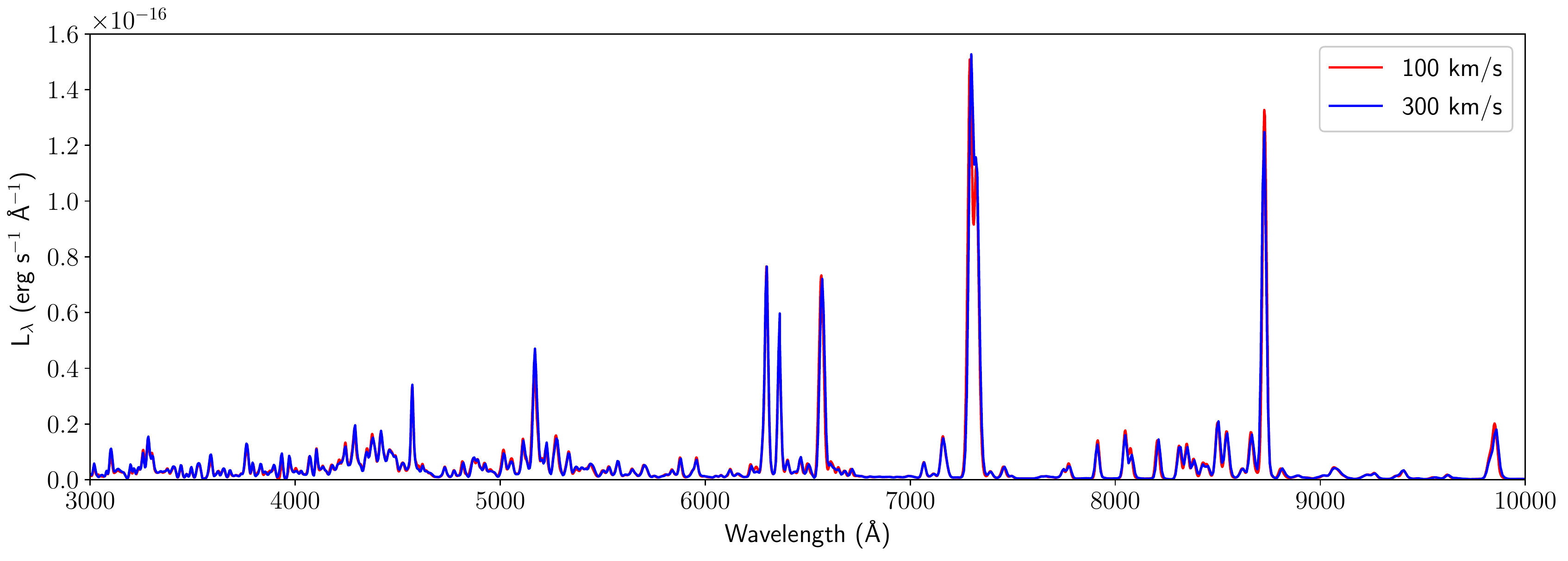} 
\caption{\arj{Sensitivity test to frequency resolution ((Old) 9 \msun\ model at 500d). A 300 \kms~resolution is fully converged.}}
\label{fig:sensfreq}
\end{figure*}

\arj{Figure \ref{fig:sensfreq} shows a comparison of models using $R (\mathbf{=\lambda/\Delta \lambda}) =1000$ (300 \kms\ bins) and $R=3000$ (100 \kms), for the 9 model at 500d. No discernible differences are obtained for either line luminosities or spectral form.
This shows that unless one needs very specific line profiles, $R=1000$ is sufficient for the modelling. Our standard resolution is $R=3000$.}
% 016_500 vs 015_500

\section{Atomic data update}
\arj{We have updated atomic data for Mg I with the new collision strength values from \citet{Barklem2017}. The old values used were from \citet{Mauas1988}}.

K I has been added to the code as a 3-level atom (with the purpose to be able to model the K I 7665,7699 ground state scattering line). Level energies and A-values from \citet{Wang1997} and \citet{Falke2006}.

\section{Relation to J12/J14 models}
\arja{\citet{J12} and \citet{Jerkstrand2014} computed Type IIP models for the 12-25 \msun~progenitor range. These models differed not only in what $M_{ZAMS}$ was used, but also in the exlosion energy (1.2 B), nucleosynthesis (larger network employed), explosion method (piston), artificial mixing, and parameterized molecule cooling was employed. It is nevertheless interesting to investigate the relation between the new models and the ones in J12/J14}.

\begin{figure*}
\includegraphics[width=1\linewidth]{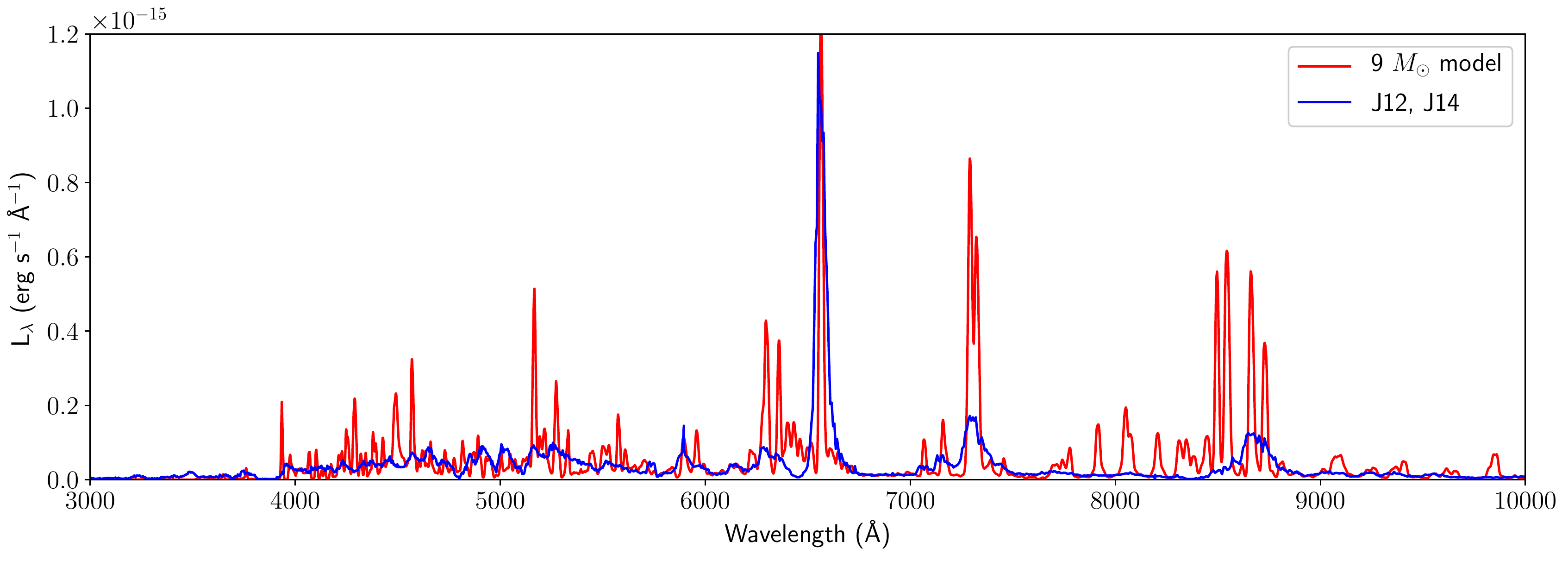} 
\caption{\arj{Comparison of 9 model  (red) with the models in J12/J14 (blue), at 300d.}}% In the bottom panel the 9 model has been smoothed to simulate higher velocities.}}
\label{fig:compJ14}
\end{figure*}

\arja{Figure \ref{fig:compJ14} compares the 9 \msun\ model computed here with the 12 \msun\ model in J14 at 300d. The spectra are rescaled with the difference in \ni~mass, and the J14 model has also been compensated at the 10\% level for its dust content. The spectra have relatively similar colors. The obvious difference is the much narrower lines in the 9 \msun\ model. For line luminosities, the [O I] + Fe I blend is stronger in the 9 \msun\ model than in the 12 \msun\ model, despite a higher O mass in the 12 \msun\ model. This effect presumably arises as the O shell in the 9 \msun\ model lies closer to the \ni\ in the unmixed setup, the 9 \msun\ model lacks molecular cooling, and there is also significant Fe I contribution to the 6364 line (but 6300 is mostly [O I])}.

\arja{The most distinct qualitative differences are the Fe I lines in the 7900-8500 \AA~region; these are not produced in the J12/J14 models. In those models, the strong mixing and clumping (treated in 1D by the method of 'virtual gridding', see \citet{J11}) of the core gives only 7\% gamma deposition in the \ni~clumps at 300d, compared to 20\% in the unmixed 9 \msun\ model. Lower velocities combine with higher energy deposition in the Fe zone to make more distinct Fe lines (happens also for e.g. [Fe II] 7155). For the O zone, the J12/J14 models use parameterized molecular cooling in the O/Si/S and O/C zones, which removes about 2/3 of the O shell optical emission, effectively giving a lower power level for [O I] than in the 9 model. As stated before, in the optically thick regime the morphology, e.g. structure and proximity to the \ni, is as important as the mass of a zone in determining its line luminosities. Only for ``all else constant'' does a higher zone mass lead to higher emission. In this comparison both zone masses and morphology are different, and the morphological differences often dominate}.
% J14 : 10% deposition in O zone..9 model similar

\arj{The lack of many trace metals in the new models may also influence the spectra, in particular the blue regions, where reduced line blocking may overestimate the blue emergent flux}.

\section{Observational comparison between subluminous and normal SNe}
\begin{figure*}
\includegraphics[width=1\linewidth]{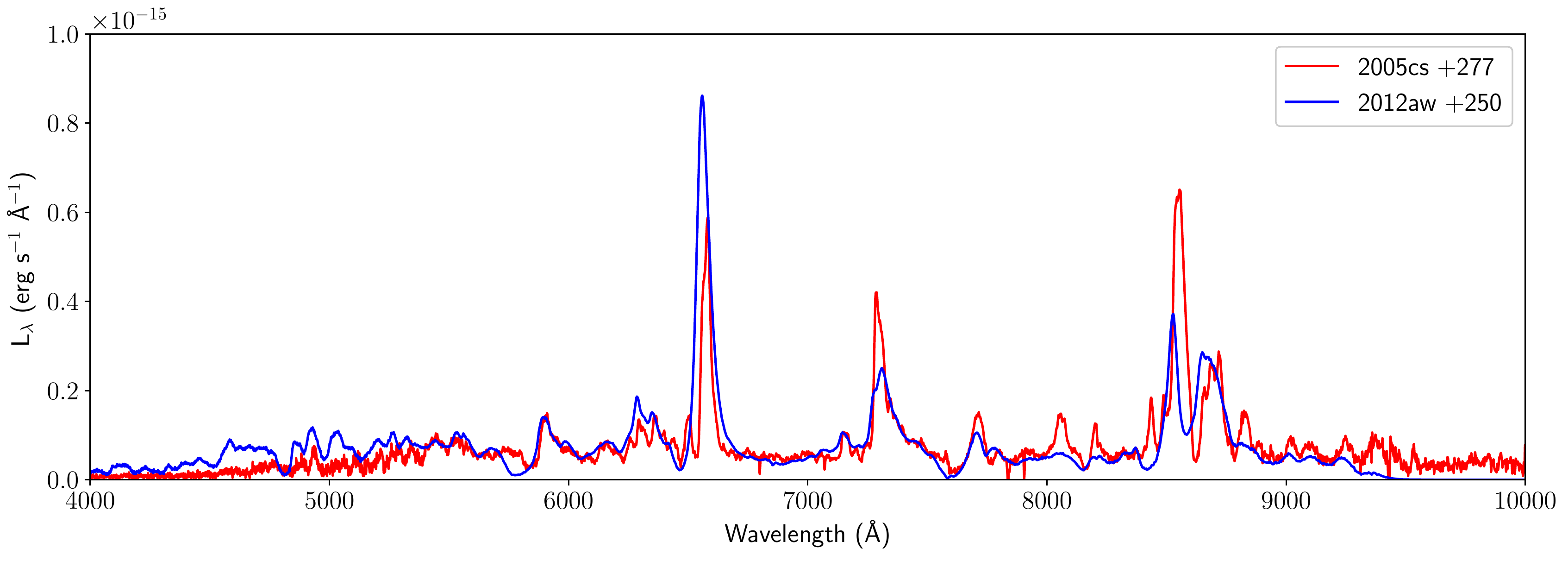} 
\caption{\arj{SN 2005cs (red) compared to SN 2012aw (blue), at about 260d. SN 2005cs has lower velocities, and more pronounced Fe I lines at 7900-8500 \AA. Note the similarity in the red wing of [Ca II].}}
\label{fig:Y}
% DC 6667 region is clean from atm bands (Matheson). 7400 region has 10-20% disturbances. The only problematic regions are 7600-7700 A and 9000-9500 A.
\end{figure*}

\arja{Figure \ref{fig:Y} compares SN 2005cs at +277d with ``normal'' IIP SN 2012aw at +250d. The main qualitative differences are the clear Fe I lines in the 8000-9500 \AA~window in SN 2005cs. There are hints of bumps at all of these also in SN 2012aw, but some combination of higher velocities and lower intrinsic brightness makes them less distinct. The higher velocities on SN 2012aw are made clear from the absorption troughs of Na I D, K I 7699, Ca II 8498 (which eats up O I 8446), and narrower emission profiles in e.g. H$\alpha$. It is clear that the velocity of the emitting region at nebular times is a factor 1.5-2 smaller in SN 2005cs than in SN 2012aw. This must invariably be linked to the (bulk of) the \ni~distribution being at a factor $\sim$ 2 lower velocities}.

%\arj{Figure \ref{fig:Z} compares SN 2005cs to SN 2008bk at +280d. The spectra are largely similar, although}
\arja{Although we do not have access to the data, we can from Fig. 13 in \citet{Lisakov2017} comment on the similarity between SN 2005cs and SN 2008bk at +280d. The spectra are largely similar, although}

\begin{itemize}
\item \arja{SN 2008bk shows a clear He I 7065, but a weaker [C I] 8727.}
\item \arja{[Ca II] 7300 is doubly peaked in SN 2008bk, and lacks the red tail of SN 2005cs}.
\item \arja{SN 2008bk shows a larger number of matching Fe I lines, including 7900 \AA, 8400 \AA. This may relate to its factor
2 higher \ni~mass.}
\end{itemize}

\arja{All in all this comparison makes it clear that it would be difficult to argue that SN 2005cs and SN 2008bk would have
any fundamental differences. But that argument also implies that our model, for some reason, is not giving very accurate output for this epoch. We note that the best matches for SN 1997D and SN 2008bk were for +350, +484 and +513d. The +207d spectrum of SN 1997D fits well from H$\alpha$ and redwards, but there are discrepancies in the blue. Thus, there may be some issues in the model that makes the blue region inaccurate at early epochs $t \lesssim 300$d.}

\begin{figure*}
\includegraphics[width=1\linewidth]{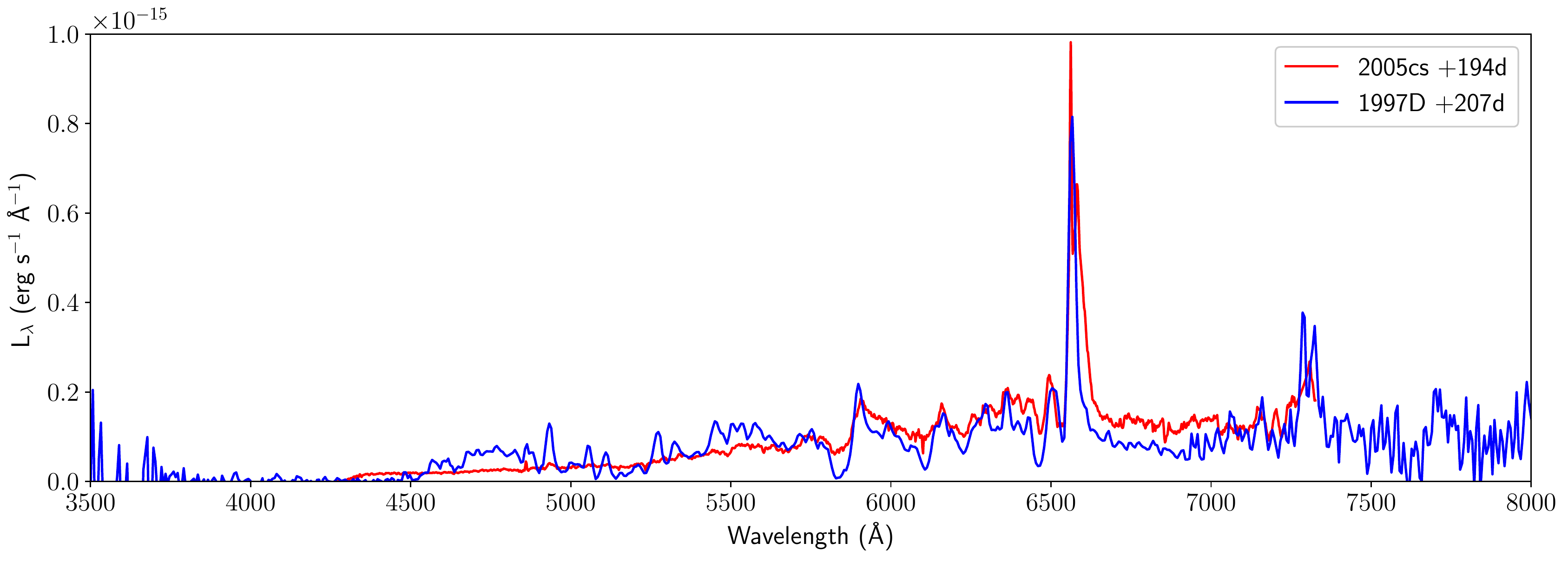} 
\caption{\arj{SN 1997D (blue) compared to SN 2005cs (red), at around 200d. }}
\label{fig:97D_05cs}
% DC 6667 region is clean from atm bands (Matheson). 7400 region has 10-20% disturbances. The only problematic regions are 7600-7700 A and 9000-9500 A.
\end{figure*}

\arja{Finally, Fig. \ref{fig:97D_05cs} compares SN 1997D and SN 2005cs at around 200d. The extreme lack of line structure in SN 2005cs at blue wavelengths is highlighted}.

\section*{Acknowledgments}
We thank the anonymous referee for careful reading and useful comments and suggestions.
We thank A. Pastorello, L. Siess, K. Nomoto, B. M\"uller and L. Tomasella for useful discussions. AJ acknowledges funding by the European Union's Framework Programme for Research and Innovation Horizon 2020 under Marie Sklodowska-Curie grant agreement No 702538.

\bibliographystyle{mn2e3}
\bibliography{bibl}

\begin{thebibliography}{102}
\expandafter\ifx\csname natexlab\endcsname\relax\def\natexlab#1{#1}\fi

\bibitem[{{Barklem} {et~al}\mbox{.}(2017){Barklem}, {Osorio}, {Fursa}, {Bray},
  {Zatsarinny}, {Bartschat}, \& {Jerkstrand}}]{Barklem2017}
{Barklem} P.~S., {Osorio} Y., {Fursa} D.~V., {Bray} I., {Zatsarinny} O.,
  {Bartschat} K., {Jerkstrand} A., 2017, ArXiv e-prints

\bibitem[{{Baron}, {Branch} \& {Hauschildt}(2007){Baron}, {Branch}, \&
  {Hauschildt}}]{Baron2007}
{Baron} E., {Branch} D., {Hauschildt} P.~H., 2007, \apj, 662, 1148

\bibitem[{{Benetti} {et~al}\mbox{.}(2001){Benetti}, {Turatto}, {Balberg},
  {Zampieri}, {Shapiro}, {Cappellaro}, {Nomoto}, {Nakamura}, {Mazzali}, \&
  {Patat}}]{Benetti2001}
{Benetti} S. {et~al.}, 2001, \mnras, 322, 361

\bibitem[{{Botticella} {et~al}\mbox{.}(2009){Botticella}, {Pastorello},
  {Smartt}, {Meikle}, {Benetti}, {Kotak}, {Cappellaro}, {Crockett}, {Mattila},
  {Sereno}, {Patat}, {Tsvetkov}, {van Loon}, {Abraham}, {Agnoletto}, {Arbour},
  {Benn}, {di Rico}, {Elias-Rosa}, {Gorshanov}, {Harutyunyan}, {Hunter},
  {Lorenzi}, {Keenan}, {Maguire}, {Mendez}, {Mobberley}, {Navasardyan}, {Ries},
  {Stanishev}, {Taubenberger}, {Trundle}, {Turatto}, \&
  {Volkov}}]{Botticella2009}
{Botticella} M.~T. {et~al.}, 2009, \mnras, 398, 1041

\bibitem[{{Brown} {et~al}\mbox{.}(2007){Brown}, {Dessart}, {Holland}, {Immler},
  {Landsman}, {Blondin}, {Blustin}, {Breeveld}, {Dewangan}, {Gehrels},
  {Hutchins}, {Kirshner}, {Mason}, {Mazzali}, {Milne}, {Modjaz}, \&
  {Roming}}]{Brown2007}
{Brown} P.~J. {et~al.}, 2007, \apj, 659, 1488

\bibitem[{{Chugai} \& {Utrobin}(2000)}]{Chugai2000}
{Chugai} N.~N., {Utrobin} V.~P., 2000, \aap, 354, 557

\bibitem[{{de Mello}, {Benetti} \& {Massone}(1997){de Mello}, {Benetti}, \&
  {Massone}}]{deMello1997}
{de Mello} D., {Benetti} S., {Massone} G., 1997, \iaucirc, 6537

\bibitem[{{Dessart} {et~al}\mbox{.}(2008){Dessart}, {Blondin}, {Brown},
  {Hicken}, {Hillier}, {Holland}, {Immler}, {Kirshner}, {Milne}, {Modjaz}, \&
  {Roming}}]{Dessart2008}
{Dessart} L. {et~al.}, 2008, \apj, 675, 644

\bibitem[{{Dessart} {et~al}\mbox{.}(2013){Dessart}, {Hillier}, {Waldman}, \&
  {Livne}}]{Dessart2013b}
{Dessart} L., {Hillier} D.~J., {Waldman} R., {Livne} E., 2013, \mnras, 433,
  1745

\bibitem[{{Doherty} {et~al}\mbox{.}(2015){Doherty}, {Gil-Pons}, {Siess},
  {Lattanzio}, \& {Lau}}]{Doherty2015}
{Doherty} C.~L., {Gil-Pons} P., {Siess} L., {Lattanzio} J.~C., {Lau} H.~H.~B.,
  2015, \mnras, 446, 2599

\bibitem[{{Eldridge}, {Mattila} \& {Smartt}(2007){Eldridge}, {Mattila}, \&
  {Smartt}}]{Eldridge2007}
{Eldridge} J.~J., {Mattila} S., {Smartt} S.~J., 2007, \mnras, 376, L52

\bibitem[{{Eldridge} \& {Tout}(2004)}]{Eldridge2004}
{Eldridge} J.~J., {Tout} C.~A., 2004, \mnras, 353, 87

\bibitem[{{Falke} {et~al}\mbox{.}(2006){Falke}, {Tiemann}, {Lisdat}, {Schnatz},
  \& {Grosche}}]{Falke2006}
{Falke} S., {Tiemann} E., {Lisdat} C., {Schnatz} H., {Grosche} G., 2006, \pra,
  74, 032503

\bibitem[{{Fransson} \& {Chevalier}(1989)}]{Fransson1989}
{Fransson} C., {Chevalier} R.~A., 1989, \apj, 343, 323

\bibitem[{{Fraser} {et~al}\mbox{.}(2011){Fraser}, {Ergon}, {Eldridge},
  {Valenti}, {Pastorello}, {Sollerman}, {Smartt}, {Agnoletto}, {Arcavi},
  {Benetti}, {Botticella}, {Bufano}, {Campillay}, {Crockett}, {Gal-Yam},
  {Kankare}, {Leloudas}, {Maguire}, {Mattila}, {Maund}, {Salgado}, {Stephens},
  {Taubenberger}, \& {Turatto}}]{Fraser2011}
{Fraser} M. {et~al.}, 2011, \mnras, 417, 1417

\bibitem[{{Garc{\'{\i}}a-Berro}, {Ritossa} \&
  {Iben}(1997){Garc{\'{\i}}a-Berro}, {Ritossa}, \& {Iben}}]{GarciaBerro1997}
{Garc{\'{\i}}a-Berro} E., {Ritossa} C., {Iben}, Jr. I., 1997, \apj, 485, 765

\bibitem[{{Hammer}, {Janka} \& {M{\"u}ller}(2010){Hammer}, {Janka}, \&
  {M{\"u}ller}}]{Hammer2010}
{Hammer} N.~J., {Janka} H.-T., {M{\"u}ller} E., 2010, \apj, 714, 1371

\bibitem[{{Hendry} {et~al}\mbox{.}(2005){Hendry}, {Smartt}, {Maund},
  {Pastorello}, {Zampieri}, {Benetti}, {Turatto}, {Cappellaro}, {Meikle},
  {Kotak}, {Irwin}, {Jonker}, {Vermaas}, {Peletier}, {van Woerden}, {Exter},
  {Pollacco}, {Leon}, {Verley}, {Benn}, \& {Pignata}}]{Hendry2005}
{Hendry} M.~A. {et~al.}, 2005, \mnras, 359, 906

\bibitem[{{Herant} \& {Benz}(1992)}]{Herant1992}
{Herant} M., {Benz} W., 1992, \apj, 387, 294

\bibitem[{{Iben}, {Ritossa} \& {Garc{\'{\i}}a-Berro}(1997){Iben}, {Ritossa}, \&
  {Garc{\'{\i}}a-Berro}}]{Iben1997}
{Iben}, Jr. I., {Ritossa} C., {Garc{\'{\i}}a-Berro} E., 1997, \apj, 489, 772

\bibitem[{{Janka} {et~al}\mbox{.}(2012){Janka}, {Hanke}, {H{\"u}depohl},
  {Marek}, {M{\"u}ller}, \& {Obergaulinger}}]{Janka2012}
{Janka} H.-T., {Hanke} F., {H{\"u}depohl} L., {Marek} A., {M{\"u}ller} B.,
  {Obergaulinger} M., 2012, Progress of Theoretical and Experimental Physics,
  2012, 01A309

\bibitem[{{Janka} \& {Mueller}(1996)}]{Janka1996}
{Janka} H.-T., {Mueller} E., 1996, \aap, 306, 167

\bibitem[{{Janka} {et~al}\mbox{.}(2008){Janka}, {M{\"u}ller}, {Kitaura}, \&
  {Buras}}]{Janka2008}
{Janka} H.-T., {M{\"u}ller} B., {Kitaura} F.~S., {Buras} R., 2008, \aap, 485,
  199

\bibitem[{{Jerkstrand} {et~al}\mbox{.}(2015{\natexlab{a}}){Jerkstrand},
  {Ergon}, {Smartt}, {Fransson}, {Sollerman}, {Taubenberger}, {Bersten}, \&
  {Spyromilio}}]{J15a}
{Jerkstrand} A., {Ergon} M., {Smartt} S.~J., {Fransson} C., {Sollerman} J.,
  {Taubenberger} S., {Bersten} M., {Spyromilio} J., 2015{\natexlab{a}}, \aap,
  573, A12, "J15a"

\bibitem[{{Jerkstrand}, {Fransson} \& {Kozma}(2011){Jerkstrand}, {Fransson}, \&
  {Kozma}}]{J11}
{Jerkstrand} A., {Fransson} C., {Kozma} C., 2011, \aap, 530, A45

\bibitem[{{Jerkstrand} {et~al}\mbox{.}(2012){Jerkstrand}, {Fransson},
  {Maguire}, {Smartt}, {Ergon}, \& {Spyromilio}}]{J12}
{Jerkstrand} A., {Fransson} C., {Maguire} K., {Smartt} S., {Ergon} M.,
  {Spyromilio} J., 2012, \aap, 546, A28

\bibitem[{{Jerkstrand} {et~al}\mbox{.}(2014){Jerkstrand}, {Smartt}, {Fraser},
  {Fransson}, {Sollerman}, {Taddia}, \& {Kotak}}]{Jerkstrand2014}
{Jerkstrand} A., {Smartt} S.~J., {Fraser} M., {Fransson} C., {Sollerman} J.,
  {Taddia} F., {Kotak} R., 2014, \mnras, 439, 3694

\bibitem[{{Jerkstrand} {et~al}\mbox{.}(2015{\natexlab{b}}){Jerkstrand},
  {Smartt}, {Sollerman}, {Inserra}, {Fraser}, {Spyromilio}, {Fransson}, {Chen},
  {Barbarino}, {Dall'Ora}, {Botticella}, {Della Valle}, {Gal-Yam}, {Valenti},
  {Maguire}, {Mazzali}, \& {Tomasella}}]{Jerkstrand2015b}
{Jerkstrand} A. {et~al.}, 2015{\natexlab{b}}, \mnras, 448, 2482

\bibitem[{{Jerkstrand} {et~al}\mbox{.}(2015{\natexlab{c}}){Jerkstrand},
  {Timmes}, {Magkotsios}, {Sim}, {Fransson}, {Spyromilio}, {M{\"u}ller},
  {Heger}, {Sollerman}, \& {Smartt}}]{Jerkstrand2015c}
{Jerkstrand} A. {et~al.}, 2015{\natexlab{c}}, \apj, 807, 110

\bibitem[{{Jones}, {Hirschi} \& {Nomoto}(2014){Jones}, {Hirschi}, \&
  {Nomoto}}]{Jones2014}
{Jones} S., {Hirschi} R., {Nomoto} K., 2014, \apj, 797, 83

\bibitem[{{Jones} {et~al}\mbox{.}(2013){Jones}, {Hirschi}, {Nomoto}, {Fischer},
  {Timmes}, {Herwig}, {Paxton}, {Toki}, {Suzuki}, {Mart{\'{\i}}nez-Pinedo},
  {Lam}, \& {Bertolli}}]{Jones2013}
{Jones} S. {et~al.}, 2013, \apj, 772, 150

\bibitem[{{Kitaura}, {Janka} \& {Hillebrandt}(2006){Kitaura}, {Janka}, \&
  {Hillebrandt}}]{Kitaura2006}
{Kitaura} F.~S., {Janka} H.-T., {Hillebrandt} W., 2006, \aap, 450, 345

\bibitem[{{Kloehr} {et~al}\mbox{.}(2005){Kloehr}, {Muendlein}, {Li}, {Yamaoka},
  \& {Itagaki}}]{Kloehr2005}
{Kloehr} W., {Muendlein} R., {Li} W., {Yamaoka} H., {Itagaki} K., 2005,
  \iaucirc, 8553

\bibitem[{{Kozma} \& {Fransson}(1992)}]{Kozma1992}
{Kozma} C., {Fransson} C., 1992, \apj, 390, 602

\bibitem[{{Kozma} \& {Fransson}(1998{\natexlab{a}})}]{Kozma1998I}
{Kozma} C., {Fransson} C., 1998{\natexlab{a}}, \apj, 496, 946

\bibitem[{{Kozma} \& {Fransson}(1998{\natexlab{b}})}]{Kozma1998II}
{Kozma} C., {Fransson} C., 1998{\natexlab{b}}, \apj, 497, 431

\bibitem[{{Lepp}, {Dalgarno} \& {McCray}(1990){Lepp}, {Dalgarno}, \&
  {McCray}}]{Lepp1990}
{Lepp} S., {Dalgarno} A., {McCray} R., 1990, \apj, 358, 262

\bibitem[{{Li} {et~al}\mbox{.}(2006){Li}, {Van Dyk}, {Filippenko},
  {Cuillandre}, {Jha}, {Bloom}, {Riess}, \& {Livio}}]{Li2006}
{Li} W., {Van Dyk} S.~D., {Filippenko} A.~V., {Cuillandre} J.-C., {Jha} S.,
  {Bloom} J.~S., {Riess} A.~G., {Livio} M., 2006, \apj, 641, 1060

\bibitem[{{Lisakov} {et~al}\mbox{.}(2017){Lisakov}, {Dessart}, {Hillier},
  {Waldman}, \& {Livne}}]{Lisakov2017}
{Lisakov} S.~M., {Dessart} L., {Hillier} D.~J., {Waldman} R., {Livne} E., 2017,
  \mnras, 466, 34

\bibitem[{{Lodders}(2003)}]{Lodders2003}
{Lodders} K., 2003, \apj, 591, 1220

\bibitem[{{Lucy} {et~al}\mbox{.}(1989){Lucy}, {Danziger}, {Gouiffes}, \&
  {Bouchet}}]{Lucy1989}
{Lucy} L.~B., {Danziger} I.~J., {Gouiffes} C., {Bouchet} P., 1989, in Lecture
  Notes in Physics, Berlin Springer Verlag, Vol. 350, IAU Colloq. 120:
  Structure and Dynamics of the Interstellar Medium, {Tenorio-Tagle} G.,
  {Moles} M., {Melnick} J., eds., p. 164

\bibitem[{{Maguire} {et~al}\mbox{.}(2012){Maguire}, {Jerkstrand}, {Smartt},
  {Fransson}, {Pastorello}, {Benetti}, {Valenti}, {Bufano}, \&
  {Leloudas}}]{Maguire2012}
{Maguire} K. {et~al.}, 2012, \mnras, 420, 3451

\bibitem[{{Mattila} {et~al}\mbox{.}(2008){Mattila}, {Smartt}, {Eldridge},
  {Maund}, {Crockett}, \& {Danziger}}]{Mattila2008}
{Mattila} S., {Smartt} S.~J., {Eldridge} J.~J., {Maund} J.~R., {Crockett}
  R.~M., {Danziger} I.~J., 2008, \apjl, 688, L91

\bibitem[{{Mauas}, {Avrett} \& {Loeser}(1988){Mauas}, {Avrett}, \&
  {Loeser}}]{Mauas1988}
{Mauas} P.~J., {Avrett} E.~H., {Loeser} R., 1988, \apj, 330, 1008

\bibitem[{{Maund} {et~al}\mbox{.}(2015){Maund}, {Fraser}, {Reilly}, {Ergon}, \&
  {Mattila}}]{Maund2015}
{Maund} J.~R., {Fraser} M., {Reilly} E., {Ergon} M., {Mattila} S., 2015,
  \mnras, 447, 3207

\bibitem[{{Maund} {et~al}\mbox{.}(2014){Maund}, {Mattila}, {Ramirez-Ruiz}, \&
  {Eldridge}}]{Maund2014}
{Maund} J.~R., {Mattila} S., {Ramirez-Ruiz} E., {Eldridge} J.~J., 2014, \mnras,
  438, 1577

\bibitem[{{Maund} \& {Smartt}(2009)}]{Maund2009}
{Maund} J.~R., {Smartt} S.~J., 2009, Science, 324, 486

\bibitem[{{Maund}, {Smartt} \& {Danziger}(2005){Maund}, {Smartt}, \&
  {Danziger}}]{Maund2005}
{Maund} J.~R., {Smartt} S.~J., {Danziger} I.~J., 2005, \mnras, 364, L33

\bibitem[{{Mayle} \& {Wilson}(1988)}]{Mayle1988}
{Mayle} R., {Wilson} J.~R., 1988, \apj, 334, 909

\bibitem[{{Meikle} {et~al}\mbox{.}(2007){Meikle}, {Mattila}, {Pastorello},
  {Gerardy}, {Kotak}, {Sollerman}, {Van Dyk}, {Farrah}, {Filippenko},
  {H{\"o}flich}, {Lundqvist}, {Pozzo}, \& {Wheeler}}]{Meikle2007}
{Meikle} W.~P.~S. {et~al.}, 2007, \apj, 665, 608

\bibitem[{{Melson}, {Janka} \& {Marek}(2015){Melson}, {Janka}, \&
  {Marek}}]{Melson2015}
{Melson} T., {Janka} H.-T., {Marek} A., 2015, \apjl, 801, L24

\bibitem[{{Miyaji} {et~al}\mbox{.}(1980){Miyaji}, {Nomoto}, {Yokoi}, \&
  {Sugimoto}}]{Miyaji1980}
{Miyaji} S., {Nomoto} K., {Yokoi} K., {Sugimoto} D., 1980, \pasj, 32, 303

\bibitem[{{Monard}(2008)}]{Monard2008}
{Monard} L.~A.~G., 2008, Central Bureau Electronic Telegrams, 1315

\bibitem[{{Moriya} \& {Eldridge}(2016)}]{Moriya2016}
{Moriya} T.~J., {Eldridge} J.~J., 2016, \mnras, 461, 2155

\bibitem[{{Moriya} {et~al}\mbox{.}(2014){Moriya}, {Tominaga}, {Langer},
  {Nomoto}, {Blinnikov}, \& {Sorokina}}]{Moriya2014}
{Moriya} T.~J., {Tominaga} N., {Langer} N., {Nomoto} K., {Blinnikov} S.~I.,
  {Sorokina} E.~I., 2014, \aap, 569, A57

\bibitem[{{M{\"u}ller}(2016)}]{Muller2016}
{M{\"u}ller} B., 2016, \pasa, 33, e048

\bibitem[{{M{\"u}ller}, {Janka} \& {Heger}(2012){M{\"u}ller}, {Janka}, \&
  {Heger}}]{Muller2012}
{M{\"u}ller} B., {Janka} H.-T., {Heger} A., 2012, \apj, 761, 72

\bibitem[{{M{\"u}ller}, {Janka} \& {Marek}(2013){M{\"u}ller}, {Janka}, \&
  {Marek}}]{Muller2013}
{M{\"u}ller} B., {Janka} H.-T., {Marek} A., 2013, \apj, 766, 43

\bibitem[{{Nomoto}(1984)}]{Nomoto1984}
{Nomoto} K., 1984, \apj, 277, 791

\bibitem[{{Nomoto}(1987)}]{Nomoto1987}
{Nomoto} K., 1987, \apj, 322, 206

\bibitem[{{Nomoto}, {Kobayashi} \& {Tominaga}(2013){Nomoto}, {Kobayashi}, \&
  {Tominaga}}]{Nomoto2013}
{Nomoto} K., {Kobayashi} C., {Tominaga} N., 2013, \araa, 51, 457

\bibitem[{{Pastorello} {et~al}\mbox{.}(2006){Pastorello}, {Sauer},
  {Taubenberger}, {Mazzali}, {Nomoto}, {Kawabata}, {Benetti}, {Elias-Rosa},
  {Harutyunyan}, {Navasardyan}, {Zampieri}, {Iijima}, {Botticella}, {di Rico},
  {Del Principe}, {Dolci}, {Gagliardi}, {Ragni}, \&
  {Valentini}}]{Pastorello2006}
{Pastorello} A. {et~al.}, 2006, \mnras, 370, 1752

\bibitem[{{Pastorello} {et~al}\mbox{.}(2009){Pastorello}, {Valenti},
  {Zampieri}, {Navasardyan}, {Taubenberger}, {Smartt}, {Arkharov},
  {B{\"a}rnbantner}, {Barwig}, {Benetti}, {Birtwhistle}, {Botticella},
  {Cappellaro}, {Del Principe}, {di Mille}, {di Rico}, {Dolci}, {Elias-Rosa},
  {Efimova}, {Fiedler}, {Harutyunyan}, {H{\"o}flich}, {Kloehr}, {Larionov},
  {Lorenzi}, {Maund}, {Napoleone}, {Ragni}, {Richmond}, {Ries}, {Spiro},
  {Temporin}, {Turatto}, \& {Wheeler}}]{Pastorello2009}
{Pastorello} A. {et~al.}, 2009, \mnras, 394, 2266

\bibitem[{{Pastorello} {et~al}\mbox{.}(2004){Pastorello}, {Zampieri},
  {Turatto}, {Cappellaro}, {Meikle}, {Benetti}, {Branch}, {Baron}, {Patat},
  {Armstrong}, {Altavilla}, {Salvo}, \& {Riello}}]{Pastorello2004}
{Pastorello} A. {et~al.}, 2004, \mnras, 347, 74

\bibitem[{{Pejcha} \& {Thompson}(2015)}]{Pejcha2015}
{Pejcha} O., {Thompson} T.~A., 2015, \apj, 801, 90

\bibitem[{{Pignata}(2013)}]{Pignata2013}
{Pignata} G., 2013, in Massive Stars: From alpha to Omega, p. 176

\bibitem[{{Pllumbi} {et~al}\mbox{.}(2015){Pllumbi}, {Tamborra}, {Wanajo},
  {Janka}, \& {H{\"u}depohl}}]{Pllumbi2015}
{Pllumbi} E., {Tamborra} I., {Wanajo} S., {Janka} H.-T., {H{\"u}depohl} L.,
  2015, \apj, 808, 188

\bibitem[{{Poelarends}(2007)}]{Poelarends2007}
{Poelarends} A.~J.~T., 2007, PhD thesis, Utrecht University

\bibitem[{{Poelarends} {et~al}\mbox{.}(2008){Poelarends}, {Herwig}, {Langer},
  \& {Heger}}]{Poelarends2008}
{Poelarends} A.~J.~T., {Herwig} F., {Langer} N., {Heger} A., 2008, \apj, 675,
  614

\bibitem[{{Pumo} {et~al}\mbox{.}(2017){Pumo}, {Zampieri}, {Spiro},
  {Pastorello}, {Benetti}, {Cappellaro}, {Manic{\`o}}, \& {Turatto}}]{Pumo2017}
{Pumo} M.~L., {Zampieri} L., {Spiro} S., {Pastorello} A., {Benetti} S.,
  {Cappellaro} E., {Manic{\`o}} G., {Turatto} M., 2017, \mnras, 464, 3013

\bibitem[{{Radice} {et~al}\mbox{.}(2017){Radice}, {Burrows}, {Vartanyan},
  {Skinner}, \& {Dolence}}]{Radice2017}
{Radice} D., {Burrows} A., {Vartanyan} D., {Skinner} M.~A., {Dolence} J.~C.,
  2017, ArXiv e-prints

\bibitem[{{Ritossa}, {Garcia-Berro} \& {Iben}(1996){Ritossa}, {Garcia-Berro},
  \& {Iben}}]{Ritossa1996}
{Ritossa} C., {Garcia-Berro} E., {Iben}, Jr. I., 1996, \apj, 460, 489

\bibitem[{{Ritossa}, {Garc{\'{\i}}a-Berro} \& {Iben}(1999){Ritossa},
  {Garc{\'{\i}}a-Berro}, \& {Iben}}]{Ritossa1999}
{Ritossa} C., {Garc{\'{\i}}a-Berro} E., {Iben}, Jr. I., 1999, \apj, 515, 381

\bibitem[{{Roy} {et~al}\mbox{.}(2011){Roy}, {Kumar}, {Benetti}, {Pastorello},
  {Yuan}, {Brown}, {Immler}, {Fatkhullin}, {Moskvitin}, {Maund}, {Akerlof},
  {Wheeler}, {Sokolov}, {Quimby}, {Bufano}, {Kumar}, {Misra}, {Pandey},
  {Elias-Rosa}, {Roming}, \& {Sagar}}]{Roy2011}
{Roy} R. {et~al.}, 2011, \apj, 736, 76

\bibitem[{{Siess}(2006)}]{Siess2006}
{Siess} L., 2006, \aap, 448, 717

\bibitem[{{Siess}(2007)}]{Siess2007}
{Siess} L., 2007, \aap, 476, 893

\bibitem[{{Spiro} {et~al}\mbox{.}(2014){Spiro}, {Pastorello}, {Pumo},
  {Zampieri}, {Turatto}, {Smartt}, {Benetti}, {Cappellaro}, {Valenti},
  {Agnoletto}, {Altavilla}, {Aoki}, {Brocato}, {Corsini}, {Di Cianno},
  {Elias-Rosa}, {Hamuy}, {Enya}, {Fiaschi}, {Folatelli}, {Desidera},
  {Harutyunyan}, {Howell}, {Kawka}, {Kobayashi}, {Leibundgut}, {Minezaki},
  {Navasardyan}, {Nomoto}, {Mattila}, {Pietrinferni}, {Pignata}, {Raimondo},
  {Salvo}, {Schmidt}, {Sollerman}, {Spyromilio}, {Taubenberger}, {Valentini},
  {Vennes}, \& {Yoshii}}]{Spiro2014}
{Spiro} S. {et~al.}, 2014, \mnras, 439, 2873

\bibitem[{{Sukhbold} {et~al}\mbox{.}(2016){Sukhbold}, {Ertl}, {Woosley},
  {Brown}, \& {Janka}}]{Sukhbold2016}
{Sukhbold} T., {Ertl} T., {Woosley} S.~E., {Brown} J.~M., {Janka} H.-T., 2016,
  \apj, 821, 38

\bibitem[{{Takahashi}, {Yoshida} \& {Umeda}(2013){Takahashi}, {Yoshida}, \&
  {Umeda}}]{Takahashi2013}
{Takahashi} K., {Yoshida} T., {Umeda} H., 2013, \apj, 771, 28

\bibitem[{{Tak{\'a}ts} \& {Vink{\'o}}(2006)}]{Takats2006}
{Tak{\'a}ts} K., {Vink{\'o}} J., 2006, \mnras, 372, 1735

\bibitem[{{Tauris}, {Langer} \& {Podsiadlowski}(2015){Tauris}, {Langer}, \&
  {Podsiadlowski}}]{Tauris2015}
{Tauris} T.~M., {Langer} N., {Podsiadlowski} P., 2015, \mnras, 451, 2123

\bibitem[{{Tominaga}, {Blinnikov} \& {Nomoto}(2013){Tominaga}, {Blinnikov}, \&
  {Nomoto}}]{Tominaga2013}
{Tominaga} N., {Blinnikov} S.~I., {Nomoto} K., 2013, \apjl, 771, L12

\bibitem[{{Tully}(1988)}]{Tully1988}
{Tully} R.~B., 1988, {Nearby galaxies catalog}

\bibitem[{{Turatto} {et~al}\mbox{.}(1998){Turatto}, {Mazzali}, {Young},
  {Nomoto}, {Iwamoto}, {Benetti}, {Cappellaro}, {Danziger}, {de Mello},
  {Phillips}, {Suntzeff}, {Clocchiatti}, {Piemonte}, {Leibundgut},
  {Covarrubias}, {Maza}, \& {Sollerman}}]{Turatto1998}
{Turatto} M. {et~al.}, 1998, \apjl, 498, L129

\bibitem[{{Umeda} {et~al}\mbox{.}(1999){Umeda}, {Nomoto}, {Yamaoka}, \&
  {Wanajo}}]{Umeda1999}
{Umeda} H., {Nomoto} K., {Yamaoka} H., {Wanajo} S., 1999, \apj, 513, 861

\bibitem[{{Utrobin} \& {Chugai}(2008)}]{Utrobin2008}
{Utrobin} V.~P., {Chugai} N.~N., 2008, \aap, 491, 507

\bibitem[{{Utrobin}, {Chugai} \& {Pastorello}(2007){Utrobin}, {Chugai}, \&
  {Pastorello}}]{Utrobin2007}
{Utrobin} V.~P., {Chugai} N.~N., {Pastorello} A., 2007, \aap, 475, 973

\bibitem[{{Van Dyk}(2013)}]{vanDyk2013}
{Van Dyk} S.~D., 2013, \aj, 146, 24

\bibitem[{{Van Dyk} {et~al}\mbox{.}(2012){Van Dyk}, {Davidge}, {Elias-Rosa},
  {Taubenberger}, {Li}, {Levesque}, {Howerton}, {Pignata}, {Morrell}, {Hamuy},
  \& {Filippenko}}]{vanDyk2012}
{Van Dyk} S.~D. {et~al.}, 2012, \aj, 143, 19

\bibitem[{{Van Dyk}, {Li} \& {Filippenko}(2003){Van Dyk}, {Li}, \&
  {Filippenko}}]{vanDyk2003}
{Van Dyk} S.~D., {Li} W., {Filippenko} A.~V., 2003, \pasp, 115, 1289

\bibitem[{{Wanajo}, {Janka} \& {M{\"u}ller}(2011){Wanajo}, {Janka}, \&
  {M{\"u}ller}}]{Wanajo2011}
{Wanajo} S., {Janka} H.-T., {M{\"u}ller} B., 2011, \apjl, 726, L15

\bibitem[{{Wanajo} {et~al}\mbox{.}(2017){Wanajo}, {M{\"u}ller}, {Janka}, \&
  {Heger}}]{Wanajo2017}
{Wanajo} S., {M{\"u}ller} B., {Janka} H.-T., {Heger} A., 2017, ArXiv e-prints

\bibitem[{{Wanajo} {et~al}\mbox{.}(2009){Wanajo}, {Nomoto}, {Janka}, {Kitaura},
  \& {M{\"u}ller}}]{Wanajo2009}
{Wanajo} S., {Nomoto} K., {Janka} H.-T., {Kitaura} F.~S., {M{\"u}ller} B.,
  2009, \apj, 695, 208

\bibitem[{{Wang} {et~al}\mbox{.}(1997){Wang}, {Li}, {Wang}, {Williams},
  {Gould}, \& {Stwalley}}]{Wang1997}
{Wang} H., {Li} J., {Wang} X.~T., {Williams} C.~J., {Gould} P.~L., {Stwalley}
  W.~C., 1997, \pra, 55, R1569

\bibitem[{{Weaver}, {Zimmerman} \& {Woosley}(1978){Weaver}, {Zimmerman}, \&
  {Woosley}}]{Weaver1978}
{Weaver} T.~A., {Zimmerman} G.~B., {Woosley} S.~E., 1978, \apj, 225, 1021

\bibitem[{{Wongwathanarat}, {M{\"u}ller} \& {Janka}(2015){Wongwathanarat},
  {M{\"u}ller}, \& {Janka}}]{Wongwathanarat2015}
{Wongwathanarat} A., {M{\"u}ller} E., {Janka} H.-T., 2015, \aap, 577, A48

\bibitem[{{Woosley} \& {Heger}(2007)}]{WH07}
{Woosley} S.~E., {Heger} A., 2007, \physrep, 442, 269

\bibitem[{{Woosley} \& {Heger}(2015)}]{Woosley2015}
{Woosley} S.~E., {Heger} A., 2015, \apj, 810, 34

\bibitem[{{Woosley} \& {Weaver}(1995)}]{Woosley1995}
{Woosley} S.~E., {Weaver} T.~A., 1995, \apjs, 101, 181

\bibitem[{{Yaron} \& {Gal-Yam}(2012)}]{Yaron2012}
{Yaron} O., {Gal-Yam} A., 2012, \pasp, 124, 668

\bibitem[{{Yoshida} {et~al}\mbox{.}(2017){Yoshida}, {Suwa}, {Umeda}, {Shibata},
  \& {Takahashi}}]{Yoshida2017}
{Yoshida} T., {Suwa} Y., {Umeda} H., {Shibata} M., {Takahashi} K., 2017, ArXiv
  e-prints

\bibitem[{{Zampieri} {et~al}\mbox{.}(2003){Zampieri}, {Pastorello}, {Turatto},
  {Cappellaro}, {Benetti}, {Altavilla}, {Mazzali}, \& {Hamuy}}]{Zampieri2003}
{Zampieri} L., {Pastorello} A., {Turatto} M., {Cappellaro} E., {Benetti} S.,
  {Altavilla} G., {Mazzali} P., {Hamuy} M., 2003, \mnras, 338, 711

\end{thebibliography}

\end{document}